\RequirePackage[T1]{fontenc}
\documentclass[12pt]{article}

\usepackage[height=8.85in,width=6.25in]{geometry}

\usepackage[utf8]{inputenc}
\usepackage{amsmath}
\usepackage{amssymb}
\usepackage{mathtools}
\numberwithin{equation}{section}
\usepackage{slashed}
\usepackage{braket}
\usepackage[svgnames]{xcolor}
\usepackage[colorlinks,citecolor=DarkGreen,linkcolor=FireBrick,urlcolor=FireBrick,linktocpage,breaklinks=true]{hyperref}
\usepackage{cite}
\usepackage[pdftex]{graphicx}
\usepackage{tikz}
\usepackage{tikz-cd}
\usepackage{times}
\usepackage{courier}
\usepackage{bm}
\usepackage{subfig}

\usepackage{xcolor}
\usepackage{mdframed}

\usetikzlibrary{decorations.pathreplacing,decorations.markings,calc}

\usepackage{mathrsfs}

\usetikzlibrary{arrows.meta}
\newcommand{\midarrow}{\tikz[baseline] \draw[-{Stealth[scale=1.5]}] (0,0) -- +(.1,0);}
\newcommand{\opmidarrow}{\tikz[baseline] \draw[{Stealth[scale=1.5]}-] (0,0) -- +(.1,0);}

\newcommand{\dmidarrow}{\tikz[baseline]  \draw[-{{Stealth}[scale=1.25]{Stealth}[scale=1.25]}] (0,0) -- +(.1,0);}

\usetikzlibrary{scopes,intersections}

\newcommand{\Coordinate}[2]%
{ \coordinate (#1) at (#2);
}

\tikzset{
  on each segment/.style={
    decorate,
    decoration={
      show path construction,
      moveto code={},
      lineto code={
        \path [#1]
        (\tikzinputsegmentfirst) -- (\tikzinputsegmentlast);
      },
      curveto code={
        \path [#1] (\tikzinputsegmentfirst)
        .. controls
        (\tikzinputsegmentsupporta) and (\tikzinputsegmentsupportb)
        ..
        (\tikzinputsegmentlast);
      },
      closepath code={
        \path [#1]
        (\tikzinputsegmentfirst) -- (\tikzinputsegmentlast);
      },
    },
  },
  mid arrow/.style={postaction={decorate,decoration={
        markings,
        mark=at position .5 with {\arrow[#1]{stealth}}
      }}},
}

\tikzset{line/.style={line width=0.25mm},
curve/.style={line,smooth,tension=1},
->-/.style={decoration={
  markings,
  mark=at position #1 with {\arrow[>={Stealth[scale=1]}]{>}}},postaction={decorate}},
-<-/.style={decoration={
  markings,
  mark=at position #1 with {\arrow[>={Stealth[scale=1]}]{<}}},postaction={decorate}},
}

\renewenvironment{figure}[1][]{
  \begin{originalfigure}[#1]
    \begin{mdframed}[linecolor=black!0,backgroundcolor=black!1]
}{
    \end{mdframed}
  \end{originalfigure}
}
\newcommand{\ie}{\begin{equation}\begin{aligned}}
\newcommand{\fe}{\end{aligned}\end{equation}}

\definecolor{dgreen}{rgb}{0, 0.55, 0}
\definecolor{grey}{rgb}{0.9,0.9,0.9}
\definecolor{dgrey}{rgb}{0.3,0.3,0.3}

\def\ZZ{\mathbb{Z}}
\def\CC{\mathbb{C}}

\newcommand{\bea}{\begin{eqnarray}}
\newcommand{\eea}{\end{eqnarray}}
\def\no{\nonumber}

\def\cO{\mathcal{O}}
\def\C{\mathsf{C}}
\def\P{\mathsf{P}}
\def\cH{\mathcal{H}}
\def\mfH{\mathfrak{H}}

\def\cP{\mathcal{P}}
\def\cC{\mathcal{C}}
\def\cCP{\mathcal{CP}}

\def\cc{\mathsf{c}}

\def\eps{\epsilon}
\def\half{{1\over 2}}

\def\cC{\mathcal{C}}
\def\cA{\mathcal{A}}

\def\T{\mathsf{T}}
\def\S{\mathsf{S}}
\usepackage{dsfont}

\newcommand{\ex}[1]{\mathrm{e}^{#1}}
\newcommand{\fr}{\frac}
\newcommand{\pa}[1]{\left(#1 \right)}
\newcommand{\ca}[1]{\mathcal{#1}}
\newcommand{\bb}[1]{\mathbb{#1}}

\newcommand{\kett}[1]{ \ket{#1}\!\rangle }
\newcommand{\bdy}[1]{ {\ket{#1}} }
\newcommand{\ydb}[1]{ {\bra{#1}} }

\def\tr{{\mathrm{Tr}}}

\begin{document}

\begin{titlepage}
\thispagestyle{empty}

\begin{flushright}
KYUSHU-HET-332
\\
RIKEN-iTHEMS-Report-25
\\

\end{flushright}

\bigskip

\begin{center}
\noindent{{\large \textbf{
New Crosscap States
}}}\\
\vspace{2cm}
Wataru Harada ${}^{1}$,
Justin Kaidi ${}^{1,2}$,
Yuya Kusuki ${}^{1,2,3}$,
Yuefeng Liu ${}^{2}$
\vspace{1cm}

${}^{1}${\small \sl 
Department of Physics, \\
Kyushu University, Fukuoka 819-0395, Japan
}

${}^{2}${\small \sl 
Institute for Advanced Study, \\
Kyushu University, Fukuoka 819-0395, Japan
}

${}^{3}${\small \sl RIKEN Interdisciplinary Theoretical and Mathematical Sciences (iTHEMS), \\Wako, Saitama 351-0198, Japan}

\vskip 2em
\end{center}

\begin{abstract}
We investigate crosscap states in two-dimensional rational conformal field theories (RCFTs), with an emphasis on the role of non-invertible symmetries. In particular, we argue for the existence of crosscap states labelled by each Verlinde line in the RCFT, extending previous constructions involving simple currents. Evidence for the existence of these new states is obtained by deriving a generalized Cardy condition incorporating both crosscaps and topological defects, which we check in some concrete examples. Finally, we briefly discuss how these crosscap states transform under the action of Verlinde lines, as well as the connection to mixed anomalies between parity and internal symmetries. 
 \end{abstract}

\end{titlepage}

\restoregeometry

\setcounter{tocdepth}{2}
\tableofcontents

\section{Introduction}

In the study of quantum many-body systems, symmetries play a pivotal role. The modern understanding of symmetry as the set of topological defects in a theory has led to various extensions of the notion of symmetry, notably including higher form \cite{Gaiotto2014} and non-invertible \cite{Bhardwaj:2017xup,Chang2019} symmetries. These developments have significantly advanced the non-perturbative understanding of quantum many-body systems.

Amid this progress, certain key structures remain relatively underexplored. One such example is the \textit{crosscap}, which appears naturally in the construction of non-orientable surfaces such as the Klein bottle. Crosscaps have a long history of study in conformal field theory and string theory \cite{Sagnotti1987, Ishibashi:1988kg, Fioravanti:1993hf,Sagnotti:1995ga,Huiszoon:2000ge,Fuchs:2000cm,Brunner:2002em,Gato-Rivera:2002whf,Fuchs:2003id}, but their interplay with non-invertible symmetries has not yet been addressed in the literature. In this brief note, we aim to fill this gap. In particular, we argue for the existence of new crosscap states $|C_a\rangle$ labelled by arbitrary Verlinde lines $a$ of an RCFT, and show that they satisfy a generalized Cardy-like condition, derived using the techniques of topological defect lines. The action of Verlinde lines on the crosscap states is also obtained, and the connection to parity anomalies is discussed.\footnote{Another recent, but orthogonal, discussion of parity anomalies can be found in \cite{Pace:2025rfu}.} Along the way, we discuss some general features of topological defects on unorientable manifolds, leaving a more complete analysis to future work. 

This note is organized as follows. First, in Section \ref{sec:review}, we review various properties of topological defects, Cardy states, and crosscap states in RCFT. 
We also discuss the interplay between parity and internal symmetries. 
In Section \ref{sec:main} we derive a generalized Cardy condition relating the inner product of crosscap states with a sum over twisted Klein bottle partition functions. 
In the invertible case, this result (mostly) reduces to one known in the literature. 
We also derive the action of Verlinde lines on the crosscap states, and briefly discuss the relation to parity anomalies.
In Section \ref{sec:examples}, we close with some concrete examples. 
Finally, we include two appendices: in Appendix \ref{app:modular} we review some of the modular properties of characters in RCFT, while in Appendix \ref{app:computation} we give the derivation of a simple formula used in Section \ref{sec:main}. 

%%%%%%%%%%%%%%%%%%%%%%%%%%%%%%%%%%%%%%%%%%%%%%%%%%%%%%%%%%%%%%%%%%%%%%%%%%%%%%%%%%%%%%%%%%%%%%
%%%%%%%%%%%%%%%%%%%%%%%%%%%%%%%%%%%%%%%%%%%%%%%%%%%%%%%%%%%%%%%%%%%%%%%%%%%%%%%%%%%%%%%%%%%%%%
\section{Boundaries, Crosscaps, and Topological Defects}
\label{sec:review}
%%%%%%%%%%%%%%%%%%%%%%%%%%%%%%%%%%%%%%%%%%%%%%%%%%%%%%%%%%%%%%%%%%%%%%%%%%%%%%%%%%%%%%%%%%%%%%
%%%%%%%%%%%%%%%%%%%%%%%%%%%%%%%%%%%%%%%%%%%%%%%%%%%%%%%%%%%%%%%%%%%%%%%%%%%%%%%%%%%%%%%%%%%%%%

We begin in this section with a review of some basic results about boundaries, crosscaps, and topological defects. Throughout this note we will consider a CFT with chiral algebra $\cA \otimes \cA$ and the charge conjugation modular invariant, whose Hilbert space has the following form,
\bea
\label{eq:Hilbertspacedef}
\cH_1 = \bigoplus_a \mfH_a \otimes \mfH_{a^+}~, 
\eea
where $\mfH_a$ is the module associated with $a$, and $a^+$ is the charge conjugate of $a$.\footnote{
In some references, the charge conjugation invariant is referred to as the ``diagonal'' invariant.
Here, to distinguish between the case of the anti-chiral representation being equal to the chiral representation or its charge conjugate, 
we call the former the diagonal invariant and the latter the charge conjugation invariant.
}

\subsection{Topological defects}

As is by now well-known, generalized symmetries in two dimensions may be represented by topological defect lines satisfying a number of elementary properties.
Consider a set of simple topological lines with the following fusion algebra,\footnote{A \textit{simple line} is one which admits no non-trivial topological operators on it. }
\bea
a \times b = \sum_c N_{ab}^c\, c~,
\eea
with the fusion coefficients $N_{ab}^c$ being non-negative integers.  We assume that the fusion rules are commutative, so that $N_{ab}^c = N_{ba}^c$. Every element $a$ has a unique orientation-reversal, or \textit{dual}, element $\bar a$ such that $N_{a \bar a}^1 = N_{\bar a a }^1 = 1$. 

We may define trivalent junctions of topological defect lines as follows,\footnote{Note that there is a difference between the \textit{dual} junction vector $\mu^\vee$, an element of $\mathrm{Hom}(\bar c, \bar a \times \bar b)$ obtained by rotating the original junction by 180 degrees, and the \textit{adjoint} junction vector $\bar \mu$, which is an element of $\mathrm{Hom}(c, a \times b)$ and gives positive-definite inner products with the original junction vector, see e.g. \cite{Huang:2021zvu}. Here we are working with the latter. }
\bea
\label{eq:Homspacebasis}
 \begin{tikzpicture}[baseline={([yshift=-1ex]current bounding box.center)},vertex/.style={anchor=base,
    circle,fill=black!25,minimum size=18pt,inner sep=2pt},scale=0.4]
   \draw[->-=0.5,thick] (-2,-2) -- (0,0);
   \draw[->-=0.5,thick] (2,-2) -- (0,0);
   \draw[->-=0.7,thick] (0,0)--(0,2) ;
     \node[below] at (-2,-2.2) {$a$};
     \node[below] at (2,-2) {$b$};
     \node[above] at (0,2) {$c$};
     \node[above] at (0,-0.2) {\footnotesize$\times$};
     \node[right] at (0,0) {\scriptsize$\mu$};
    \end{tikzpicture} \in\, \mathrm{Hom}(a\times b, c)~, \hspace{0.4 in}
     \begin{tikzpicture}[baseline={([yshift=-1ex]current bounding box.center)},vertex/.style={anchor=base,
    circle,fill=black!25,minimum size=18pt,inner sep=2pt},scale=0.4]
   \draw[->-=0.6,thick] (0,-2) -- (0,0);
   \draw[->-=0.6,thick]  (0,0) -- (2,2);
   \draw[->-=0.6,thick] (0,0)--(-2,2) ;
     \node[above] at (2,2) {$b$};
     \node[above] at (-2,2) {$a$};
     \node[below] at (0,-2) {$c$};
       \node[right] at (0,-0.2) {\scriptsize$\overline \mu$};
      \node[] at (0,-0.4) {\footnotesize$\times$};
    \end{tikzpicture}
    \in\, \mathrm{Hom}(c,a\times b)~,
\eea
which take values in the complex vector spaces $\mathrm{Hom}(a\times b, c)$ and $\mathrm{Hom}(c, a\times b)$ of dimension $N_{ab}^c$. 
We mark one of the lines by an $\times$ to make it explicit which hom-space the junction belongs to, although it is already implicit in the direction of the arrows. Basis vectors of the hom-spaces are chosen such that they satisfy the following completeness and orthogonality relations, 
\bea
\label{eq:basisconventions}
\begin{tikzpicture}[baseline={([yshift=-1ex]current bounding box.center)},vertex/.style={anchor=base,
    circle,fill=black!25,minimum size=18pt,inner sep=2pt},scale=0.4]
   \draw[->-=0.5,thick] (0,0) -- (0,4);
   \draw[->-=0.5,thick] (2,0) -- (2,4);
     \node[below] at (0,-0.2) {$a$};
     \node[below] at (2,0) {$b$};
    \end{tikzpicture} 
    =\sum_{c} \sum_{\mu=1}^{N^c_{ab}} \sqrt{d_c \over d_a d_b} \,\,\,
    \begin{tikzpicture}[baseline={([yshift=-1ex]current bounding box.center)},vertex/.style={anchor=base,
    circle,fill=black!25,minimum size=18pt,inner sep=2pt},scale=0.4]
   \draw[->-=0.7,thick] (0,0) -- (1,1);
   \draw[->-=0.7,thick] (2,0) -- (1,1);
    \draw[->-=0.6,thick] (1,1) -- (1,3);
     \draw[->-=0.7,thick] (1,3) -- (0,4);
       \draw[->-=0.7,thick] (1,3) -- (2,4);
     \node[below] at (0,-0.2) {$a$};
     \node[below] at (2,0) {$b$};
      \node[above] at (0,4) {$a$};
     \node[above] at (2,4) {$b$};
      \node[right] at (1,2) {$c$};
      \node[above] at (1,0.6) {\footnotesize$\times$};
      \node[below] at (1,3.4) {\footnotesize$\times$};
      
      \node[left] at (1,1.1) {\scriptsize$\mu$};
       \node[left] at (1,2.8) {\scriptsize$\overline\mu$};
    \end{tikzpicture} ~, 
    \hspace{0.5 in}
\begin{tikzpicture}[baseline={([yshift=-2ex]current bounding box.center)},vertex/.style={anchor=base,
    circle,fill=black!25,minimum size=18pt,inner sep=2pt},scale=0.4]
  
    \draw[thick] (0,0) to [out = 180, in = 180,distance = 1.2 cm]  node[rotate=90]{\midarrow} (0,2);
     \draw[thick] (0,0) to [out = 0, in = 0,distance=1.2 cm]  node[rotate=90]{\midarrow}(0,2);
   \draw[thick, ->-=0.5] (0,-1.5) -- (0,0); 
   \draw[thick, ->-=0.9] (0,2) -- (0,3.5);

    \node[left] at (-0.9,1) {\footnotesize $a$};
        \node[right] at (0.9,1) {\footnotesize $ b$};
    \node[below] at (0,-1.5) {\footnotesize $c$};
    \node[above] at (0,3.5) {\footnotesize $d$};
      \node[below] at (0,0.3) {\footnotesize$\times$};
       \node[above] at (0,1.7) {\footnotesize$\times$};
       
       \node[left] at (0,2.2) {\scriptsize$\mu$};
       \node[left] at (0,-0.3) {\scriptsize$\overline\nu$};
      
\end{tikzpicture}
\quad=\delta_{cd} \delta_{\mu \nu} \sqrt{{d_a d_b \over d_c}} \,\, 
\begin{tikzpicture}[baseline={([yshift=-1ex]current bounding box.center)},vertex/.style={anchor=base,
    circle,fill=black!25,minimum size=18pt,inner sep=2pt},scale=0.4]
  
   \draw[thick, ->-=0.5] (0,-1.5) -- (0,3.5); 
    \node[below] at (0,-1.5) {\footnotesize $c$};
    
\end{tikzpicture}~,
\eea
where $d_a$ is the quantum dimension of the line $a$. Our conventions for the F- and G-symbols are as follows, 
\bea
   \begin{tikzpicture}[baseline={([yshift=-1ex]current bounding box.center)},vertex/.style={anchor=base,
    circle,fill=black!25,minimum size=18pt,inner sep=2pt},scale=0.4]
   \draw[->-=0.2,->-=0.6,->-=0.95,thick] (0,0) -- (4,4);
   \draw[->-=0.7,thick] (2.6,0) -- (1.3,1.3);
    \draw[->-=0.6,thick] (6,0) -- (3,3);
          \node[below] at (0,-0.2) {$a$};
     \node[below] at (3,0) {$b$};
      \node[below] at (6,-0.2) {$c$};
     \node[above] at (4,4) {$d$};
      \node[right] at (2,2) {$e$};
     \node at (1.65,1.65) {\footnotesize$+$};
      \node at (3.3,3.3) {\footnotesize$+$};
      \node[left] at (1.3,1.3) {\scriptsize$\mu$};
       \node[left] at (3,3) {\scriptsize$\nu$};
    \end{tikzpicture} 
    = 
    \sum_{f} \sum_{\sigma = 1}^{N_{bc}^f} \sum_{\rho = 1}^{N_{af}^d} (F_{abc}^d)_{(e\mu\nu)(f \rho\sigma)}\,\,\,
       \begin{tikzpicture}[baseline={([yshift=-1ex]current bounding box.center)},vertex/.style={anchor=base,
    circle,fill=black!25,minimum size=18pt,inner sep=2pt},scale=0.4]
   \draw[->-=0.45,->-=0.95,thick] (0,0) -- (4,4);
   \draw[->-=0.6,thick] (3,0) -- (4.5,1.5);
    \draw[->-=0.3,->-=0.8,thick] (6,0) -- (3,3);
          \node[below] at (0,-0.2) {$a$};
     \node[below] at (3,0) {$b$};
      \node[below] at (6,-0.2) {$c$};
     \node[above] at (4,4) {$d$};
      \node[right] at (2.5,1.8) {$f$};
     \node at (4.2,1.8) {\footnotesize$+$};
      \node at (3.3,3.3) {\footnotesize$+$};
      \node[left] at (3,3) {\scriptsize $\rho$};
      \node[right] at (4.4,1.5) {\scriptsize $\sigma$};
    \end{tikzpicture} ~,
    \\
       \begin{tikzpicture}[baseline={([yshift=-1ex]current bounding box.center)},vertex/.style={anchor=base,
    circle,fill=black!25,minimum size=18pt,inner sep=2pt},scale=0.5]
   \draw[->-=0.2,->-=0.6,->-=0.95,thick] (4,0) -- (0,4);
   \draw[->-=0.7,thick] (2.5,1.5) -- (5,4);
   \draw[->-=0.7,thick] (1,3) -- (2,4);
   \node[below] at (4,0) {$d$};
    \node[above] at (0,4) {$a$};
    \node[above] at (2,4) {$b$};
    \node[above] at (5,4) {$c$};
      \node at (2.8,1.2) {\footnotesize$+$};
      \node at (1.3,2.7) {\footnotesize$+$};
        \node[left] at (2.5,1.3) {\scriptsize $\overline\nu$};
      \node[left] at (1,2.8) {\scriptsize $\overline\mu$};
       \node[right] at (1.7,2.6) {$e$};
           \end{tikzpicture} 
    = 
    \sum_{f} \sum_{\sigma = 1}^{N_{bc}^f} \sum_{\rho = 1}^{N_{af}^d} (G^{abc}_d)_{(e\overline\mu\overline\nu)(f \overline\rho\overline\sigma)}\,\,\,
       \begin{tikzpicture}[baseline={([yshift=-1ex]current bounding box.center)},vertex/.style={anchor=base,
    circle,fill=black!25,minimum size=18pt,inner sep=2pt},scale=0.5]
   \draw[->-=0.2,->-=0.6,->-=0.95,thick] (4,0) -- (0,4);
   \draw[->-=0.8,->-=0.2,thick] (2.5,1.5) -- (5,4);
   \draw[->-=0.7,,thick] (3.5,2.5) -- (2,4);
   \node[below] at (4,0) {$d$};
    \node[above] at (0,4) {$a$};
    \node[above] at (2,4) {$b$};
    \node[above] at (5,4) {$c$};
      \node at (2.8,1.2) {\footnotesize$+$};
      \node at (3.3,2.3) {\footnotesize$+$};
        \node[left] at (2.5,1.3) {\scriptsize $\overline\rho$};
      \node[right] at (3.5,2.5) {\scriptsize $\overline\sigma$};
       \node[right] at (2,2.4) {$f$};
           \end{tikzpicture} ~,
\eea
which are related by 
\bea
(G^{abc}_d)_{(e\overline\mu\overline\nu)(f \overline\rho\overline\sigma)} = (F_{abc}^d)_{(f \rho\sigma)(e\mu\nu)}^{-1}~.
\eea

One is always free to make a redefinition of the hom-space basis vectors via unitary matrices, 
\bea
\label{eq:junctiongaugetrans}
 \begin{tikzpicture}[baseline={([yshift=-1ex]current bounding box.center)},vertex/.style={anchor=base,
    circle,fill=black!25,minimum size=18pt,inner sep=2pt},scale=0.4]
   \draw[->-=0.5,thick] (-2,-2) -- (0,0);
   \draw[->-=0.5,thick] (2,-2) -- (0,0);
   \draw[->-=0.7,thick] (0,0)--(0,2) ;
     \node[below] at (-2,-2.2) {$a$};
     \node[below] at (2,-2) {$b$};
     \node[above] at (0,2) {$c$};
     \node[above] at (0,-0.2) {\footnotesize$\times$};
     \node[right] at (0,0) {\scriptsize$\mu$};
    \end{tikzpicture} \longrightarrow \sum_{\nu=1}^{N_{ab}^c}\,\, (U_{ab}^c)_{\mu \nu}
     \begin{tikzpicture}[baseline={([yshift=-1ex]current bounding box.center)},vertex/.style={anchor=base,
    circle,fill=black!25,minimum size=18pt,inner sep=2pt},scale=0.4]
   \draw[->-=0.5,thick] (-2,-2) -- (0,0);
   \draw[->-=0.5,thick] (2,-2) -- (0,0);
   \draw[->-=0.7,thick] (0,0)--(0,2) ;
     \node[below] at (-2,-2.2) {$a$};
     \node[below] at (2,-2) {$b$};
     \node[above] at (0,2) {$c$};
     \node[above] at (0,-0.2) {\footnotesize$\times$};
     \node[right] at (0,0) {\scriptsize$\nu$};
    \end{tikzpicture}, \hspace{0.4 in}
     \begin{tikzpicture}[baseline={([yshift=-1ex]current bounding box.center)},vertex/.style={anchor=base,
    circle,fill=black!25,minimum size=18pt,inner sep=2pt},scale=0.4]
   \draw[->-=0.6,thick] (0,-2) -- (0,0);
   \draw[->-=0.6,thick]  (0,0) -- (2,2);
   \draw[->-=0.6,thick] (0,0)--(-2,2) ;
     \node[above] at (2,2) {$a$};
     \node[above] at (-2,2) {$b$};
     \node[below] at (0,-2) {$c$};
       \node[right] at (0,-0.2) {\scriptsize$\overline \mu$};
      \node[] at (0,-0.4) {\footnotesize$\times$};
    \end{tikzpicture}
   \longrightarrow \sum_{\nu=1}^{N_{ab}^c}\,\, (V_c^{ab})_{\overline\mu \overline\nu}
     \begin{tikzpicture}[baseline={([yshift=-1ex]current bounding box.center)},vertex/.style={anchor=base,
    circle,fill=black!25,minimum size=18pt,inner sep=2pt},scale=0.4]
   \draw[->-=0.6,thick] (0,-2) -- (0,0);
   \draw[->-=0.6,thick]  (0,0) -- (2,2);
   \draw[->-=0.6,thick] (0,0)--(-2,2) ;
     \node[above] at (2,2) {$a$};
     \node[above] at (-2,2) {$b$};
     \node[below] at (0,-2) {$c$};
       \node[right] at (0,-0.2) {\scriptsize$\overline \nu$};
      \node[] at (0,-0.4) {\footnotesize$\times$};
    \end{tikzpicture}.\hspace{0.2 in}
\eea
We take $(V_c^{ab})_{\overline\mu \overline\nu} = (U_{ab}^c)^{-1}_{\nu \mu}$ such that the conventions in (\ref{eq:basisconventions}) remain unchanged. 
Upon such a gauge transformation, the F-symbols are modified via 
\bea
\label{eq:Fgaugeredun}
 (F_{abc}^d)_{(e\mu\nu)(f \rho\sigma)}\rightarrow  \sum_{\alpha = 1}^{N_{ab}^e} \sum_{\beta = 1}^{N_{ec}^d} \sum_{\gamma=1}^{N_{af}^d} \sum_{\delta = 1}^{N_{bc}^f}  (U_{ab}^e)^{-1}_{\mu \alpha} (U_{ec}^d)^{-1}_{\nu \beta}  (F_{abc}^d)_{(e\alpha\beta)(f \gamma\delta)} (U_{af}^d)_{\gamma \rho} (U_{bc}^f)_{\delta \sigma} ~,
\eea
and hence the F-symbols themselves are not, in general, gauge invariant.

Of particular interest to us in this paper are trivalent junctions involving $a$, $\bar a$, and the identity. For a given orientation of the vertex, there is always a unique such junction up to gauge transformation, i.e. $N_{a \bar a}^1 = N_{{ \bar a} a}^1 = 1$. As follows from (\ref{eq:basisconventions}) and $d_a = d_{ \bar a}$, we have 
\bea
\begin{tikzpicture}[baseline={([yshift=-0.5ex]current bounding box.center)},vertex/.style={anchor=base,
    circle,fill=black!25,minimum size=18pt,inner sep=2pt},scale=0.4]
  
    \draw[thick] (0,0) to [out = 180, in = 180,distance = 1.25 cm] node[rotate=90]{\midarrow} (0,2);
     \draw[thick] (0,0) to [out = 0, in = 0,distance=1.25 cm] node[rotate=90]{\midarrow} (0,2);
   \draw[thick, dashed] (0,-1) -- (0,0); 
   \draw[thick, dashed] (0,2) -- (0,3); 

  \filldraw  (0,0) circle (2pt);
  \filldraw  (0,2) circle (2pt);
  
    \node[left] at (-1,1.1) {\footnotesize $ \bar a$};
        \node[right] at (1,1) {\footnotesize $ a$};
       
\end{tikzpicture} = \,\,\, d_a ~,
\eea
and using this together with an F-move shows that
\bea
\label{eq:FSisotopydef}
\begin{tikzpicture}[scale=0.8,baseline=20]
    \begin{scope}[ thick, every node/.style={sloped,allow upside down}]
         \draw[ postaction={decorate}, decoration={markings,
      % place a > arrow at halfway (0.5) along the path
      mark=at position 0.4 with {\arrow[scale=0.9]{Stealth}}, mark=at position 0.99 with {\arrow[scale=0.9]{Stealth[reversed]}} }] (-1,-0.5+0.5) .. controls (-1,1.8+0.5) and (0,1.4+0.5) .. (0,0.5+0.5);
      \draw[ postaction={decorate}, decoration={markings,
      % place a > arrow at halfway (0.5) along the path
      mark=at position 0.7 with {\arrow[scale=0.9]{Stealth}} }] (0,0.5+0.5) .. controls (0,-0.3+0.5) and (1,-0.8+0.5) .. (1,1.6+0.5);
      \filldraw  (-0.45,1.21+0.5) circle (1pt);
      \filldraw  (0.45,-0.15+0.5) circle (1pt);
        \node[ below] at (-1,-0.4+0.5) {$a$};
        \node[ right] at (-0.1,0.9+0.5) {$\bar{a}$};
        \node[right] at (1,1.5+0.5) {$a$};
        
       \draw[dashed] (0.46,0.2)   -- (0.46,0.2-0.5)  ;
       \draw[dashed] (-0.46,1.8)   -- (-0.46,1.8+0.5)  ;
    \end{scope}
\end{tikzpicture}
\hspace{0.1 in}= \quad d_a \, (F_{a \bar a a}^a)_{11} \hspace{0.1 in} 
\begin{tikzpicture}[scale=0.8,baseline=20]
\begin{scope}[ thick, every node/.style={sloped,allow upside down}]
   \draw(0,-0.7+0.5) -- node{\midarrow} (0,1.5+0.5);
   \node[right] at (0,0) {$a$};
\end{scope}
\end{tikzpicture}~.
\eea
As usual, this coefficient is subject to a gauge redundancy---in particular, (\ref{eq:Fgaugeredun}) simplifies in this case to 
\bea
\label{eq:simplifiedFtrans}
(F_{a \bar a a}^a)_{11} \rightarrow {U_{ \bar a a}^1 \over U_{a  \bar a }^1} (F_{a \bar a a }^a)_{11} ~. 
\eea
When $a\neq  \bar a$, one may always do a gauge transformation to set this F-symbol to $d_a^{-1}$. 
On the other hand, when $a = \bar a$, this particular F-symbol is a gauge-invariant quantity which we denote by $\kappa_a := d_a \, (F_{a \bar a a}^a)_{11}$. The quantity $\kappa_a$ is known as the \textit{Frobenius-Schur (FS) indicator} of $a$, and can be shown to take values $\pm 1$, 
\bea
\label{eq:FSdeftop}
\kappa_a = \left\{ \begin{matrix} \pm 1 & & a =  \bar a \\ 1 && a \neq \bar a\end{matrix} \right.~. 
\eea
Useful discussions of the Frobenius-Schur indicator can be found in e.g. \cite{Kitaev:2005hzj,Simon:2022ohj}. 

An example of topological defect lines are Verlinde lines in an RCFT, which will be the focus of the rest of this note. In this case, the dual lines are given by the charge conjugate lines, i.e. $\bar a = a^+$, and the action on states of the untwisted Hilbert space $\cH_1$ in (\ref{eq:Hilbertspacedef}) is given by
\bea
\label{eq:invertiblegactions}
\widehat b\, |a, a^+\rangle  = {\S_{ba} \over \S_{1a}}\, | a, a^+\rangle~,
\eea
where $\S$ is the modular S-matrix. For example, in the invertible case we have $\S_{ga} / \S_{1a}= \ex{2 \pi i Q_g(a)}$, where $Q_g(a) := h_a + h_g - h_{g a}$ mod $1$ and $h_i$ are the respective conformal dimensions. 

More generally, the Verlinde line $b$ maps an operator $\cO$ in the $a$-twisted Hilbert space to one in the Hilbert space $\bigoplus_{x \prec b a \bar b}\cH_x$, and each projection of this action onto a fixed Hilbert space $\cH_x$ is given by the following lasso diagram,
\bea
\label{eq:lassodiagram}
\begin{tikzpicture}[scale=0.7,baseline=15]
\begin{scope}[ thick, every node/.style={sloped,allow upside down}]
%\draw[->-=0.25, ->- = 0.55, ->- = 0.9](0,0) -- (0,3);
\draw[->- = 0.7](0,0) -- (0,1);
\draw[->-=0.7] (0,1) -- (0,2);
\draw[->- = 0.8](0,2) -- (0,3);
\draw(0,1) to[out=0,in=0,distance=0.6in] node{\opmidarrow}  (0,-1);
\draw  (0,-1) to[out=180,in=180,distance=0.8in] node{\opmidarrow}   (0,2);
\node[above] at (0,3) {$x$};
\node[left] at (-1.3,0) {$b$};
\node[right] at (0,1.4) {$c$};
\node[right] at (0.1,0.4) {$a$};
\node[below] at (0,0) {$\footnotesize\cO$};
\node[] at (0,1.2) {\scriptsize$\times$};
\node[] at (0,1.8) {\scriptsize$\times$};

\filldraw (0,0) circle (0.3ex);
\filldraw (0,1) circle (0.2ex);
\filldraw (0,2) circle (0.2ex);

\node[left] at (0,1) {\scriptsize $\mu$};
\node[right] at (0,2) {\scriptsize $\overline \nu$};
\end{scope}
\end{tikzpicture}
\hspace{0.2 in}=\hspace{0.3 in}
\begin{tikzpicture}[scale=0.6,baseline=15]
\begin{scope}[ thick, every node/.style={sloped,allow upside down}]
%\draw[->-=0.25, ->- = 0.55, ->- = 0.9](0,0) -- (0,3);
\draw[->- = 0.7](0,0) -- (0,3);
\filldraw (0,0) circle (0.3ex);
\node[above] at  (0,3) {$x$};
\node[below] at (0,0) {$\footnotesize\cO'$};
\end{scope}
\end{tikzpicture}\hspace{0.4 in},
\eea
where $\cO'$ is an operator in the $x$-twisted sector, and our conventions are that $b$ encircles the operator counterclockwise. For every choice of internal line $c$ and junction vectors $\mu$, $\overline \nu$, we obtain a different such action.

\subsubsection{Generalized symmetries and parity}
\label{sec:parityanomaly}

Let us now define the following discrete symmetries, 
\bea
\label{eq:CPCPactions}
&\cC:& \mfH_a \otimes \mfH_b \rightarrow \mfH_{a^+} \otimes \mfH_{b^+} ~, \hspace{0.6in} x \rightarrow x
\no\\
&\cP:& \mfH_a \otimes \mfH_b \rightarrow \mfH_{b} \otimes \mfH_{a} ~, \hspace{0.76in} x \rightarrow -x
\no\\
&\cCP:& \mfH_a \otimes \mfH_b \rightarrow \mfH_{b^+} \otimes \mfH_{a^+}  ~, \hspace{0.6in} x \rightarrow -x~
\eea
where $\cC$ is charge conjugation and $\cP$ is spacetime parity along the $x$-axis. From this we easily compute the action of $\cC$, $\cP$, and $\cCP$ on the $a$-twisted Hilbert spaces,
\bea
\label{eq:Hilbertspaceaction}
\cC: \cH_a \rightarrow \cH_{a^+}~, \hspace{0.5 in}\cP: \cH_a \rightarrow \cH_{a}~, \hspace{0.5 in}\cCP: \cH_a \rightarrow \cH_{a^+}~.
\eea
For example, since the twisted Hilbert space $\cH_a = \bigoplus_{b,c} N_{bc}^a\, \mfH_b \otimes \mfH_{c}$, we have
\bea
\cP :\,\,\, \cH_a \longrightarrow \bigoplus_{b,c} N_{bc}^a\,  \mfH_{c} \otimes \mfH_b= \bigoplus_{b,c} N_{cb}^a\,  \mfH_{c} \otimes \mfH_b= \cH_a~,
\eea
where we have used that $N_{bc}^a =N_{cb}^{a}$. Of course, this does not mean that the action of $\cP$ on any individual state in $\cH_a$ is trivial; the precise action will be discussed further below.

We may also consider commutators involving $a$ and the discrete symmetries, with the results
\bea
\label{eq:CPCPcomms}
\cC\, \widehat a = \widehat a^+\, \cC~, \hspace{0.5in} \cP\, \widehat a =\widehat a^+\, \cP~, \hspace{0.5 in} \cCP\, \widehat a = \widehat a\, \cCP~,
\eea
when acting on the untwisted Hilbert space.\footnote{
The commutators between $\cP$ and $a$ depend on whether one considers the charge conjugation invariant  
or the diagonal invariant.  
If one is interested in the diagonal invariant, one can replace $\cP$ with $\cC\cP$ in the following discussion.
}
These follow from the actions given in (\ref{eq:invertiblegactions}) and (\ref{eq:CPCPactions}), e.g.
\bea
\cP\, \widehat a \,|b, b^+\rangle = {\S_{a b} \over \S_{1b}}\,\kappa_b |b^+, b \rangle = {\S_{a^+ b^+} \over \S_{1b^+}}\,\kappa_b|b^+, b \rangle= \widehat a^+ \,\cP\, |b, b^+ \rangle ~,
\eea
where $\kappa_b$ is the coefficient arising from the action of $\cP$ on $|b,b^+\rangle$, to be discussed further momentarily, 
and we have noted that $\S_{a^+b^+}= \S_{ab}$. 
Let us emphasize that, as we see above, the action of the discrete symmetries is different depending on whether $a$ is treated as a defect or an operator.  

%Indeed, upon changing our quantization to swap between $a$ being a defect or an operator,  $\cP$ and $ \cC\cP$ are interchanged, so we expect that the action of $\cP$ (resp. $\cC \cP$) on the operator $a$ is the same as the action of $\cC \cP$ (resp. $\cP$) on the defect $a$, which is what we have observed above.

In order to understand meshes of defects on unorientable manifolds, we will need to understand the action of parity on trivalent junctions. 
Say that we have 
\bea
\label{eq:firstPeq}
\cP: \hspace{0.05in}
 \begin{tikzpicture}[baseline={([yshift=-1ex]current bounding box.center)},vertex/.style={anchor=base,
    circle,fill=black!25,minimum size=18pt,inner sep=2pt},scale=0.35]
   \draw[->-=0.5,thick] (-2,-2) -- (0,0);
   \draw[->-=0.5,thick] (2,-2) -- (0,0);
   \draw[->-=0.7,thick] (0,0)--(0,2) ;
     \node[below] at (-2,-2.2) {$a$};
     \node[below] at (2,-2) {$b$};
     \node[above] at (0,2) {$c$};
     \node[above] at (0,-0.2) {\footnotesize$\times$};
     \node[right] at (0,0) {\scriptsize$\mu$};
    \end{tikzpicture} \longrightarrow \sum_{\nu=1}^{N_{ab}^c}\,\, (P_{ab}^c)_{\mu \nu}
     \begin{tikzpicture}[baseline={([yshift=-1ex]current bounding box.center)},vertex/.style={anchor=base,
    circle,fill=black!25,minimum size=18pt,inner sep=2pt},scale=0.35]
   \draw[->-=0.5,thick] (-2,-2) -- (0,0);
   \draw[->-=0.5,thick] (2,-2) -- (0,0);
   \draw[->-=0.7,thick] (0,0)--(0,2) ;
     \node[below] at (-2,-2) {$b$};
     \node[below] at (2,-2.2) {$a$};
     \node[above] at (0,2) {$c$};
     \node[above] at (0,-0.2) {\footnotesize$\times$};
     \node[right] at (0,0) {\scriptsize$\nu$};
    \end{tikzpicture}~,
\hspace{0.2 in}\begin{tikzpicture}[baseline={([yshift=-1ex]current bounding box.center)},vertex/.style={anchor=base,
    circle,fill=black!25,minimum size=18pt,inner sep=2pt},scale=0.35]
   \draw[->-=0.6,thick] (0,-2) -- (0,0);
   \draw[->-=0.6,thick]  (0,0) -- (2,2);
   \draw[->-=0.6,thick] (0,0)--(-2,2) ;
     \node[above] at (2,2) {$b$};
     \node[above] at (-2,2) {$a$};
     \node[below] at (0,-2) {$c$};
       \node[right] at (0,-0.2) {\scriptsize$\overline \mu$};
      \node[] at (0,-0.4) {\footnotesize$\times$};
    \end{tikzpicture} \longrightarrow  \sum_{\nu=1}^{N_{ab}^c}\,\, (P_c^{ab})_{\overline \mu \overline \nu}
       \begin{tikzpicture}[baseline={([yshift=-1ex]current bounding box.center)},vertex/.style={anchor=base,
    circle,fill=black!25,minimum size=18pt,inner sep=2pt},scale=0.35]
   \draw[->-=0.6,thick] (0,-2) -- (0,0);
   \draw[->-=0.6,thick]  (0,0) -- (2,2);
   \draw[->-=0.6,thick] (0,0)--(-2,2) ;
     \node[above] at (2,2) {$a$};
     \node[above] at (-2,2) {$b$};
     \node[below] at (0,-2) {$c$};
       \node[right] at (0,-0.2) {\scriptsize$\overline \nu$};
      \node[] at (0,-0.4) {\footnotesize$\times$};
    \end{tikzpicture}
\eea
where parity has mapped an element of the vector space $\mathrm{Hom}(a \otimes b, c)$ to an element of $\mathrm{Hom}(b \otimes a, c)$. Consistency with (\ref{eq:basisconventions}) requires that $(P_c^{ab})_{\overline \mu \overline \nu} = (P_{ab}^c)^{-1}_{ \nu\mu}$, so we focus on just $(P_{ab}^c)_{\mu \nu}$ for now. These coefficients are subject to a gauge redundancy, since shifting as in (\ref{eq:junctiongaugetrans}) changes 
\bea
(P_{ab}^c)_{\mu \nu} \,\,\longrightarrow\,\, \sum_{\rho,\sigma=1}^{N_{ab}^c}(U_{ab}^c)^{-1}_{\mu \rho} (P_{ab}^c)_{\rho \sigma}(U_{ba}^c)_{\sigma \nu}~.
\eea
Since $(P_{ab}^c)_{\mu \nu}$ is itself a unitary matrix, for generic $b \neq a$ we may in principle work in a gauge such that $(P_{ab}^c)_{\mu \nu} = \delta_{\mu \nu}$. One must, however, be careful to note that these gauge transformations may also lead to shifts in the F-symbols---for example, in the case of $b= a^+$ and $c=1$, we have
\bea
\label{eq:easyPtransf}
P_{a a^+}^1 \,\,\longrightarrow\,\,{U_{ a^+ a}^1 \over U_{a a^+}^1} P_{a a^+}^1 ~, 
\eea
and we note that the ratio $U_{ a^+ a}^1 / U_{a a^+}^1$ was already fixed in (\ref{eq:simplifiedFtrans}) when we chose our convention that $(F_{a a^+a}^a)_{11} = \kappa_a / d_a$.  As such, the action of $\cP$ on such junctions is already fixed in our conventions---namely to $P_{a a^+}^1 = \kappa_a$, as will be shown below.  Let us emphasize that this does \textit{not} mean that $P_{a a^+}^1$ must be a gauge invariant piece of data. For $a\neq  a^+$ it can be modified by basis change, but at the cost of changing our conventions for  $(F_{a a^+a }^a)_{11}$.\footnote{For $a =  a^+$ it \textit{is} gauge invariant, as discussed in the following paragraph.} In explicit examples below, we will always choose our gauge such that the F-symbols match with standard expressions in the literature, for which we may not necessarily have $(P_{ab}^c)_{\mu \nu} = \delta_{\mu \nu}$. 
 
 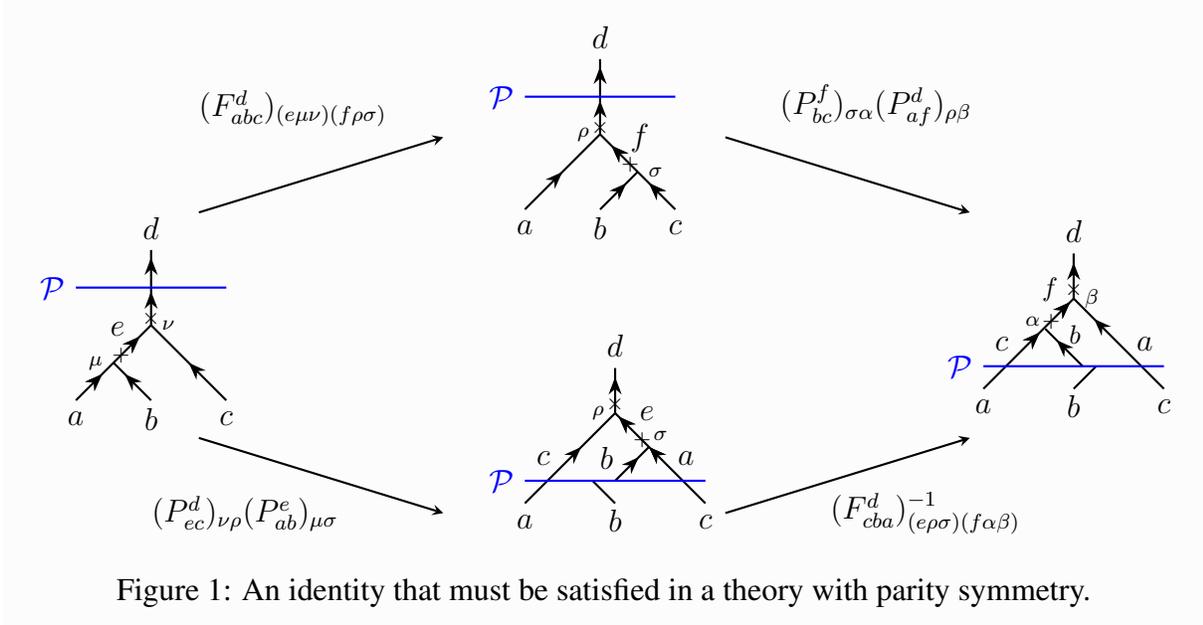
\begin{figure}
\begin{tikzpicture}[baseline=0,scale = 0.5, baseline=-10]
\begin{scope}[xshift=-0.5in]
\draw [ thick,->-=0.35,->-=0.85] (0,-1) -- (2,1);
\draw [ thick,->-=0.7] (2,-1) --(1,0);
\draw [ thick,->-=0.5] (4,-1) -- (2,1);
\draw [ thick,->-=0.9] (2,1) -- (2,2);
\draw [ thick,->-=0.8] (2,2) -- (2,3);
\draw [ thick,blue] (0,2) to (4,2);
\node[below] at (0,-1) {$a$};
\node[below] at (2,-0.9) {$b$};
\node[below] at (4,-1) {$c$};
\node[above] at (2,3) {$d$};
\node[left] at (1.6,0.9) {$e$};
\node[left] at (1,0) {\scriptsize $\mu$};
\node[right] at (2,1) {\scriptsize $\nu$};
\node[blue, left] at (0,2) {$\cP$};
\node[] at (1.2,0.2) {\scriptsize $+$};
\node[] at (2,1.17) {\scriptsize $\times$};
\end{scope}

\begin{scope}[xshift=4.2in, yshift = 2in]
\draw [ thick,->-=0.5] (0,-1) -- (2,1);
\draw [ thick,->-=0.7] (2,-1) --(3,0);
\draw [ thick,->-=0.35,->-=0.85] (4,-1) -- (2,1);
\draw [ thick,->-=0.9] (2,1) -- (2,2);
\draw [ thick,->-=0.8] (2,2) -- (2,3);
\draw [ thick,blue] (0,2) to (4,2);
\node[below] at (0,-1) {$a$};
\node[below] at (2,-0.9) {$b$};
\node[below] at (4,-1) {$c$};
\node[above] at (2,3) {$d$};
\node[left] at (3.6,0.9) {$f$};
\node[right] at (3,0) {\scriptsize $\sigma$};
\node[left] at (2,1) {\scriptsize $\rho$};
\node[blue, left] at (0,2) {$\cP$};

\node[] at (2.8,0.2) {\scriptsize $+$};
\node[] at (2,1.17) {\scriptsize $\times$};
\end{scope}

\begin{scope}[xshift=9in, yshift = 0.2in,scale=1.2]
\draw [ thick] (0,-1) -- (0.5,-0.5);
\draw [ thick] (2,-1) --(2.5,-0.5);
\draw [ thick] (4,-1) -- (3.5,-0.5);

\draw [ thick,->-=0.5,->-=0.95] (0.5,-0.5) -- (2,1);
\draw [ thick,->-=0.7] (2.2,-0.5) -- (1.37,0.33);
\draw [ thick,->-=0.7] (3.5,-0.5) -- (2,1);

\draw [ thick,->-=0.8] (2,1) -- (2,2);
\draw [ thick,blue] (0,-0.5) to (4,-0.5);
\node[below] at (0,-1) {$a$};
\node[below] at (2,-0.9) {$b$};
\node[below] at (4,-1) {$c$};
\node[above] at (2,2) {$d$};
\node[left] at (1.9,1.2) {\footnotesize $f$};
\node[left] at (1.5,0.5) {\scriptsize $\alpha$};
\node[right] at (2,1) {\scriptsize $\beta$};

\node[left] at (0.8,0) { $c$};
\node[left] at (2.4,0.2) {\footnotesize $b$};
\node[left] at (4,0) { $a$};
\node[blue, left] at (0,-0.5) {$\cP$};

\node[] at (1.5,0.5) {\scriptsize $+$};
\node[] at (2,1.2) {\scriptsize $\times$};
\end{scope}

\begin{scope}[xshift=4.2in, yshift = -1in,scale=1.2]
\draw [ thick] (0,-1) -- (0.5,-0.5);
\draw [ thick] (2,-1) --(1.5,-0.5);
\draw [ thick] (4,-1) -- (3.5,-0.5);

\draw [ thick,->-=0.5] (0.5,-0.5) -- (2,1);
\draw [ thick,->-=0.7] (2,-0.5) -- (2.75,0.25);
\draw [ thick,->-=0.5,->-=0.95] (3.5,-0.5) -- (2,1);

\draw [ thick,->-=0.8] (2,1) -- (2,2);
\draw [ thick,blue] (0,-0.5) to (4,-0.5);
\node[below] at (0,-1) {$a$};
\node[below] at (2,-0.9) {$b$};
\node[below] at (4,-1) {$c$};
\node[above] at (2,2) {$d$};
\node[right] at (2.3,1) {$e$};

\node[left] at (3.4,0.5) {\scriptsize $\sigma$};
\node[left] at (2,1) {\scriptsize $\rho$};

\node[left] at (0.8,0) { $c$};
\node[left] at (2.2,0) { $b$};
\node[left] at (4,0) { $a$};
\node[blue, left] at (0,-0.5) {$\cP$};

\node[] at (2.6,0.4) {\scriptsize $+$};
\node[] at (2,1.2) {\scriptsize $\times$};

\end{scope}

\draw[thick,-stealth] (2,-2) -- (8.5,-4);

\draw[thick,-stealth] (16,-4) -- (6.5+16,-2);
\draw[thick,-stealth] (2,4) -- (8.5,6);
\draw[thick,-stealth] (16,6) -- (6.5+16,4);

\node[left] at (6, -4) {$(P_{ec}^d)_{\nu \rho}(P_{ab}^e)_{\mu \sigma}$};
\node[right] at (18.5, -4) {$(F_{cba}^d)^{-1}_{(e \rho \sigma)(f \alpha \beta)}$};
\node[above] at (4.5, 6) {$(F_{abc}^d)_{(e \mu \nu)(f \rho \sigma)}$};
\node[above] at (20, 6) {$(P_{bc}^f)_{\sigma \alpha} (P_{af}^d)_{\rho \beta}$};

\end{tikzpicture}
\caption{An identity that must be satisfied in a theory with parity symmetry.}
 \label{fig:PFidentity}
\end{figure}

On the other hand, when $b=a$, the gauge transformation acts as a similarity transformation on $(P_{aa}^c)_{\mu\nu}$. This means that the eigenvalues of $(P_{aa}^c)_{\mu\nu}$ \textit{are} gauge invariant quantities. Since $\cP$ is an involution, these eigenvalues should simply be signs, 
and these signs may be constrained by the coherence condition given in Figure \ref{fig:PFidentity}, explicitly 
\bea
\sum_{\rho =1}^{N_{ec}^d}\sum_{\sigma = 1}^{N_{ab}^e} (P_{ec}^d)_{\nu \rho} (P_{ab}^e)_{\mu\sigma}(F_{cba}^d)^{-1}_{(e \rho \sigma)(f \alpha \beta)} = \sum_{\rho=1}^{N_{af}^d}\sum_{\sigma=1}^{N_{bc}^f} (F_{abc}^d)_{(e \mu \nu)(f \rho \sigma)} (P_{bc}^f)_{\sigma \alpha} (P_{af}^d)_{\rho \beta}~.
\label{eq:PFidentity}
\eea

Since the coefficients $P_{ab}^c$ describe the non-trivial action of parity on trivalent junctions of the categorical symmetry, these coefficients (or rather the gauge invariant portions of them) capture mixed anomalies between $\cP$ and the categorical symmetry. As will be described below, one way to measure these anomalies is to act with an element $a$ on the so-called ``crosscap state'' $|C_1\rangle$, which produces gauge-invariant factors of $\mathrm{Tr}(P_{aa}^b)$ for appropriate $b$. This method was already hinted at in the invertible case in \cite{Cho2015}, and will be developed in more detail in the current work.

Let us mention that in some cases (\ref{eq:PFidentity}) can be used to solve for $(P_{aa}^c)_{\mu\nu}$ uniquely, while in other cases it cannot. 
A more thorough analysis of the full set of conditions satisfied by $P_{ab}^c$, and of non-invertible symmetries on unorientable manifolds more generally, will be given in future work. We should also mention that other, more complete discussions of non-invertible symmetries on unorientable manifolds have appeared in works such as \cite{Fuchs:2003id,Kapustin:2015uma,Bhardwaj:2016dtk,Inamura:2021wuo}, albeit in a slightly different language. 
 In the current note, we content ourselves with the bare minimum ingredients needed to understand our new crosscap states.

%%%%%%%%%%%%%%%%%%%%%%%%%%%%%%%%%%%%%%%%%%%%%%%%%%%%%%%%%%%%%%%%%%%%%%%%%%%%%%%%%%%%%%%%%%%%%%
\subsection{Boundary states and crosscaps}
%%%%%%%%%%%%%%%%%%%%%%%%%%%%%%%%%%%%%%%%%%%%%%%%%%%%%%%%%%%%%%%%%%%%%%%%%%%%%%%%%%%%%%%%%%%%%%

\subsubsection{Cardy states}
 We next review boundary and crosscap states.  Conformal boundaries in 2d are defined by the following boundary conditions on the stress tensor,
\begin{equation}\label{eq:T=T}
\left. \left(T(z) - \overline{T}(\bar{z})\right)\right|_{\text{bdy}} = 0~.
\end{equation}
Although boundaries generally break conformal symmetry, conformal boundaries preserve $\mathrm{Vir}\subset \mathrm{Vir}\otimes \mathrm{Vir}$.
The mode decomposition of the above boundary condition is expressed as follows,
\begin{equation}\label{eq:L=L}
(L_n - \overline{L}_{-n})| {B}\rangle =0~,
\end{equation}
and if the theory has a symmetry $\ca{A} \otimes \ca{A}$ generated by $W^{(r)}$,
we further impose the following conditions,
\begin{equation}\label{eq:W=W}
\pa{W^{(r)}_n-(-1)^{s_r} \overline{W^{(r)}}_{-n}}\bdy{B}=0~,
\end{equation}
where $s_r$ is the spin of the current $W^{(r)}$ and $W^{(r)}_n$ are its modes, which ensures the preservation of the diagonal $\cA \subset \cA \otimes \cA$.

The solutions to the boundary condition (\ref{eq:W=W}) can be written as linear combinations of Ishibashi states \cite{Ishibashi:1988kg},
\begin{equation}
\kett{a} := \sum_N \ket{a,N} \otimes U\ket{a,N}~,
\end{equation}
where we have denoted an orthonormal basis for $\mfH_a$ by $\ket{a,N}$, and the operation $U: \mfH_a \to \mfH_{a^+}$ acts on the generators as
\begin{equation}
U W_n^{(r)}U^{-1} = (-1)^{s_r} {W^{(r)}_{-n}}^\dagger~.
\end{equation}

Physical boundary states are constrained by the following bootstrap equation, called the {\it Cardy condition} \cite{Cardy2004},
\begin{equation}\label{eq:open}
\ydb{B_a}\ex{-\fr{\pi i}{\tau}\pa{L_0 + \overline{L}_0 -\fr{\cc}{12} }} \bdy{B_b}=\tr_{\ca{H}_{a;b}} \ex{2\pi i \tau \pa{L_0-\fr{\cc}{24}}}~,
\end{equation}
where $\mathcal{H}_{a;b}$ is the Hilbert space of states defined on a segment with boundary conditions $a$ and $b$ at its left and right ends, respectively,
and $\cc$ is the central charge.
Since the conformal boundaries preserve the diagonal symmetry $\ca{A}$,
the Hilbert space $\ca{H}_{a;b}$ can be split into a sum of irreducible representations of $\ca{A}$,
\begin{equation}
\ca{H}_{a;b} := \bigoplus_c n_{ab}^c\, \mfH_c~,
\end{equation}
where $n_{ab}^c$ are non-negative integers.
Denoting the boundary states by $\bdy{B_a} = \sum_b B_a^b\kett{b} $, the Cardy condition can be rewritten in terms of the characters $\chi_a(\tau)$ as follows,
\begin{equation}
 \sum_c \pa{B_a^c}^* B_b^c\, \chi_c \pa{-\fr{1}{\tau}}=\sum_c n_{ab}^c\, \chi_c (\tau) ~.
\end{equation}
The requirement that $n_{a b}^c $ be an integer imposes strong constraints on the coefficients $B_a^b$.
In particular, in RCFT we may use the modular S transformation
\begin{equation}
\chi_a\pa{-\fr{1}{\tau}} = \sum_b \S_{ab} \chi_b(\tau)~
\end{equation}
to rewrite the Cardy condition as,
\begin{equation}\label{eq:open2}
n_{a b}^d = \sum_c (B_a^c)^* B_b^c\, \S_{cd} \in \bb{Z}_{\geq0}~.
\end{equation}
The solutions to the Cardy condition are known as {\it Cardy states} \cite{Cardy2004}, and are given by
\begin{equation}\label{eq:CardyState}
\bdy{B_a} := \sum_b \fr{\S_{a b}}{\sqrt{\S_{1b}}}\kett{b}~,
\end{equation}
where the labels $a,b,\dots$ run over all Verlinde lines. 
That they are solutions follows from identifying $n_{ab}^c  = N_{ab^+}^c$ and making use of the Verlinde formula \cite{Verlinde1988},
\begin{equation}\label{eq:Verlinde}
N_{ab}^c= \sum_i \fr{\S_{a i} \S_{b i} {\S_{ci}}^*   }{\S_{1i}} ~.
\end{equation}
In addition, the boundary conditions should satisfy the Cardy-Lewellen sewing constraints \cite{Lewellen1992}, which, however, will not be particularly important in the discussion below.

Since any positive integer linear combination of boundary states also satisfies the bootstrap equations,
there exist infinitely many boundary states in a single theory.
Among them, we refer to states $\bdy{B_a}$ as \textit{simple} if $\ca{H}_{a;a}$ has a unique vacuum.\footnote{A simple boundary state is also referred to in the literature as \textit{fundamental} or \textit{elementary}.}
The Cardy states (\ref{eq:CardyState}) form a set of simple boundaries \cite{Fuchs:2002cm}.

%%%%%%%%%%%%%%%%%%%%%%%%%%%%%%%%%%%%%%%%%%%%%%%%%%%%%%%%%%%%%%%%%%%%%%%%%%%%%%%%%%%%%%%%%%%%%%
\subsubsection{Crosscap States}
%%%%%%%%%%%%%%%%%%%%%%%%%%%%%%%%%%%%%%%%%%%%%%%%%%%%%%%%%%%%%%%%%%%%%%%%%%%%%%%%%%%%%%%%%%%%%%
\label{sec:crosscapsubsec}

A \textit{crosscap} is an object obtained by excising a disc and identifying antipodal points, 
which we represent as a disc with a cross mark, 
\bea
\begin{tikzpicture}[scale=0.6,baseline=0,block/.style={white, rectangle, draw, fill=white}]
\begin{scope}[very thick, every node/.style={sloped,allow upside down}]

\draw[black] (0,0) circle (50 pt);

\draw[blue,stealth-stealth] (-1.27,-1.27) -- (1.27,1.27);
\draw[red,stealth-stealth] (-1.27,1.27) -- (1.27,-1.27);
\node[block] at (0,0) {};
\node[] at (0,0) {Identify};
\end{scope}
\end{tikzpicture}
\hspace{0.3 in}\Leftrightarrow \hspace{0.3 in}
  \begin{tikzpicture}[font=\sffamily\small,very thick,scale=0.7,baseline=20]

  (0,0) -- +(2.5,0) -- +(2.5,2.5) -- +(0,2.5) -- cycle;
  \coordinate (X) at ($(  35:1.5cm  and 0.5cm)$);
  \coordinate (Y) at ($(-135:1.5cm  and 0.5cm)$);
  \coordinate (A) at ($( -45:1.5cm  and 0.5cm)$);
  \coordinate (B) at ($( 145:1.5cm  and 0.5cm)$);
  %  \draw (-1.5cm,0) edge[out=90,in=90,looseness=2.5] (1.5cm,0);
  \draw[] (-1.5cm,0) -- (-1.5cm,2cm);
    \draw[] (1.5cm,0) -- (1.5cm,2cm);
   % \draw[blue,dashed] ($(  125:1.5cm  and 0.5cm)$) edge[out=90,in=180,looseness=0.55] (0,2.19);
    %\draw[blue] (0,2.19) edge[out=0,in=90,looseness=0.55] ($(-55:1.5cm  and 0.5cm)$) ;
    \draw[] (0,0) ellipse (1.5cm and 0.5cm);
    %($(  125:2cm  and 0.7cm)$);
    \draw[blue] (X)--(Y);
    \draw[red] (A)--(B);
  
\end{tikzpicture}\hspace{0.2 in}.
\eea
One may then obtain a \textit{crosscap state} by performing the path integral on a disc with the insertion of a crosscap, i.e. the path integral on $\bb{RP}^2$ with a disc removed.

In a similar way to conformal boundaries, the basic crosscap state $|C_1\rangle$ can be defined by the following conditions \cite{Ishibashi:1988kg},\footnote{
As will be seen later, different crosscaps can be defined depending on how the parity symmetry is implemented on the Hilbert space.
Among these, we will refer to the simplest one---where no additional symmetry (such as an outer automorphism of the current algebra) is involved---as the \textit{basic crosscap state}.
}
\bea
\label{eq:L=-L}
\pa{L_n - (-1)^n\overline{L}_{-n}}\bdy{C_1}&=&0~,
\no\\
\pa{W^{(r)}_n-(-1)^{s_r+n} \overline{W^{(r)}}_{-n}}\bdy{C_1}&=&0~,
\eea
which requires preservation of the diagonal symmetry. 
The solutions to the gluing condition (\ref{eq:L=-L}) are given by a linear combination of  crosscap Ishibashi states,
\begin{equation}
\kett{a}_C :=  \ex{\pi i \pa{L_0-h_a}}\kett{a}~,
\end{equation}
where we note that the inner products of $|a\rangle\!\rangle$ and $|a\rangle\!\rangle_C$ are given by 
\bea
\label{eq:Ishibashiinnerproducts}
\langle\!\langle a | \ex{2 \pi i \tau H} | b \rangle\!\rangle &=& \delta_{ab}\, \chi_a(2 \tau)~, 
\no\\
{}_C\langle\!\langle a | \ex{2 \pi i \tau H} | b \rangle\!\rangle_C &=& \delta_{ab}\, \chi_a(2 \tau)~, 
\no\\
\langle\!\langle a | \ex{2 \pi i \tau H} | b \rangle\!\rangle_C &=& \delta_{ab}\, \widehat \chi_a(2 \tau)~. 
\eea
The twisted characters $\widehat \chi_a(\tau)$ appearing in the final line are defined as follows, 
\bea
\label{eq:twistedchardef}
\widehat{\chi}_a(\tau) := \ex{-\pi i \pa{h_a-\fr{\cc}{24}}} \chi_a \pa{\tau+\fr{1}{2}}~.
\eea

Similar to boundaries, the amplitude between crosscap states is also subject to a constraint descending from modular invariance.\footnote{
Just as there is the Cardy-Lewellen sewing condition for boundaries,
there also exists a sewing condition for crosscaps \cite{Fioravanti:1993hf}.
Concretely, it is the conformal bootstrap equation for a two-point function on $\bb{RP}^2$.
This imposes a constraint that is independent of the modular bootstrap equations for the Klein bottle and M{\"o}bius strip.
In this brief note, however, we will not discuss this additional constraint,
and will focus only on the modular bootstrap equation.
}
In particular, as will be reviewed in Section \ref{sec:Kleinbottle}, the dual channel of the amplitude is given by the Klein bottle partition function.
This gives the following Cardy-like condition \cite{Sagnotti:1996eb},
\bea
\label{eq:Cardylikecond1}
  \langle C_1 | \ex{- {\pi i \over 2 \tau}\left( L_0 + \overline{L}_0 - {\mathsf{c}\over 12}\right)}|C_1\rangle = \mathrm{Tr}_\cH \, \cP \ex{2 \pi i \tau\left( L_0 + \overline{L}_0 - {\mathsf{c}\over 12}\right) }~,
\eea
where $\cP$ is the parity operation defined in (\ref{eq:CPCPactions}).
Alternatively, writing the crosscap state as $\bdy{C_1}=\sum_a \Gamma^a \kett{a}_C$,
the Cardy-like condition can be expanded in terms of characters as,
\begin{equation}
 \sum_a | \Gamma^a|^2\, \chi_a \pa{-\fr{1}{2\tau}} = \sum_a k_a \chi_a (2\tau)~,
\end{equation}
where the coefficients $k_a$ of the Klein bottle partition function are required to be integers.
Since the operator $\half (1 + \cP)$ acts as a projection operator onto parity-even states \cite{Sagnotti:1996eb}, we have
\begin{equation}\label{eq:int1}
\fr{h_{aa}+k_a}{2} \in \bb{Z}_{\geq0}~, \hspace{0.2 in} 0\leq\fr{h_{aa}+k_a}{2}  \leq h^{aa}~,
\end{equation}
where $h_{ab}$ is the multiplicity of the irreducible representation $\mfH_a \otimes \mfH_b$ in $\cH_1$.

Furthermore, the requirement of consistency between the crosscap and Cardy states gives rise to the following equation,
\begin{equation}\label{eq:Mobius}
\ydb{ B_a}\ex{-\fr{\pi i}{4\tau}\pa{L_0 + \overline{L}_0 -\fr{\cc}{12} }} \bdy{C_1}=\tr_{\ca{H}_{a;a^+}} \ \cP \ex{2\pi i \tau \pa{L_0-\fr{\cc}{24}}}
~,
\end{equation}
with the right-hand side being the M{\"o}bius strip amplitude.
Denoting by $m^b_a$ the coefficients of the M{\"o}bius strip in an expansion in terms of twisted characters (\ref{eq:twistedchardef}),
we can re-express (\ref{eq:Mobius}) as
\begin{equation}
\sum_b (B_a^b)^* \Gamma^b\, \widehat{\chi}_b \pa{-\fr{1}{4\tau}} = \sum_b m_a^b\, \widehat{\chi}_b (\tau) ~.
\end{equation}
Similar to (\ref{eq:int1}), the coefficients $m_a^b$ are constrained by
\begin{equation}\label{eq:int2}
\fr{N_{a a}^b+m_a^b}{2} \in \bb{Z}_{\geq0}~, \ \ \ \ \ 0 \leq  \fr{N_{a a}^b+m_a^b}{2}  \leq N_{a a}^b~.
\end{equation}

The consistency conditions (\ref{eq:int1}) and (\ref{eq:int2}) impose strong constraints on the coefficients $\Gamma^a$. One solution is to take $ \Gamma^a =\eta_1 {\P_{1a} / \sqrt{\S_{1a}}}$, so that
\begin{equation}
\label{eq:C1exp}
\bdy{C_1} := \eta_1 \sum_b \fr{\P_{1b}}{\sqrt{\S_{1b}}}\kett{b}_C~. 
\end{equation}
Here $\eta_1=\pm1 $ is a sign and $\P$ is the matrix capturing the modular transformation of the twisted characters,\footnote{The matrix $\P$ is not to be confused with the parity operator $\cP$. }
\begin{equation}
\label{eq:Pandchihat}
\widehat{\chi_a}\pa{-\fr{1}{4\tau}} = \sum_b \P_{ab}\, \widehat{\chi}_b(\tau)~, 
\end{equation}
which can be expressed in terms of the modular $\S$ and $\T$ matrices as
\begin{equation}
\P= \T^{\half} \S \T^2 \S \T^{\half}~,
\end{equation}
where we recall that $\T_{ab} := \delta_{ab} \ex{2\pi i \pa{h_a-\fr{\cc}{24}}}$ and we have $\T^{1\over 2}_{ab} := \delta_{ab} \ex{\pi i \pa{h_a-\fr{\cc}{24}}}$. As we will discuss below, there is an ambiguity in our definition of $\T^{1\over 2}$, which is related to the appearance of the sign $\eta_1$ above.

That (\ref{eq:C1exp}) solves the positivity and integrality constraints is a non-trivial fact. In particular, for this solution one finds that the Klein bottle coefficients $k_a$ are given by
\begin{equation}
\label{eq:FSindicatordef1}
k_a = \sum_b \fr{\S_{ab} \P_{1b} \P_{1b}}{\S_{1b}}~,
\end{equation}
which is the definition of the Frobenius-Schur (FS) indicator of the representation $\mfH_a$ introduced in the RCFT context in \cite{Bantay1997}. This definition of the FS indicator gives $1$ for real representations, $-1$ for pseudoreal representations, and $0$ for complex representations, which immediately implies that (\ref{eq:int1}) is satisfied \cite{Pradisi1995}. Note that $k_a$ defined here is slightly different from $\kappa_a$ defined in (\ref{eq:FSdeftop}). The two match for self-dual lines, but for non-self dual lines, the latter is set to 1, whereas the former is always zero. 

Additional crosscap states labelled by simple currents (i.e. invertible symmetries) were introduced in \cite{Huiszoon1999} and studied further in \cite{Fuchs2000a,Brunner:2002em}, with the explicit form 
\bea
\label{eq:simplecurrentcrosscaps}
\bdy{C_a} := \eta_a \sum_b \fr{\P_{ab}}{\sqrt{\S_{1b}}}\kett{b}_C~,
\eea
where $a$ labels a simple current and $\eta_a$ is a sign. Correlation functions between such states can again be interpreted as ``twisted'' Klein bottle partition functions \cite{Brunner:2002em},\footnote{This formula differs slightly from the one in \cite{Brunner:2002em} due to our choice of a different convention for the inner product in (\ref{eq:invinnerproduct}). 
The current definition will prove more convenient for later computations. Also, note that at first sight, (\ref{eq:scccamp}) may seem inconsistent with the commutation relation $\ca{P} \hat{b} = \hat{b}^+ \ca{P}$ in (\ref{eq:CPCPcomms}) for the untwisted sector, since using this relation and cyclicity of the trace we could obtain an equality of the form  $\mathrm{Tr}_{\cH_1} ( \cP \,\widehat b\dots) = \mathrm{Tr}_{\cH_1} ( \cP \,\widehat b^+\dots)$, which one might naively think is not always satisfied. However, inserting $\cP$ into the trace projects onto spin zero states, for which the action of $\widehat b$ and $\widehat b^+$ is indeed the same.}
\bea
\label{eq:scccamp}
\langle C_{b} | \ex{-{\pi i \over 2 \tau}(L_0 + \overline{L}_0 - {\mathsf{c}\over 12})}|C_{a}\rangle = P_a^{ab^+,b}\,\mathrm{Tr}_{\cH_{a b^+}} \cP \,\widehat b\, \ex{2 \pi i \tau (L_0 + \overline{L}_0 - {\mathsf{c}\over 12})}~,
\eea
where $\widehat b$ is the symmetry operator associated with $b$.
A modern understanding of this formula in terms of topological defects will be presented below---in particular, we will show that the crosscap $|C_a\rangle$ can be obtained from the basic crosscap $|C_1\rangle$ by wrapping the Verlinde line $a$ along the orientation-reversing cycle of the crosscap, pictorially,
\vspace{-0.2 in}\bea
\label{eq:crosscappictorialdef}
|C_a\rangle =  \,\,\, \begin{tikzpicture}[scale=0.6,baseline=0]
\begin{scope}[ thick, every node/.style={sloped,allow upside down}]
   \draw  (0,0) circle (3ex);
   \draw[thick] (-0.4,-0.4) -- (0.4,0.4);
   \draw[thick] (-0.4,0.4) -- (0.4,-0.4);
  
  \draw[blue,-<-=0.5] (0,0.566) to[out = 45, in = -45,distance = 1.3in] (0,-0.566);

     \node[blue, above] at (2.2,-0.2) {$a$};

\end{scope}
\end{tikzpicture}
\,\,\, = \,\, \kappa_a\,\,\, 
\begin{tikzpicture}[scale=0.6,baseline=0]
\begin{scope}[ thick, every node/.style={sloped,allow upside down}]
   \draw  (0,0) circle (3ex);
   \draw[thick] (-0.4,-0.4) -- (0.4,0.4);
   \draw[thick] (-0.4,0.4) -- (0.4,-0.4);
  
  \draw[blue,->-=0.5] (0,0.566) to[out = 45, in = -45,distance = 1.3in] (0,-0.566);

     \node[blue, above] at (2.4,-0.2) {$ a^+$};

\end{scope}
\end{tikzpicture}~.
\eea
The second equality follows by first using (\ref{eq:FSisotopydef}) to insert a segment of $a^+$ into the $a$ line, moving one of the $a$-$a^+$-$1$ vertices past the crosscap using $P_{a a^+}^1$, and then using the conjugate of (\ref{eq:FSisotopydef}) to obtain a line of only $a^+$. Each of these steps produces a factor of $\kappa_a$, giving the overall factor of $\kappa_a^3 = \kappa_a$. 

The expression for the crosscaps $|C_a\rangle$ in (\ref{eq:simplecurrentcrosscaps}) has been proven to satisfy the generalized Cardy condition (\ref{eq:scccamp}) when $a$ is a simple current \cite{Huiszoon1999,Fuchs2000a,Brunner:2002em}. However, we will give strong evidence (though not fully prove) that  (\ref{eq:simplecurrentcrosscaps})  gives legitimate crosscap states for non-invertible $a$ as well. In particular, in the non-invertible case there is an analogous generalized Cardy condition---derived in Section \ref{sec:crosscapnoninv}, with the final result given in (\ref{eq:generalizedCardylikeform})---where the right-hand side is now not a single (twisted) Klein bottle partition function, but rather a sum thereof, and we will prove that the expression in (\ref{eq:simplecurrentcrosscaps}) satisfies a subset of these generalized Cardy conditions (upon appropriate choice of signs $\eta_a$). The remaining subset will be checked in some concrete examples.

As before, the generalized Cardy-like conditions give rise to constraints on the coefficients appearing in the expansion of $|C_a\rangle$. In particular, we may compute
\bea
\langle C_{1} | \ex{-{\pi i \over 2 \tau}(L_0 + \overline{L}_0 - {\mathsf{c}\over 12})}|C_{a}\rangle = \eta_1 \eta_a \sum_{c} {\P_{1 c}\P_{a c} \over \S_{1c}} \chi_c\left(-{1\over 2\tau} \right) =   \eta_1 \eta_a \sum_{c} Y_{c 1}^{a} \,\chi_c\left(2 \tau \right) ~,
\eea
with
\bea
Y_{ab}^{c} := \sum_\ell {\S_{a\ell} \P_{b \ell} \P^*_{c \ell} \over \S_{1\ell} }~,
\eea
and where we note that we have used $\S^{-1}$ to obtain the third equality---the reason for using $\S^{-1}$ as opposed to $\S$ is related to our choice of the direction of $b$ in the lasso action (\ref{eq:lassodiagram}), as is explained in more detail in Appendix \ref{app:modular}. 
From this, we then see that the coefficients $Y_{c1}^{a}$ must be integers. Indeed, the $Y_{ab}^{c}$ are claimed to be integer for \textit{any} choice of indices \cite{Pradisi1995, Sagnotti:1996eb}. Furthermore, one has the equality $Y_{a1}^1 = k_a$, as well as Bantay's relations \cite{Bantay1997}, 
\bea
\label{eq:Bantayrel}
Y_{c1}^a = N_{cc}^a \,\,\,\mathrm{mod}\,\,2~, \hspace{0.3in} |Y_{c1}^a| \leq N_{cc}^a~. 
\eea
When $N_{cc}^a = 1$, these imply that $Y_{c1}^a$ is equal to $N_{cc}^a$ up to a sign. 

\subsubsection{Subtleties about signs}
Before moving on, an important point is that there is an ambiguity in the definition of $\T^{\half}$. In particular, given any one choice, we may obtain another legitimate $\T^{\half}$ by shifting $\T^{\half} \rightarrow \boldsymbol{\eta}\T^{\half} $, where each component $\eta_a$ of $\boldsymbol{\eta} = \mathrm{diag}(\eta_1, \eta_2, \dots)$ is a sign $\eta_a = \pm 1$. Shifting $\T^{\half}$ in this way leads to a shift $\P_{ab} \rightarrow \eta_a \eta_b \P_{ab}$, which preserves the symmetricity and unitarity of $\P$, and also preserves the property $\P^2 = \mathsf{C}$ (with $\mathsf{C}$ the charge conjugation matrix), so long as we choose $\eta_{a^+ } = \eta_a$. However, in order to preserve (\ref{eq:Pandchihat}), this change of sign must be compensated by $\widehat{\chi_a}(\tau) \rightarrow \eta_a \widehat{\chi_a}(\tau)$. Furthermore, by the third line of (\ref{eq:Ishibashiinnerproducts}), this must also be accompanied by $|a\rangle\!\rangle_C \rightarrow \eta_a |a\rangle\!\rangle_C$. Altogether, this leads to $|C_a \rangle \rightarrow \eta_a |C_a\rangle$. 

Of course, it is always possible for us to choose conventions for which we have $\T^{1\over 2}_{ab} = \delta_{ab} \ex{\pi i \pa{h_a-\fr{c}{24}}}$ with no additional signs, but this convention may lead to non-trivial signs $\eta_a$ in our definition of $|C_a\rangle$, as we have allowed for above. 
Indeed, the Cardy-like condition in (\ref{eq:scccamp}) can be used to show that some of these signs \textit{must} be non-trivial.
To see this, let us set $b=1$ and compute the right-hand side of (\ref{eq:scccamp}), or the right-hand side of (\ref{eq:generalizedCardylikeform}) in the non-invertible case (the two agree when $b=1$). 
This is done by simply noting that parity  $\cP$ acts on a state  $|b, c; \mu\rangle\in \cH_{a} = \bigoplus_{b,c} N_{b^+ a}^c \mfH_b \otimes \mfH_c$ as
\bea
\cP |b, c ; \mu\rangle = \sum_{\nu = 1}^{N_{bc}^{a}} (P_{bc}^{a})_{\mu \nu} | c, b; \nu\rangle ~,
\eea
where $ (P_{bc}^{a})_{\mu \nu} $ are the coefficients introduced in (\ref{eq:firstPeq}). 
The trace in the Klein bottle partition function then gives contributions only from $b=c$, which implies that $ a \prec c^2$.\footnote{The notation $z \prec x y$ means that $z$ is included in the fusion of $x$ and $y$ with non-zero coefficient $N_{xy}^z$. The fact that $a \prec c^2$ then follows from the fact that $1 \prec x  y$ if and only if $x =  y^+$. Indeed, since $b=c \prec c^+ a$, we have that $1 \prec c^+ c \prec (c^+)^2 a$, which in turn implies that $a^+ \prec (c^+)^2$. Taking the conjugate gives the desired result. } 
We thus conclude that 
\bea
\mathrm{Tr}_{\cH_{ a}} \cP \, \ex{2 \pi i \tau (L_0 + \overline{L}_0 - {\mathsf{c}\over 12})} = \sum_{c} \mathrm{Tr}P_{cc}^{a}\,   \chi_c(2 \tau)~,
\eea
where $\mathrm{Tr}P_{cc}^{a}$ is the trace of the $N_{cc}^{a} \times N_{cc}^{a}$ dimensional matrix $P_{cc}^{a}$; in particular, it is zero if $N_{cc}^{a} = 0$. 
On the other hand, the left-hand side of (\ref{eq:scccamp}) is given by
\bea
\langle C_{1} | \ex{-{\pi i \over 2 \tau}(L_0 + \overline{L}_0 - {\mathsf{c}\over 12})}|C_{a}\rangle = \eta_1 \eta_a  \sum_{c} Y^{a}_{c1}  \chi_c(2 \tau)~.
\eea
Comparing the two, we conclude that for any $a$ which is a ``perfect square'' (by which we mean that it is contained in the fusion of some element $c$ with itself), we have the following constraint,\footnote{Note that this also shows that $|Y_{c1}^{a}| = |\mathrm{Tr} P^{a}_{cc}|$. Since $P^{a}_{cc}$ is an $N_{cc}^{a} \times N_{cc}^{a} $ matrix with eigenvalues $\pm 1$, we clearly have that $|\mathrm{Tr} P^{a}_{cc}| \leq N_{cc}^{a}$, and this gives an alternative proof of Bantay's second relation (\ref{eq:Bantayrel}).}
\bea
\label{eq:etaconstraint}
\eta_a = \eta_1\, \mathrm{sgn} \left( Y_{c1}^{a}\mathrm{Tr}\, P_{cc}^{a} \right)~, \hspace{0.5 in}a \prec c^2~.
\eea

This result can be reproduced via an alternative computation---namely, we act with $c$ on the crosscap state $|C_1\rangle$. 
Using (\ref{eq:crosscappictorialdef}), this action can be computed pictorially as follows,

\bea
\begin{tikzpicture}[scale=0.6,baseline=0]
\begin{scope}[ thick, every node/.style={sloped,allow upside down}]
   \draw  (0,0) circle (3ex);
   \draw[thick] (-0.4,-0.4) -- (0.4,0.4);
   \draw[thick] (-0.4,0.4) -- (0.4,-0.4);
  
   \draw[blue,->-=0]  (0,0) circle (10ex);
   \node[blue, right] at (2,0) {$c$};
\end{scope}
\end{tikzpicture}
&=&
\begin{tikzpicture}[scale=0.6,baseline=0]
\begin{scope}[ thick, every node/.style={sloped,allow upside down}]
   \draw  (0,0) circle (3ex);
   \draw[thick] (-0.4,-0.4) -- (0.4,0.4);
   \draw[thick] (-0.4,0.4) -- (0.4,-0.4);

   \draw[blue,-<-=0.6] (0.95,0.8) to [out = -40, in = 40] (0.95,-0.8); 
   \draw[blue] (0.95,0.8) to [out = 180-40, in = 45] (0.4,0.4);
   \draw[blue] (0.95,-0.8) to [out = 180+40, in = -45] (0.4,-0.4);
   \draw[blue,->-=0.5] (0,-1.9) to [out=0,in=0,distance=1in] (0,1.9); 
   \draw[blue] (0,-1.9) to [out=180,in=225,distance=0.5in] (-0.4,-0.4); 
    \draw[blue] (0,1.9) to [out=180,in=135,distance=0.5in] (-0.4,0.4); 
   \node[blue, right] at (2,0) {$c$};
\end{scope}
\end{tikzpicture}
= 
\sum_{a }\sum_{\mu = 1}^{N_{cc}^{a}} \sqrt{d_a\over d_c^2}\begin{tikzpicture}[scale=0.6,baseline=0]
\begin{scope}[ thick, every node/.style={sloped,allow upside down}]
   \draw  (0,0) circle (3ex);
   \draw[thick] (-0.4,-0.4) -- (0.4,0.4);
   \draw[thick] (-0.4,0.4) -- (0.4,-0.4);
  
   \draw[blue,->-=0.5,-<-=0.15] (0.4,0.4) to[out = 45, in = 135,distance = 1.1in] (-0.4,0.4);
   \draw[blue,-<-=0.5,->-=0.15] (0.4,-0.4) to[out = -45, in = 225,distance = 1.1in] (-0.4,-0.4);
   
   \draw[blue,-<-=0.5] (0.85,1.5) to[out=20, in = -20, distance = 1 in] (0.85,-1.5);
   \node[blue, right] at (1.6,0) {$a$};
   
   \node[blue,left] at (-0.8,1.1) {$c$};
    \node[blue,left] at (-0.8,-1.1) {$c$};
    \node[above] at (0.87,1.5) {\scriptsize $ \bar\mu$};
    \node[below] at (0.85,-1.5) {\scriptsize $\mu$};
    \node[right,blue] at (0.8,1.6) {\footnotesize $\times$};
     \node[right,blue] at (0.8,-1.58) {\footnotesize $\times$};
\end{scope}
\end{tikzpicture}
\no\\
&\vphantom{.}& \hspace{-0.9 in}= \sum_{a }\sum_{\mu,\nu=1}^{N_{cc}^{a}} \sqrt{d_a\over d_c^2} (P_{cc}^{a})_{\mu \nu}  \,\,\, \begin{tikzpicture}[scale=0.6,baseline=0]
\begin{scope}[ thick, every node/.style={sloped,allow upside down}]
   \draw  (0,0) circle (3ex);
   \draw[thick] (-0.4,-0.4) -- (0.4,0.4);
   \draw[thick] (-0.4,0.4) -- (0.4,-0.4);
  
  \draw[blue,-<-=0.55] (0,0.566) to[out = 60, in = 120,distance = 0.6in] (2.6,0.566);
    \draw[blue,->-=0.55] (0,-0.566) to[out = -60, in = -120,distance = 0.6in]  (2.6,-0.566);
    \draw[blue] (2.6,0.566) to[out = 180, in = 180,distance = 0.3 in] node {\opmidarrow}(2.6,-0.566);
      \draw[blue] (2.6,0.566) to[out = 0, in = 0,distance = 0.3 in] node {\opmidarrow}(2.6,-0.566);

   \node[blue, above] at (1.3,1.6) {$a$};
     \node[blue, below] at (1.3,-1.6) {$a$};
   \node[blue,left] at (2.1,0) {$c$};
    \node[blue,right] at (3.1,0) {$c$};
    
    \node[below,blue] at (2.4,-0.4) {\footnotesize$+$};
    \node[above,blue] at (2.4,0.4) {\footnotesize$+$};
    
     \node[above] at (2.8,0.566) {\scriptsize $\nu$};
    \node[below] at (2.8,-0.566) {\scriptsize $ \bar\mu$};
\end{scope}
\end{tikzpicture}
= \sum_{a}  \mathrm{Tr}P_{cc}^{a} \,\,\, \begin{tikzpicture}[scale=0.6,baseline=0]
\begin{scope}[ thick, every node/.style={sloped,allow upside down}]
   \draw  (0,0) circle (3ex);
   \draw[thick] (-0.4,-0.4) -- (0.4,0.4);
   \draw[thick] (-0.4,0.4) -- (0.4,-0.4);
  
  \draw[blue,-<-=0.5] (0,0.566) to[out = 45, in = -45,distance = 1.3in] (0,-0.566);

     \node[blue, above] at (2.3,-0.2) {$a$};

\end{scope}
\end{tikzpicture}\,\,\,\,\,
\eea
where in the first step we have passed the line through the crosscap, in the second we have used the first equation of (\ref{eq:basisconventions}), in the third we have passed a vertex through the crosscap, and in the final equality we have made use of the second equation of (\ref{eq:basisconventions}). 
As such, we arrive at 
\bea
\label{eq:Paaxrel}
\widehat c\, |C_1 \rangle {=} \sum_{a} \mathrm{Tr} P_{cc}^{a} |C_a\rangle~, 
\eea
which gives the action of $c$ on the crosscap state in terms of the trace of $P_{cc}^{a}$. 

On the other hand, we may work out the action of $c$ on $|C_1 \rangle$ using the explicit form of the crosscap in terms of the crosscap Ishibashi states,
\bea
\widehat c\, | C_1 \rangle &=& \eta_1 \sum_b {\P_{1b} \over \sqrt{\S_{1b}}} \,\widehat c\, |b \rangle\!\rangle_C \,\,=\,\, \eta_1 \sum_b { \S_{cb} \P_{1b}\over {\S_{1b}}^{3/2}} \,|b \rangle\!\rangle_C 
\no\\
& =& \eta_1 \sum_{a,b,d} {\S_{cb} \P_{1b} \P_{ab}^* \over {\S_{1b}}}{\P_{ad} \over \sqrt{ \S_{1d}}} \,|d \rangle\!\rangle_C  \,\,=\,\,  \eta_1 \sum_a \eta_a Y_{c1}^{a} \,|C_a \rangle~.
\eea
Comparing the two then reproduces (\ref{eq:etaconstraint}).

One corollary of (\ref{eq:etaconstraint}) is that for $a=1$, which means that $c$ must be self-dual to have $a \prec c^2$, we obtain\footnote{Note that $N_{cc}^1=1$ by simplicity and self-duality of $c$, and that $k_c = \kappa_c$ when $c$ is self-dual.}
\bea
P_{cc}^1 = Y_{c1}^1 = \kappa_c~. 
\eea
This is consistent with the results described below (\ref{eq:easyPtransf}).
We may draw it as
\bea
\label{eq:PactionNequal1}
\cP: \hspace{0.1in}
 \begin{tikzpicture}[baseline={([yshift=-1ex]current bounding box.center)},vertex/.style={anchor=base,
    circle,fill=black!25,minimum size=18pt,inner sep=2pt},scale=0.4]
   \draw[->-=0.5,thick] (-2,-2) -- (0,0);
   \draw[->-=0.5,thick] (2,-2) -- (0,0);
   \draw[thick,dashed] (0,0)--(0,2) ;
     \node[below] at (-2,-2) {$c$};
     \node[below] at (2,-2) {$c$};
     \node[above] at (0,2) {$1$};
    \end{tikzpicture} \,\,\,\longrightarrow \,\,\, \kappa_c
     \begin{tikzpicture}[baseline={([yshift=-1ex]current bounding box.center)},vertex/.style={anchor=base,
    circle,fill=black!25,minimum size=18pt,inner sep=2pt},scale=0.4]
   \draw[->-=0.5,thick] (-2,-2) -- (0,0);
   \draw[->-=0.5,thick] (2,-2) -- (0,0);
   \draw[thick,dashed] (0,0)--(0,2) ;
     \node[below] at (-2,-2) {$c$};
     \node[below] at (2,-2) {$c$};
     \node[above] at (0,2) {$1$};
    \end{tikzpicture}~.
\eea

\subsection{Relation to parity anomalies}
\label{sec:parityanomalies}

In \cite{Cho2015}, it was suggested that the mixed anomaly between parity and a $\ZZ_2$ internal symmetry $g$ could be captured by the action of $g$ on the standard crosscap state $|C_1\rangle$. 
From (\ref{eq:Paaxrel}), we see that in an RCFT this action is given by
\bea
\widehat g |C_1 \rangle = \kappa_g |C_{g^2} \rangle=\kappa_g |C_{1} \rangle~.
\eea
That this is an anomaly follows from the fact that $\cP$ acts non-trivially on the trivalent junctions of $g$, as in (\ref{eq:PactionNequal1}). Said in an alternative way, since classically we have $g^2 = 1$, we see that when $\kappa_g = -1$, the multiplication rule is realized projectively on the crosscap states. 

An important point, however, is that for $\ZZ_2$ symmetries, $\kappa_g$ is non-trivial if and only if the symmetry is itself anomalous. In this case, one can always redefine the parity operation $\cP\rightarrow \hat{g} \cP$ such that the new parity operator has no mixed anomaly with $g$. From the anomaly inflow point of view, the inflow action for such a situation is\footnote{These are the only non-trivial generators of $\Omega_3^O(B\ZZ_2)$. Note that the other potential inflow terms $a \cup a \cup w_1$ and $w_1^3$ vanish. }
\bea
\label{eq:inflowaction}
\int a\cup a \cup a + \int a \cup w_1 \cup w_1 ~,
\eea
with $w_1$ the first Stiefel-Whitney class and $a$ the gauge field for the $\ZZ_2$ symmetry. From this, it is clear that one can remove the second, mixed anomaly term by redefining $a \rightarrow a + w_1$ (since $w_1^3 = 0$). This shift leads to a shift in the boundary background gauge field $A \rightarrow A + w_1$, which physically corresponds to wrapping a $g$ defect along all cycles with intersection number 1 with the orientation-reversing cycles---in other words, it corresponds to redefining $\cP\rightarrow \hat{g} \cP$. 
From the point of view of the crosscaps, this shift sends $|C_1 \rangle\rightarrow |C_g\rangle$ (potentially up to a sign). The general formula for the action of $a$ on a crosscap state $|C_b\rangle$ will be derived in Section \ref{sec:crosscapnoninv} and is given by (\ref{eq:actiononcc}), from which one straightforwardly obtains,
\bea
\widehat g |C_g \rangle =  |C_g \rangle~.
\eea
 Thus we see that on the minimal crosscap state constructed from $\hat g \cP$, the multiplication rule for $g$ is realized linearly, and there is indeed no anomaly.
To summarize then, when we fix a formula such as (\ref{eq:C1exp}) and refer to the result as the ``basic crosscap'', we are implicitly fixing a choice of $\cP$, and it is sensible to discuss whether this $\cP$ has an anomaly or not. But when there is a $\ZZ_2$ symmetry $g$, it is also possible to consider $\hat g\cP$, and this will always be anomaly-free. 

More generally, the action of an element $c$ on $|C_1\rangle$ is given by the sum in (\ref{eq:Paaxrel}), with coefficients given by $\mathrm{Tr}(P_{cc}^{a})$. 
As described before, the eigenvalues of $P_{ab}^c$ are gauge invariant for $b=a$, so the trace appearing here is a gauge-invariant quantity capturing a mixed anomaly between $\cP$ and $c$. In practice, one may often fix these quantities by solving (\ref{eq:PFidentity}), as is done for some concrete examples in Section \ref{sec:examples}. 
Since there are several distinct notions of anomaly for a non-invertible symmetry, see e.g. \cite{Choi:2023xjw}, here we should clarify that we are using the word to refer to the obstruction to gauging both symmetries simultaneously.

\section{Generalized Cardy Condition and Action of Verlinde Lines}
\label{sec:main}
In this section, we derive a generalized Cardy-like condition for crosscaps labelled by arbitrary elements of the fusion algebra. 
We do this by manipulating topological lines on the Klein bottle, beginning with the invertible case and then generalizing to the non-invertible case. 
We also describe the action of Verlinde lines on the crosscap states.

\subsection{Klein bottles, crosscaps, and invertible symmetries}
\label{sec:Kleinbottle}

In the current work we will focus on Klein bottles constructed from the parity operator $\cP$, drawn below\footnote{Note that many previous studies of topological lines on unorientable manifolds \cite{Kapustin:2015uma,Bhardwaj:2016dtk,Inamura:2021wuo} use $\cC \cP$ instead of $\cP$. In particular, $\cP$ does not commute with generic elements of the internal symmetry, and so in our case  the total symmetry is not necessarily a direct product of the internal symmetry and an orientation-reversing $\ZZ_2$. }

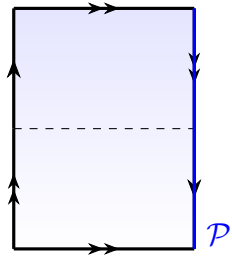
\begin{center}\begin{tikzpicture}[baseline=50,scale=0.8]
\begin{scope}[very thick, every node/.style={sloped,allow upside down}]
\shade[top color=blue, bottom color=white,opacity = 0.1] (0,0)--(3,0)--(3,4)--(0,4)--(0,0);
\draw  (0,4)--node {\dmidarrow}(3,4);

\draw(0,0) --node {\dmidarrow}  (3,0);

\draw (3,2)-- node {\midarrow}(3,0);
\draw (3,4)-- node {\dmidarrow}(3,2);

\draw (0,0)-- node {\dmidarrow}(0,2);
\draw (0,2)-- node {\midarrow}(0,4);
\draw[thin, dashed] (0,2) -- (3,2);
\draw[blue] (3,0)--(3,4);
\node[blue, right] at (3,0.25) {$\cP$};

\end{scope}
\end{tikzpicture}~.
\end{center}
The blue line represents the fact that, when traversing a cycle intersecting it an odd number of times, orientation is reversed. 
One should not think of it as a topological defect that can be deformed freely.

As is well known, the Klein bottle can be realized as a sphere with the insertion of two crosscaps. This may be seen via the following manipulations \cite{Blumenhagen:2009zz,Blumenhagen:2013fgp},
\bea
\label{eq:Kb}
\begin{tikzpicture}[scale=0.8,baseline=35]
\shade[top color=blue, bottom color=white,opacity = 0.1] (0,0)--(3,0)--(3,4)--(0,4)--(0,0);
\begin{scope}[very thick, every node/.style={sloped,allow upside down}]
\draw  (0,4)--node {\dmidarrow}(3,4);

\draw(0,0) --node {\dmidarrow}  (3,0);

\draw (3,2)-- node {\midarrow}(3,0);
\draw (3,4)-- node {\dmidarrow}(3,2);

\draw (0,0)-- node {\dmidarrow}(0,2);
\draw (0,2)-- node {\midarrow}(0,4);
\draw[thin, dashed] (0,2) -- (3,2);
%parity
\draw[blue] (3,0)--(3,4);
\end{scope}
\end{tikzpicture}
\hspace{0.1 in}=\hspace{0.1 in}
%center
\begin{tikzpicture}[scale=0.8,baseline=35]
\shade[top color=blue, bottom color=white,opacity = 0.1] (0,2)--(2.5,2)--(2.5,4)--(0,4)--(0,2);
\shade[top color=blue, bottom color=white,opacity = 0.1] (2.5,0)--(5,0)--(5,2)--(2.5,2)--(2.5,0);
\begin{scope}[very thick, every node/.style={sloped,allow upside down}]
\draw(2.5,0) -- node {\dmidarrow} (5,0);
\draw (5,0)-- node {\opmidarrow}(5,2);
\draw (2.5,4)-- node {\dmidarrow}(2.5,2);
\draw  (0,4)-- node {\dmidarrow}(2.5,4);
\draw (2.5,0)-- node {\dmidarrow}(2.5,2);
\draw (0,4)-- node {\opmidarrow}(0,2);
\draw(0,2) --node {\midarrow} (2.5,2);
\draw(2.5,2) --node {\midarrow} (5,2);
\draw[blue] (2.5,2)--(2.5,4);
\draw[blue] (5,0)--(5,2);
\end{scope}
\end{tikzpicture}
\hspace{0.1 in}=\hspace{0.1 in}
\begin{tikzpicture}[scale=0.8,baseline=20]
\begin{scope}[very thick, every node/.style={sloped,allow upside down}]
\shade[top color=blue, bottom color=white,opacity = 0.1] (0,0)--(5,0)--(5,2)--(0,2)--(0,0);
\draw(2.5,0) -- node {\midarrow} (5,0);
\draw (5,0)-- node {\midarrow}(5,2);
\draw  (0,0)-- node {\midarrow}(2.5,0);
\draw[thin,dashed] (2.5,2)--(2.5,0);
\draw (0,0)-- node {\midarrow}(0,2);
\draw(0,2) --node {\dmidarrow} (2.5,2);
\draw(2.5,2) --node {\dmidarrow} (5,2);
\draw[blue] (0,2) --(2.5,2);
\draw[blue] (0,0) --(2.5,0);
\end{scope}
\end{tikzpicture}
~,
\eea
where in the first step we have simply shifted the bottom half of the Klein bottle, and in the second step we have reflected the right half along the horizontal axis.
The blue lines in the crosscap picture again reflect the fact that cycles passing through the crosscap lead to a change in orientation, schematically,
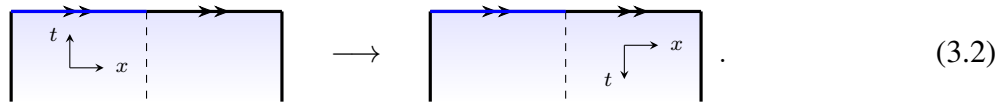
\bea
\begin{tikzpicture}[scale=0.6,baseline=15]
\begin{scope}[very thick, every node/.style={sloped,allow upside down}]
\shade[top color=blue, bottom color=white,opacity = 0.1] (0,0)--(0,2)--(6,2)--(6,0)--(0,0);

\draw (6,0)-- (6,2);
\draw (0,0)-- (0,2);
\draw(0,2) --node {\dmidarrow} (3,2);
\draw(3,2) -- node {\dmidarrow}(6,2);
\draw[thin, dashed] (3,2)--(3,0);

\draw[blue] (0,2)--(3,2);

\draw[thin, -stealth] (1.3,0.75) -- (1.3,1.5);
\draw[thin, -stealth] (1.3,0.75) -- (2.05,0.75);

\node[left] at (1.3,1.5) {\scriptsize $t$};
\node[right] at (2.05,0.75) {\scriptsize $x$};

\end{scope}
\end{tikzpicture}
\hspace{0.2 in}\longrightarrow \hspace{0.2 in}
\begin{tikzpicture}[scale=0.6,baseline=15]
\begin{scope}[very thick, every node/.style={sloped,allow upside down}]
\shade[top color=blue, bottom color=white,opacity = 0.1] (0,0)--(0,2)--(6,2)--(6,0)--(0,0);

\draw (6,0)-- (6,2);
\draw (0,0)-- (0,2);
\draw(0,2) --node {\dmidarrow} (3,2);
\draw(3,2) -- node {\dmidarrow}(6,2);
\draw[thin, dashed] (3,2)--(3,0);

\draw[blue] (0,2)--(3,2);

\draw[thin, -stealth] (4.3,1.25) -- (4.3, 0.5);
\draw[thin, -stealth] (4.3,1.25) -- (5.05,1.25);

\node[left] at (4.3, 0.5) {\scriptsize $t$};
\node[right] at (5.05,1.25) {\scriptsize $x$};

\end{scope}
\end{tikzpicture}~~.
\eea

Now consider inserting a topological defect $a$ along the orientation-preserving cycles of the Klein bottle. There are two classes of orientation-preserving cycles, which map to the following two configurations in the crosscap picture,
\bea
\begin{tikzpicture}[scale=0.8,baseline=35]
\begin{scope}[very thick, every node/.style={sloped,allow upside down}]
\shade[top color=blue, bottom color=white,opacity = 0.1] (0,0)--(3,0)--(3,4)--(0,4)--(0,0);
\draw[dashed, thin] (0,2) -- (3,2);
\draw[red] (1.5,0) -- node {\opmidarrow}(1.5,4);
\draw  (0,4)--node {\dmidarrow}(3,4);

\draw(0,0) --node {\dmidarrow}  (3,0);

\draw (3,2)-- node {\midarrow}(3,0);
\draw (3,4)-- node {\dmidarrow}(3,2);

\draw (0,0)-- node {\dmidarrow}(0,2);
\draw (0,2)-- node {\midarrow}(0,4);

\draw[blue] (3,0)--(3,4);

\node[red,right] at (1.5,1) {$a$};
\end{scope}
\end{tikzpicture}
\hspace{0.3 in}=\hspace{0.3 in} %insertion of =%
%RHS
\begin{tikzpicture}[scale=0.8,baseline=20]
\begin{scope}[very thick, every node/.style={sloped,allow upside down}]
\shade[top color=blue, bottom color=white,opacity = 0.1] (0,0)--(5,0)--(5,2)--(0,2)--(0,0);
\draw[red] (1.25,0) -- node {\opmidarrow}(1.25,2);
\draw[red] (3.75,2) -- node {\opmidarrow}(3.75,0);
\draw(2.5,0) -- node {\midarrow} (5,0);
\draw (5,0)-- node {\midarrow}(5,2);
\draw  (0,0)-- node {\midarrow}(2.5,0);
\draw[thin, dashed] (2.5,2)--(2.5,0);
\draw (0,0)-- node {\midarrow}(0,2);
\draw(0,2) --node {\dmidarrow} (2.5,2);
\draw(2.5,2) --node {\dmidarrow} (5,2);

\draw[blue] (0,2) -- (2.5,2);
\draw[blue] (0,0) -- (2.5,0);

\node[red,left] at (1.25,1) {$a$};
\node[red,right] at (3.75,1) {$a$};
\end{scope}
\end{tikzpicture},
\\
\begin{tikzpicture}[scale=0.8,baseline=35]
\begin{scope}[very thick, every node/.style={sloped,allow upside down}]
\shade[top color=blue, bottom color=white,opacity = 0.1] (0,0)--(3,0)--(3,4)--(0,4)--(0,0);
\draw[dashed, thin] (0,2) -- (3,2);
\draw[red] (0,1) -- node {\opmidarrow}(3,1);
\draw[red] (0,3) -- node {\opmidarrow}(3,3);
\draw  (0,4)--node {\dmidarrow}(3,4);

\draw(0,0) --node {\dmidarrow}  (3,0);

\draw (3,2)-- node {\midarrow}(3,0);
\draw (3,4)-- node {\dmidarrow}(3,2);

\draw (0,0)-- node {\dmidarrow}(0,2);
\draw (0,2)-- node {\midarrow}(0,4);

\draw[blue] (3,0)--(3,4);

\node[red,below] at (1.5,1) {$a$};
\node[red,above] at (1.5,3) {$a$};
\end{scope}
\end{tikzpicture}
\hspace{0.3 in}=\hspace{0.3 in}
\begin{tikzpicture}[scale=0.8,baseline=20]
\begin{scope}[very thick, every node/.style={sloped,allow upside down}]
\shade[top color=blue, bottom color=white,opacity = 0.1] (0,0)--(5,0)--(5,2)--(0,2)--(0,0);
\draw[red] (0,1) -- node {\opmidarrow}(5,1);
\draw(2.5,0) -- node {\midarrow} (5,0);
\draw (5,0)-- node {\midarrow}(5,2);
\draw  (0,0)-- node {\midarrow}(2.5,0);
\draw[thin,dashed] (2.5,2)--(2.5,0);
\draw (0,0)-- node {\midarrow}(0,2);
\draw(0,2) --node {\dmidarrow} (2.5,2);
\draw(2.5,2) --node {\dmidarrow} (5,2);

\draw[blue] (0,2) -- (2.5,2);
\draw[blue] (0,0) -- (2.5,0);
\node[red,below] at (2.25,1) {$a$};
\end{scope}
\end{tikzpicture}.
\eea
\noindent
Note when $a$ passes once through $\cP$ it emerges as $a$ itself, consistent with the action of $\cP$ on the $a$-twisted Hilbert spaces given in (\ref{eq:Hilbertspaceaction}). 

When $a=g$ is invertible, the first configuration can be used to define the following crosscap states, 
\bea
|C_g\rangle
\hspace{0.1 in}:=\hspace{0.1 in}
%crosscap def
\begin{tikzpicture}[scale=0.8,baseline=20]
\begin{scope}[very thick, every node/.style={sloped,allow upside down}]
    %bottom crosscap
  \draw (0,-0.5) ellipse [x radius=1.3cm, y radius=0.5cm];
  \draw (-1.12, -0.25) -- ( 1.12, -0.75);
  \draw (-1.12, -0.75) -- ( 1.12, -0.25);
    % center dash
  \draw[thick, dashed,dgreen] (-0.5,2.5) -- (-0.5,-1);
    % left red arrow
  \draw[red]  (-1.3,2.5)--node {\midarrow} (-1.3,-0.5);
     % right red arrow
  \draw[red] ( 1.3,-0.5)--node {\midarrow} ( 1.3,2.5) ;
     % g
  \node[red,left] at (-1.3,1.1) {$g$};
  \node[red,right] at (1.3,1.1) {$g$};
  \node[dgreen] at (-0.5,-1) {$\times$};
  \end{scope}
\end{tikzpicture}
%crosscap def
\hspace{0.1 in}=\hspace{0.1 in}
%RHS
\begin{tikzpicture}[scale=0.8,baseline=20]
\shade[top color=white, bottom color=blue,opacity = 0.1] (0,0)--(6,0)--(6,2)--(0,2)--(0,0);
\begin{scope}[very thick, every node/.style={sloped,allow upside down}]
%red arrow
\draw[red] (1.5,0) -- node {\opmidarrow}(1.5,2);
\draw[red] (4.5,2) -- node {\opmidarrow}(4.5,0);
%bottom part
\draw  (3,0) --node {\midarrow} (6,0);
\draw  (0,0)--node {\midarrow} (3,0);
%vertiiale line
\draw (6,0)--(6,2);
\draw (0,0)--(0,2);
%upper part
%\draw(0,2) --node {\dmidarrow} (3,2);
%\draw(3,2) --node {\dmidarrow} (6,2);
%dash line
\draw[thin, dashed] (3,2)--(3,0);
% g
\node[red,left] at (1.5,1) {$g$};
\node[red,right] at (4.5,1) {$g$};
\draw[blue] (0,0)--(3,0);
\end{scope}
\end{tikzpicture}~\,\,\,,
\eea
as well as their conjugates%\footnote{BPZ conjugation in a Euclidean theory effectively amounts to time and parity reversal of the boundary state \cite{}.}
\bea
\label{eq:firstcrosscapdef}
\langle C_g|
\hspace{0.1 in}:=\hspace{0.1 in}
%crosscap def%%%%%%%%%%%%%%%%%%%%%%%%%%%%%%%%%%%%%%
\begin{tikzpicture}[scale=0.8,baseline=20]
\begin{scope}[very thick, every node/.style={sloped,allow upside down}]
    %upper crosscap
  \draw (0,2.5) ellipse [x radius=1.3cm, y radius=0.5cm];
  \draw (-1.12, 2.75) -- ( 1.12, 2.25);
  \draw (-1.12, 2.25) -- ( 1.12, 2.75);
    % center dash
  \draw[thick, dashed,dgreen] (-0.5,2.0) -- (-0.5,-0.5);
  %left right black line
  \draw (-1.3,2.5) -- (-1.3,-0.5);
  \draw ( 1.3,-0.5) -- ( 1.3,2.5);
    % left red arrow
  \draw[red]  (-1.3,-0.5)--node {\opmidarrow} (-1.3,2.5);
     % right red arrow
  \draw[red]  ( 1.3,2.5)--node {\opmidarrow} ( 1.3,-0.5);
     % g1,g2
  \node[red,left] at (-1.3,1.1) {$g$};
  \node[red,right] at (1.3,1.1) {$g$};
  \node[dgreen] at (-0.5,2.0) {$\times$};
  \end{scope}
\end{tikzpicture}
%crosscap def%%%%%%%%%%%%%%%%%%%%%%%%%%%%%%%%%%%%%%%%%%%%
\hspace{0.1 in}=\hspace{0.1 in}
%RHS
\begin{tikzpicture}[scale=0.8,baseline=20]
\begin{scope}[very thick, every node/.style={sloped,allow upside down}]
\shade[top color=blue, bottom color=white,opacity = 0.1] (0,0)--(0,2)--(6,2)--(6,0)--(0,0);

%red arrow
\draw[red] (1.5,0) -- node {\opmidarrow}(1.5,2);
\draw[red] (4.5,2) -- node {\opmidarrow}(4.5,0);
%bottom part
%\draw(3,0) -- node {\midarrow} (6,0);
%\draw  (0,0)-- node {\midarrow}(3,0);
%vertiiale line
\draw (6,0)-- (6,2);
\draw (0,0)-- (0,2);
%upper part
\draw(0,2) --node {\dmidarrow} (3,2);
\draw(3,2) -- node {\dmidarrow}(6,2);
\draw[thin, dashed] (3,2)--(3,0);
% g
\node[red,left] at (1.5,1) {$g$};
\node[red,right] at (4.5,1) {$g$};
\draw[blue] (0,2)--(3,2);
\end{scope}
\end{tikzpicture}~\,\,\,.
\eea
The green dashed line is a choice of basepoint, which we use to keep track of the relative orientation of the two lines along the boundary. 
Note that we are restricting to invertible defects here since in this case the above states live in a single Hilbert space,
namely the untwisted Hilbert space $\cH_1$. If we were to replace $g$ with a general $a$ at this point, then the states would live in 
$\bigoplus_{y \prec a a^+} \cH_y$.

We may define the inner product between such states via the following configuration,
\bea
\label{eq:invinnerproduct}
%innerproduct of crosscap
\begin{tikzpicture}[scale=0.8,baseline=20]
\begin{scope}[very thick, every node/.style={sloped,allow upside down}]
    %upper crosscap
  \draw (0,2.5) ellipse [x radius=1.3cm, y radius=0.5cm];
  \draw (-1.12, 2.75) -- ( 1.12, 2.25);
  \draw (-1.12, 2.25) -- ( 1.12, 2.75);
    %bottom crosscap
  \draw (0,-0.5) ellipse [x radius=1.3cm, y radius=0.5cm];
  \draw (-1.12, -0.25) -- ( 1.12, -0.75);
  \draw (-1.12, -0.75) -- ( 1.12, -0.25);
    % center dash
  \draw[thick, dashed, dgreen] (-0.5,2.0) -- (-0.5,-1);
    % left red arrow
  \draw[red] (-1.3, 1.0) --node {\opmidarrow} (-1.3,2.5);
  \draw[red] (-1.3,-0.5) --node {\opmidarrow} (-1.3,1.0);  
     % right red arrow
  \draw[red] ( 1.3,2.5) --node {\opmidarrow} ( 1.3, 1.0);
  \draw[red] ( 1.3,1.0) --node {\opmidarrow} ( 1.3,-0.5);
  % curved red line
\draw[red] (-1.3,1.0) to[out=-90,in=-90,distance=0.2 in ]node {\opmidarrow}(1.3,1.0); 
% curved red line
     % g on the left
  \node[red,left] at (-1.3,1.7) {$h$};
  \node[red,left] at (-1.3,0.2) {$g$};
     % g on the right
  \node[red,right] at (1.3,1.7) {$h$};
  \node[red,right] at (1.3,0.2) {$g$};
  
 \node[dgreen]  at (-0.5,2.0) {$\times$};
 \node[dgreen]  at (-0.5,-1) {$\times$};
 
   \node[red, above] at (0.15,0.6) {$g h^{-1}$};
  \node[red]  at (-1.3,0.6) {\footnotesize $\times$};
 \node[red]  at (1.3,0.6) {\footnotesize $\times$};
 
 %\node[red,right] at (-1.27,0.74) {\footnotesize $\times$};
  \end{scope}
\end{tikzpicture}
%innerproduct of crosscap
\hspace{0.1 in}=\hspace{0.1 in}
%RHS%%%%%%%%%%%%%%%%%%%%%%%%%%%%%%%%%%%%%%%%%%%%%%%%%%%%%%%%%%%%%%%%%%%%%%%
\begin{tikzpicture}[scale=0.8,baseline=20]
\shade[top color=blue, bottom color=white,opacity = 0.1] (0,-0.5)--(6,-0.5)--(6,2.5)--(0,2.5)--(0,-0.5);
\begin{scope}[very thick, every node/.style={sloped,allow upside down}]
%red arrow left
\draw[red] (1.5,-0.5) -- node {\opmidarrow}(1.5,1.0);
\draw[red] (1.5,1.0) -- node {\opmidarrow}(1.5,2.5);
% red arrow right
\draw[red] (4.5,2.5) -- node {\opmidarrow}(4.5,1.0);
\draw[red] (4.5,1.0) -- node {\opmidarrow}(4.5,-0.5);
% red arrow horizon
\draw[red] (1.5,1.0) -- node {\opmidarrow}(4.5,1.0);
%bottom part
\draw  (3,-0.5) --node {\midarrow}(6,-0.5);
\draw  (0,-0.5)--node {\midarrow}(3,-0.5);
%vertiiale line
\draw (6,-0.5)--node {\midarrow}(6,2.5);
\draw (0,-0.5)--node {\midarrow}(0,2.5);
%upper part
\draw(0,2.5) --node {\dmidarrow} (3,2.5);
\draw(3,2.5) --node {\dmidarrow} (6,2.5);
%dash line
\draw[thin, dashed] (3,2.5)--(3,-0.5);
% g part on the left
\node[red,left] at (1.5,1.7) {$h$};
\node[red,left] at (1.5,0.2) {$g$};
% g part on the right
\node[red,right] at (4.5,1.7) {$h$};
\node[red,right] at (4.5,0.2) {$g$};
% g on the center
\node[red,above] at (3.0,1.0) {$g h^{-1}$};

\draw[blue] (0,2.5)--(3,2.5);
\draw[blue] (0,-0.5)--(3,-0.5);

\node[red,below] at (1.5,1.1) {\footnotesize $\times$};
\node[red,below] at (4.5,1.1) {\footnotesize $\times$};
\end{scope}
\end{tikzpicture}~,
\eea
where we note that, for invertible topological defects, the internal line is uniquely determined given the two crosscap states.\footnote{We could also consider defining the inner product with the internal line running along the back of the cylinder, instead of the front. The inner products computed in this convention are related to the ones in the current convention by exchanging all lines with their charge conjugates. }
Passing to the Klein bottle picture then gives
\bea
\begin{tikzpicture}[scale=0.8,baseline=20]
\shade[top color=blue, bottom color=white,opacity = 0.1] (0,-0.5)--(6,-0.5)--(6,2.5)--(0,2.5)--(0,-0.5);
\begin{scope}[very thick, every node/.style={sloped,allow upside down}]
%red arrow left
\draw[red] (1.5,-0.5) -- node {\opmidarrow}(1.5,1.0);
\draw[red] (1.5,1.0) -- node {\opmidarrow}(1.5,2.5);
% red arrow right
\draw[red] (4.5,2.5) -- node {\opmidarrow}(4.5,1.0);
\draw[red] (4.5,1.0) -- node {\opmidarrow}(4.5,-0.5);
% red arrow horizon
\draw[red] (1.5,1.0) -- node {\opmidarrow}(4.5,1.0);
%bottom part
\draw  (3,-0.5) --node {\midarrow}(6,-0.5);
\draw  (0,-0.5)--node {\midarrow}(3,-0.5);
%vertiiale line
\draw (6,-0.5)--node {\midarrow}(6,2.5);
\draw (0,-0.5)--node {\midarrow}(0,2.5);
%upper part
\draw(0,2.5) --node {\dmidarrow} (3,2.5);
\draw(3,2.5) --node {\dmidarrow} (6,2.5);
%dash line
\draw[thin, dashed] (3,2.5)--(3,-0.5);
% g part on the left
\node[red,left] at (1.5,1.7) {$h$};
\node[red,left] at (1.5,0.2) {$g$};
% g part on the right
\node[red,right] at (4.5,1.7) {$h$};
\node[red,right] at (4.5,0.2) {$g$};
% g on the center
\node[red,above] at (3.0,1.0) {$g h^{-1}$};

\draw[blue] (0,2.5)--(3,2.5);
\draw[blue] (0,-0.5)--(3,-0.5);

\node[red,below] at (1.5,1.1) {\footnotesize $\times$};
\node[red,below] at (4.5,1.1) {\footnotesize $\times$};
\end{scope}\end{tikzpicture}
\hspace{0.1 in}&=& P_{g}^{gh^{-1},h}\hspace{0.1 in}
%RHS%%%%%%%%%%%%%%%%%%%%%%%%%%%%%%%%%%%%%%%%%%%%%%%%%%%%%%%%%%%%%%%%%%%
\begin{tikzpicture}[scale=0.8,baseline=35]
\shade[top color=blue, bottom color=white,opacity = 0.1] (0,2)--(3,2)--(3,4)--(0,4)--(0,2);
\shade[top color=blue, bottom color=white,opacity = 0.1] (3,0)--(6,0)--(6,2)--(3,2)--(3,0);
\begin{scope}[very thick, every node/.style={sloped,allow upside down}]
%red line on the left
\draw[red] (1,2)-- node {\opmidarrow}(1,3);
\draw[red] (1,3)-- node {\opmidarrow}(1,4);
\draw[red] (1,3)-- node {\opmidarrow}(3,3);
%red line on the left
\draw[red] (5,0)-- node {\opmidarrow}(5,1);
\draw[red] (5,1)-- node {\opmidarrow}(5,2);
\draw[red] (3,1)-- node {\opmidarrow}(5,1);
%left box (up down right left)
\draw  (0,4)--node {\dmidarrow} (3,4);
\draw(0,2) --node {\midarrow} (3,2);
\draw (3,4)-- node {\dmidarrow}(3,2);
\draw (0,2)-- node {\midarrow}(0,4);
%right box (up down right left)
\draw(3,2) -- node {\midarrow} (6,2);
\draw(3,0) --node {\dmidarrow}  (6,0);
\draw (6,2)-- node {\midarrow}(6,0);
\draw (3,0)-- node {\dmidarrow}(3,2);
%parity
\draw[blue] (3,2)--(3,4);
\draw[blue] (0,2)--(0,4);
% g on the left box
\node[red,left] at (1,2.5) {$g$};
\node[red,left] at (1,3.5) {$h$};
\node[red,above] at (2,3) {$g h^{-1}$};
% g on the right box
\node[red,right] at (5,0.5) {$h$};
\node[red,right] at (5,1.5) {$g$};
\node[red,above] at (4,1) {$g h^{-1}$};

\node[red,below] at (5,1.55) {\footnotesize $\times$};
\node[red,below] at (1,3.1) {\footnotesize $\times$};

\end{scope}
\end{tikzpicture}
\no\\
%RHS%%%%%%%%%%%%%%%%%%%%%%%%%%%%%%%%%%%%%%%%%%%%%%%%%%%%%%%%%%%%%%%%
\hspace{0.1 in}&=& P_{g}^{gh^{-1},h}\hspace{0.1 in}
%LHS%%%%%%%%%%%%%%%%%%%%%%%%%%%%%%%%%%%%%%%%%%%%%%%%%%%%%%%%%%%%%%%%%
\begin{tikzpicture}[scale=0.8,baseline=35]
\shade[top color=blue, bottom color=white,opacity = 0.1] (0,0)--(3,0)--(3,4)--(0,4)--(0,0);
\begin{scope}[very thick, every node/.style={sloped,allow upside down}]
%redline
\draw[red] (1.5,0) -- node {\opmidarrow}(1.5,1);
\draw[red] (1.5,1) -- node {\opmidarrow}(1.5,3);
\draw[red] (1.5,3) -- node {\opmidarrow}(1.5,4);
%redline horizon
\draw[red] (0,1) -- node {\opmidarrow}(1.5,1);
\draw[red] (1.5,3) -- node {\opmidarrow}(3,3);
%box (up down right left)
\draw  (0,4)--node {\dmidarrow}(3,4);

\draw(0,0) --node {\dmidarrow}  (3,0);

\draw (3,2)-- node {\midarrow}(3,0);
\draw (3,4)-- node {\dmidarrow}(3,2);

\draw (0,0)-- node {\dmidarrow}(0,2);
\draw (0,2)-- node {\midarrow}(0,4);
%parity
\draw[blue] (3,0)--(3,4);
%g part
\node[red,right] at (1.5,0.5) {$h$};
\node[red,right] at (1.5,2) {$g$};
\node[red,left] at (1.5,3.5) {$h$};
\node[red,above] at (0.75,0) {$g h^{-1} $};
\node[red,above] at (2.3,3) {$g h^{-1} $};

\draw[dashed,thin] (0,2) -- (3,2);

\node[red,above] at (1.5,1) {\footnotesize $\times$};
\node[red,below] at (1.5,3) {\footnotesize $\times$};

\end{scope}
\end{tikzpicture}~,
\eea
where the factor of $P_{g}^{gh^{-1},h}$ appears when we flip the right-hand side of the crosscap, due to the action of orientation-reversing transformations on trivalent junctions, i.e. (\ref{eq:firstPeq}). 
Using an S$^{-1}$-transformation on the Klein bottle partition function, we obtain the following identity, 
\bea
\langle C_h | \ex{- {i \pi \over 2 \tau} (L_0 + \overline L_0 - {\mathsf{c} \over 12})} | C_g \rangle = P_{g}^{gh^{-1},h}\, \mathrm{Tr}_{\cH_{g h^{-1} }}\cP \,\widehat h\, \ex{2 \pi i \tau   (L_0 + \overline L_0 - {\mathsf{c} \over 12})} ~,
\eea
which is the generalized Cardy condition for crosscap states labelled by simple currents.

Note that in many cases, the overall sign $P_{gh^{-1},h}^g$ on the right-hand side can be dropped. For example, consider the case in which $g,h \in \ZZ_2$. This implies that $P_{g}^{gh^{-1},h} = \pa{\kappa_h}^{\delta_{g,1}}$. Since invertible symmetries have non-zero FS indicators if and only if they are $\ZZ_2$ symmetries with non-trivial `t Hooft anomaly, and since $\ZZ_2$ symmetries with `t Hooft anomalies must have spins \cite{Chang2019},
\bea
s \in {\ZZ \over 2} + {1\over 4}~,
\eea
we then conclude that there are no spin-0 states in $\cH_{g h^{-1}} = \cH_{h}$ when $\kappa_{h}$ is non-trivial, and hence that the trace with $\cP$ inserted vanishes whenever the sign is non-trivial. 

\subsection{Crosscaps and non-invertible symmetries}
\label{sec:crosscapnoninv}

We now generalize the above discussion to crosscap states labelled by non-invertible symmetries.  In order to do so, it is first useful to note that a series of F-moves can be used to rewrite,
\bea
\label{eq:rearrangement}
\vphantom{.}\hspace{-0.4 in}\begin{tikzpicture}[scale=0.8,baseline=20]
\begin{scope}[very thick, every node/.style={sloped,allow upside down}]
    %upper crosscap
  \draw (0,2.5) ellipse [x radius=1.3cm, y radius=0.5cm];
  \draw (-1.12, 2.75) -- ( 1.12, 2.25);
  \draw (-1.12, 2.25) -- ( 1.12, 2.75);
    %bottom crosscap
  \draw (0,-0.5) ellipse [x radius=1.3cm, y radius=0.5cm];
  \draw (-1.12, -0.25) -- ( 1.12, -0.75);
  \draw (-1.12, -0.75) -- ( 1.12, -0.25);
    % center dash
  \draw[thick, dashed, dgreen] (-0.5,2.0) -- (-0.5,-1);
    % left red arrow
  \draw[red] (-1.3, 1.0) --node {\opmidarrow} (-1.3,2.5);
  \draw[red] (-1.3,-0.5) --node {\opmidarrow} (-1.3,1.0);  
     % right red arrow
  \draw[red] ( 1.3,2.5) --node {\opmidarrow} ( 1.3, 1.0);
  \draw[red] ( 1.3,1.0) --node {\opmidarrow} ( 1.3,-0.5);
  % curved red line
\draw[red] (-1.3,1.0) to[out=-90,in=-90,distance=0.2 in ]node {\opmidarrow}(1.3,1.0); 
% curved red line
     % g on the left
  \node[red,left] at (-1.3,1.7) {$h$};
  \node[red,left] at (-1.3,0.2) {$g$};
     % g on the right
  \node[red,right] at (1.3,1.7) {$h$};
  \node[red,right] at (1.3,0.2) {$g$};
  
 \node[dgreen]  at (-0.5,2.0) {$\times$};
 \node[dgreen]  at (-0.5,-1) {$\times$};
 
   \node[red, above] at (0.15,0.6) {$g h^{-1}$};
  \node[red]  at (-1.3,0.6) {\footnotesize $\times$};
 \node[red]  at (1.3,0.6) {\footnotesize $\times$};
 
 %\node[red,right] at (-1.27,0.74) {\footnotesize $\times$};
  \end{scope}
\end{tikzpicture}
%innerproduct of crosscap
\hspace{0.1 in}=\hspace{0.1 in}
%RHS%%%%%%%%%%%%%%%%%%%%%%%%%%%%%%%%%%%%%%%%%%%%%%%%%%%%%%%%%%%%%%%%%%%%%%%
\begin{tikzpicture}[scale=0.7,baseline=20]
\shade[top color=blue, bottom color=white,opacity = 0.1] (0,-0.5)--(6,-0.5)--(6,2.5)--(0,2.5)--(0,-0.5);
\begin{scope}[very thick, every node/.style={sloped,allow upside down}]
%red arrow left
\draw[red] (1.5,-0.5) -- node {\opmidarrow}(1.5,1.0);
\draw[red] (1.5,1.0) -- node {\opmidarrow}(1.5,2.5);
% red arrow right
\draw[red] (4.5,2.5) -- node {\opmidarrow}(4.5,1.0);
\draw[red] (4.5,1.0) -- node {\opmidarrow}(4.5,-0.5);
% red arrow horizon
\draw[red] (1.5,1.0) -- node {\opmidarrow}(4.5,1.0);
%bottom part
\draw  (3,-0.5) --node {\midarrow}(6,-0.5);
\draw  (0,-0.5)--node {\midarrow}(3,-0.5);
%vertiiale line
\draw (6,-0.5)--node {\midarrow}(6,2.5);
\draw (0,-0.5)--node {\midarrow}(0,2.5);
%upper part
\draw(0,2.5) --node {\dmidarrow} (3,2.5);
\draw(3,2.5) --node {\dmidarrow} (6,2.5);
%dash line
\draw[thin, dashed] (3,2.5)--(3,-0.5);
% g part on the left
\node[red,left] at (1.5,1.7) {$h$};
\node[red,left] at (1.5,0.2) {$g$};
% g part on the right
\node[red,right] at (4.5,1.7) {$h$};
\node[red,right] at (4.5,0.2) {$g$};
% g on the center
\node[red,above] at (3.0,1.0) {$g h^{-1}$};

\draw[blue] (0,2.5)--(3,2.5);
\draw[blue] (0,-0.5)--(3,-0.5);

\node[red,below] at (1.5,1.1) {\footnotesize $\times$};
\node[red,below] at (4.5,1.1) {\footnotesize $\times$};
\end{scope}
\end{tikzpicture}
%MID%%%%%%%%%%%%%%%%%%%%%%%%%%%%%%%%%%%%%%%%%%%%%%%%%%%%%%%%%%%%%%%%%%
\hspace{0.1 in}=\hspace{0.05 in}\kappa_h
%RHS%%%%%%%%%%%%%%%%%%%%%%%%%%%%%%%%%%%%%%%%%%%%%%%%%%%%%%%%%%%%%%%%%%
\begin{tikzpicture}[scale=0.7,baseline=20]
\begin{scope}[very thick, every node/.style={sloped,allow upside down}]
\shade[top color=blue, bottom color=white,opacity = 0.1] (0,-0.5)--(0,2.5)--(6,2.5)--(6,-0.5)--(0,-0.5);
%red curved
\draw[red] (4.5,2.5) to[out=-90,in=-90,distance=0.6 in ]node {\opmidarrow}(1.5,2.5); 
\draw[red] (1.5,-0.5) to[out=90,in=90,distance=0.6 in ] node {\opmidarrow}(4.5,-0.5); 
%bottom part
\draw  (3,-0.5) --node {\midarrow}(6,-0.5);
\draw  (0,-0.5)--node {\midarrow}(3,-0.5);
%vertiiale line
\draw (6,-0.5)--node {\midarrow}(6,2.5);
\draw (0,-0.5)--node {\midarrow}(0,2.5);
%upper part
\draw(0,2.5) --node {\dmidarrow} (3,2.5);
\draw(3,2.5) --node {\dmidarrow} (6,2.5);
%dash line
\draw[thin, dashed] (3,2.5)--(3,-0.5);
% g part on the left
\node[red,left] at (1.5,1.7) {$h$};
\node[red,left] at (1.5,0.2) {$g$};
% g part on the right
\node[red,right] at (4.5,1.7) {$h$};
\node[red,right] at (4.5,0.2) {$g$};
% g on the center
%\node[red,above] at (3.0,1.0) {$g_1 g_2^{-1}$};

\draw[blue] (0,2.5)--(3,2.5);
\draw[blue] (0,-0.5)--(3,-0.5);
\end{scope}
\end{tikzpicture}~;
\eea
\newline
\newline
\noindent
details of this (straightforward) computation are given in Appendix \ref{app:computation}. On the right-hand side, the lines $g$ and $h$ wrap once around the orientation-reversing cycles of the respective crosscaps. This then leads us to the following, alternative definition of crosscap states,\footnote{Without $\kappa_g$, defining the bra-vector would break the conjugate symmetry of the inner product.}
\bea
\label{eq:secondcrosscapdef}
|C_g\rangle
\hspace{0.1 in}:=\hspace{0.2 in}
%crosscap def%%%%%%%%%%%%%%%%%%%%%%%%%%%%%%%%%%%%%%%%%%%%%%
\begin{tikzpicture}[scale=0.8,baseline=20]
\begin{scope}[very thick, every node/.style={sloped,allow upside down}]
    %bottom crosscap
  \draw (0,-0.5) ellipse [x radius=1.3cm, y radius=0.5cm];
  \draw (-1.12, -0.25) -- ( 1.12, -0.75);
  \draw (-1.12, -0.75) -- ( 1.12, -0.25);
  
    % center dash
 % \draw[thin, dashed] (-0.5,2.5) -- (-0.5,-1);
  %left right black line
  \draw (-1.3,2.5) -- (-1.3,-0.5);
  \draw ( 1.3,-0.5) -- ( 1.3,2.5);
   %curved red line 
 \draw[red] (-1.12,-0.75) to[out=70, in=90, distance=0.4 in ]node {\opmidarrow}(1.3,1.0); 
 \draw[red, dashed] (1.3,0.9) to[out=-110,in=70,distance=0.1 in ]node {\opmidarrow}(1.12,-0.25); 
  
     % g
      \node[red,right] at (0,1.3) {$g$};
      
          \node[dgreen]  at (-0.5,-0.95) {$\times$};
        \draw[dgreen,dashed,thin]  (-0.5,-0.95)-- (-0.5, 2);

  \end{scope}
\end{tikzpicture}\,\,\,%crosscap def%%%%%%%%%%%%%%%%%%%%%%%%%%%%%%%%%%%%%%%%%%%%
\,\,\,~,\hspace{0.75 in}
\langle C_g|
\hspace{0.1 in}:=\hspace{0.1 in} \kappa_g \,\,\, 
%crosscap def%%%%%%%%%%%%%%%%%%%%%%%%%%%%%%%%%%%%%%
\begin{tikzpicture}[scale=0.8,baseline=20]
\begin{scope}[very thick, every node/.style={sloped,allow upside down}]
    %upper crosscap
  \draw (0,2.5) ellipse [x radius=1.3cm, y radius=0.5cm];
  \draw (-1.12, 2.75) -- ( 1.12, 2.25);
  \draw (-1.12, 2.25) -- ( 1.12, 2.75);
    % center dash
  %\draw[thin, dashed] (-0.5,2.0) -- (-0.5,-0.5);
  %left right black line
  \draw (-1.3,2.5) -- (-1.3,-0.5);
  \draw ( 1.3,-0.5) -- ( 1.3,2.5);
 %curved red line 
 \draw[red] (1.3,1.0) to[out=-90,in=-70,distance=0.4 in ]node {\opmidarrow}(-1.12,2.25); 
\draw[red, dashed] (1.3,1.0) to[out=90,in=-50,distance=0.1 in ]node {\midarrow}(1.0,2.75); 
    % left red arrow
 % \draw[red] (-1.3,2.5) --node {\midarrow} (-1.3,-0.5);
     % right red arrow
 % \draw[red] ( 1.3,-0.5) --node {\midarrow} ( 1.3,2.5);
     % g
  \node[red,below] at (0.2,0.8) {$g$};
  
   \node[dgreen]  at (-0.6,2.05) {$\times$};
 \draw[dgreen,dashed,thin]  (-0.6,2.1)-- (-0.6, -0.25);
 
  \end{scope}
\end{tikzpicture}
~.
\eea
Inspired by the above discussion, we now try to define similar states for generic topological defects $a$, 
\bea
|C_a\rangle
\hspace{0.1 in}:=\hspace{0.2 in}
%crosscap def%%%%%%%%%%%%%%%%%%%%%%%%%%%%%%%%%%%%%%%%%%%%%%
\begin{tikzpicture}[scale=0.8,baseline=20]
\begin{scope}[very thick, every node/.style={sloped,allow upside down}]
    %bottom crosscap
  \draw (0,-0.5) ellipse [x radius=1.3cm, y radius=0.5cm];
  \draw (-1.12, -0.25) -- ( 1.12, -0.75);
  \draw (-1.12, -0.75) -- ( 1.12, -0.25);
    % center dash
 % \draw[thin, dashed] (-0.5,2.5) -- (-0.5,-1);
  %left right black line
  \draw (-1.3,2.5) -- (-1.3,-0.5);
  \draw ( 1.3,-0.5) -- ( 1.3,2.5);
   %curved red line 
 \draw[red] (-1.12,-0.75) to[out=70, in=90, distance=0.4 in ]node {\opmidarrow}(1.3,1.0); 
 \draw[red, dashed] (1.3,0.9) to[out=-110,in=70,distance=0.1 in ]node {\opmidarrow}(1.12,-0.25); 
     % g
  \node[red,above] at (0.4,1.05) {$a$};
  
     \node[dgreen]  at (-0.5,-0.95) {$\times$};
        \draw[dgreen,dashed,thin]  (-0.5,-0.95)-- (-0.5, 2);
  \end{scope}
\end{tikzpicture}\,\,\,%crosscap def%%%%%%%%%%%%%%%%%%%%%%%%%%%%%%%%%%%%%%%%%%%%
\,\,\,~,\hspace{0.75 in}
\langle C_a|
\hspace{0.1 in}:=\hspace{0.1 in}\kappa_a \,\,\, 
%crosscap def%%%%%%%%%%%%%%%%%%%%%%%%%%%%%%%%%%%%%%
\begin{tikzpicture}[scale=0.8,baseline=20]
\begin{scope}[very thick, every node/.style={sloped,allow upside down}]
    %upper crosscap
  \draw (0,2.5) ellipse [x radius=1.3cm, y radius=0.5cm];
  \draw (-1.12, 2.75) -- ( 1.12, 2.25);
  \draw (-1.12, 2.25) -- ( 1.12, 2.75);
    % center dash
  %\draw[thin, dashed] (-0.5,2.0) -- (-0.5,-0.5);
  %left right black line
  \draw (-1.3,2.5) -- (-1.3,-0.5);
  \draw ( 1.3,-0.5) -- ( 1.3,2.5);
 %curved red line 
 \draw[red] (1.3,1.0) to[out=-90,in=-70,distance=0.4 in ]node {\opmidarrow}(-1.12,2.25); 
\draw[red, dashed] (1.3,1.0) to[out=90,in=-50,distance=0.1 in ]node {\midarrow}(1.0,2.75); 
    % left red arrow
 % \draw[red] (-1.3,2.5) --node {\midarrow} (-1.3,-0.5);
     % right red arrow
 % \draw[red] ( 1.3,-0.5) --node {\midarrow} ( 1.3,2.5);
     % g
  \node[red,below] at (0.2,0.8) {$a$};
  
     \node[dgreen]  at (-0.6,2.05) {$\times$};
 \draw[dgreen,dashed,thin]  (-0.6,2.1)-- (-0.6, -0.25);
  \end{scope}
\end{tikzpicture}
~,
\eea
which may alternatively be drawn as in (\ref{eq:crosscappictorialdef}). 
There are two reasons why we are choosing to generalize the crosscap states in the form (\ref{eq:secondcrosscapdef}) as opposed to (\ref{eq:firstcrosscapdef}). 
First, if we generalized the latter, one would have to decompose it into elements of different Hilbert spaces, as already described above. 
Second, if we worked with  (\ref{eq:firstcrosscapdef}), computing amplitudes between $\langle{{C}_b}| $ and $|{{C}_a} \rangle$ would require specifying additional data, 
namely an internal line $y \prec a  b^+$ so that we can consistently connect the two as in (\ref{eq:invinnerproduct}). 

We may now compute the inner product between two such states by a series of F-moves.
In particular, the inner product between $\langle C_a|$ and $|C_b\rangle$ is given by
\bea
&\vphantom{.}& \hspace{0in}\begin{tikzpicture}[scale=0.7,baseline=20]
\begin{scope}[very thick, every node/.style={sloped,allow upside down}]
\shade[top color=blue, bottom color=white,opacity = 0.1] (0,-0.5)--(0,2.5)--(6,2.5)--(6,-0.5)--(0,-0.5);
%red curved
\draw[red] (4.5,2.5) to[out=-90,in=-90,distance=0.6 in ]node {\midarrow}(1.5,2.5); 
\draw[red] (1.5,-0.5) to[out=90,in=90,distance=0.6 in ] node {\opmidarrow}(4.5,-0.5); 
%bottom part
\draw  (3,-0.5) --node {\midarrow}(6,-0.5);
\draw  (0,-0.5)--node {\midarrow}(3,-0.5);
%vertiiale line
\draw (6,-0.5)--node {\midarrow}(6,2.5);
\draw (0,-0.5)--node {\midarrow}(0,2.5);
%upper part
\draw(0,2.5) --node {\dmidarrow} (3,2.5);
\draw(3,2.5) --node {\dmidarrow} (6,2.5);
%dash line
\draw[thin, dashed] (3,2.5)--(3,-0.5);
% g part on the left
\node[red,left] at (1.5,1.7) {$a^+$};
\node[red,left] at (1.5,0.2) {$b$};
% g part on the right
\node[red,right] at (4.5,1.7) {$a^+$};
\node[red,right] at (4.5,0.2) {$b$};
% g on the center
%\node[red,above] at (3.0,1.0) {$g_1 g_2^{-1}$};

\draw[blue] (0,2.5)--(3,2.5);
\draw[blue] (0,-0.5)--(3,-0.5);
\end{scope}
\end{tikzpicture}
= \,\,\, \sum_c \sum_{\mu = 1}^{N^a_{bc^+}} \sqrt{ d_c \over d_a d_b} \,\,\,
\begin{tikzpicture}[scale=0.7,baseline=20]
\begin{scope}[very thick, every node/.style={sloped,allow upside down}]
\shade[top color=blue, bottom color=white,opacity = 0.1] (0,-0.5)--(0,2.5)--(6,2.5)--(6,-0.5)--(0,-0.5);

\draw[red] (1.5,-0.5) -- node {\opmidarrow} (1.5,1) ;
\draw[red] (1.5,1) -- node {\midarrow} (1.5,2.5) ;
\draw[red] (4.5,-0.5) -- node {\midarrow} (4.5,1) ;
\draw[red] (4.5,1) -- node {\opmidarrow} (4.5,2.5) ;

\draw[red] (1.5,1) -- node {\opmidarrow}(4.5,1); 
%bottom part
\draw  (3,-0.5) --node {\midarrow}(6,-0.5);
\draw  (0,-0.5)--node {\midarrow}(3,-0.5);
%vertiiale line
\draw (6,-0.5)--node {\midarrow}(6,2.5);
\draw (0,-0.5)--node {\midarrow}(0,2.5);
%upper part
\draw(0,2.5) --node {\dmidarrow} (3,2.5);
\draw(3,2.5) --node {\dmidarrow} (6,2.5);
%dash line
\draw[thin, dashed] (3,2.5)--(3,-0.5);

\node[red,left] at (1.5,1.7) {$a^+$};
\node[red,left] at (1.5,0.2) {$b$};
\node[red,right] at (4.5,1.7) {$a^+$};
\node[red,right] at (4.5,0.2) {$b$};
\node[red,above] at (2.7,1.0) {$c$};
\node[red,left] at (2.2,1.0) {\footnotesize $\times$};
\node[red,right] at (3.8,1.0) {\footnotesize $\times$};
\node[left] at (1.5,1.0) {\scriptsize$\overline\mu$};
\node[right] at (4.5,1.0) {\scriptsize$ \mu$};

\draw[blue] (0,2.5)--(3,2.5);
\draw[blue] (0,-0.5)--(3,-0.5);
\end{scope}
\end{tikzpicture}
\no\\
&\vphantom{.}& \hspace{0.5 in}
\,\,\,= \,\,\, \sum_c \sum_{\mu = 1}^{N^a_{bc^+}} \sqrt{ d_c \over d_a d_b} \,\,\,
\begin{tikzpicture}[scale=0.7,baseline=20]
\begin{scope}[very thick, every node/.style={sloped,allow upside down}]
\shade[top color=blue, bottom color=white,opacity = 0.1] (0,-0.5)--(0,2.5)--(6,2.5)--(6,-0.5)--(0,-0.5);

\draw[red] (1.5,-0.5) -- node {\opmidarrow} (1.5,1) ;
\draw[red] (1.5,1) -- node {\midarrow} (1.5,1.75) ;
\draw[red] (1.5,1.75) -- node {\opmidarrow} (1.5,2.5) ;

\draw[red] (4.5,-0.5) -- node {\midarrow} (4.5,1) ;
\draw[red] (4.5,1) -- node {\opmidarrow} (4.5,1.75) ;
\draw[red] (4.5,1.75) -- node {\midarrow} (4.5,2.5) ;

\draw[red] (1.5,1) -- node {\opmidarrow}(4.5,1); 
%bottom part
\draw  (3,-0.5) --node {\midarrow}(6,-0.5);
\draw  (0,-0.5)--node {\midarrow}(3,-0.5);
%vertiiale line
\draw (6,-0.5)--node {\midarrow}(6,2.5);
\draw (0,-0.5)--node {\midarrow}(0,2.5);
%upper part
\draw(0,2.5) --node {\dmidarrow} (3,2.5);
\draw(3,2.5) --node {\dmidarrow} (6,2.5);
%dash line
\draw[thin, dashed] (3,2.5)--(3,-0.5);

\node[red,left] at (1.3,2.1) {$a$};
\node[red,left] at (1.6,1.5) {$a^+$};
\node[red,left] at (1.4,0.2) {$b$};
\node[red,right] at (4.5,2.1) {$a$};
\node[red,right] at (4.5,1.5) {$a^+$};
\node[red,right] at (4.5,0.2) {$b$};
\node[red,above] at (2.7,1.0) {$c$};
\node[red,left] at (2.2,1.0) {\footnotesize $\times$};
\node[red,right] at (3.8,1.0) {\footnotesize $\times$};
\node[left] at (1.5,1.0) {\scriptsize$\overline\mu$};
\node[right] at (4.5,1.0) {\scriptsize$ \mu$};

\draw[blue] (0,2.5)--(3,2.5);
\draw[blue] (0,-0.5)--(3,-0.5);

\filldraw[red] (1.5, 1.75) circle (0.2 ex);
\filldraw[red] (4.5, 1.75) circle (0.2 ex);

\end{scope}
\end{tikzpicture}
\\
&\vphantom{.}& \hspace{0.5 in}=\sum_c \sum_{\mu, \nu, \rho = 1}^{N^a_{bc^+}} \sqrt{d_a d_b \over d_c} (F_{b a^+ a}^b)_{(c \mu \nu) 1}(F_{b a^+ a}^b)^{-1}_{1 (c \mu \rho)}
\begin{tikzpicture}[scale=0.7,baseline=20]
\begin{scope}[very thick, every node/.style={sloped,allow upside down}]
\shade[top color=blue, bottom color=white,opacity = 0.1] (0,-0.5)--(0,2.5)--(6,2.5)--(6,-0.5)--(0,-0.5);

\draw[red] (1.5,-0.5) -- node {\opmidarrow} (1.5,1) ;
\draw[red] (1.5,1) -- node {\opmidarrow} (1.5,2.5) ;
\draw[red] (4.5,-0.5) -- node {\midarrow} (4.5,1) ;
\draw[red] (4.5,1) -- node {\midarrow} (4.5,2.5) ;

\draw[red] (1.5,1) -- node {\opmidarrow}(4.5,1); 
%bottom part
\draw  (3,-0.5) --node {\midarrow}(6,-0.5);
\draw  (0,-0.5)--node {\midarrow}(3,-0.5);
%vertiiale line
\draw (6,-0.5)--node {\midarrow}(6,2.5);
\draw (0,-0.5)--node {\midarrow}(0,2.5);
%upper part
\draw(0,2.5) --node {\dmidarrow} (3,2.5);
\draw(3,2.5) --node {\dmidarrow} (6,2.5);
%dash line
\draw[thin, dashed] (3,2.5)--(3,-0.5);

\node[red,left] at (1.5,1.7) {$a$};
\node[red,left] at (1.4,0.3) {$b$};
\node[red,right] at (4.5,1.7) {$a$};
\node[red,right] at (4.5,0.3) {$b$};
\node[red,above] at (2.7,1.0) {$c$};
\node[red,above] at (1.5,0.3) {\footnotesize $\times$};
\node[red,above] at (4.5,0.3) {\footnotesize $\times$};
\node[left] at (1.5,1.0) {\scriptsize$\rho$};
\node[right] at (4.5,1.0) {\scriptsize$\overline \nu$};

\draw[blue] (0,2.5)--(3,2.5);
\draw[blue] (0,-0.5)--(3,-0.5);

\filldraw[red] (1.5, 0.25) circle (0.2 ex);
\filldraw[red] (4.5, 0.25) circle (0.2 ex);

\end{scope}
\end{tikzpicture}~,
\no
\eea
where to obtain the final equality, we connected an identity line from each $a$-$a^+$ vertex to the $b$ line directly below it, used an F- and G-move, and then used the second equation of (\ref{eq:basisconventions}) twice. Using the identity
\bea
\label{eq:FFsumidentity}
\sum_{\mu = 1}^{N^b_{ac}} (F_{ba^+ a}^b)_{(c \mu \nu) 1} (F_{b a^+ a}^b)^{-1}_{1 (c \mu \rho)} = {d_c \over d_a d_b} \delta_{\nu \rho}~
\eea
and transitioning to the Klein bottle, we then obtain the following generalized Cardy condition,
\bea
\label{eq:generalizedCardylikeform}
\langle C_a | \ex{- {\pi i \over 2 \tau} (L_0 + \overline{L}_0 - {\mathsf{c}\over 12})} | C_b\rangle =\sum_{c } \sum_{\mu,\nu = 1}^{N_{ ac}^b} \sqrt{d_c \over d_a d_b} \,(P_{b}^{ca})_{\bar \mu \bar \nu}\, \mathrm{Tr}_{\cH_c} \cP \,\widehat a\, \ex{2 \pi i \tau(L_0 + \overline{L}_0 - {\mathsf{c}\over 12})} \Big|_{(b,\mu,\bar \nu)},
\eea
where the restriction $(b, \mu, \bar \nu)$ denotes the particular configuration on the Klein bottle,
\bea
 \mathrm{Tr}_{\cH_c} \cP \,\widehat a\, \ex{2 \pi i \tau(L_0 + \overline{L}_0 - {\mathsf{c}\over 12})} \Big|_{(b,\mu,\bar \nu)} \,\,\, := \,\,\, 
\begin{tikzpicture}[scale=0.8,baseline=35]
\shade[top color=blue, bottom color=white,opacity = 0.1] (0,0)--(3,0)--(3,4)--(0,4)--(0,0);
\begin{scope}[very thick, every node/.style={sloped,allow upside down}]
%redline
\draw[red] (1.5,0) -- node {\opmidarrow}(1.5,1);
\draw[red] (1.5,1) -- node {\opmidarrow}(1.5,3);
\draw[red] (1.5,3) -- node {\opmidarrow}(1.5,4);
%redline horizon
\draw[red] (0,1) -- node {\opmidarrow}(1.5,1);
\draw[red] (1.5,3) -- node {\opmidarrow}(3,3);
%box (up down right left)
\draw  (0,4)--node {\dmidarrow}(3,4);

\draw(0,0) --node {\dmidarrow}  (3,0);

\draw (3,2)-- node {\midarrow}(3,0);
\draw (3,4)-- node {\dmidarrow}(3,2);

\draw (0,0)-- node {\dmidarrow}(0,2);
\draw (0,2)-- node {\midarrow}(0,4);
%parity
\draw[blue] (3,0)--(3,4);
%g part
\node[red,below] at (1.5,0) {$a$};
\node[red,right] at (1.5,1.8) {$b$};
\node[red,above] at (1.5,4) {$a$};
\node[red,above] at (0.75,0.3) {$c$};
\node[red,above] at (2.3,3) {$c$};

\node[right] at (1.5,1) {\scriptsize $\overline \nu$};
\node[left] at (1.5,3) {\scriptsize $\mu$};

\draw[dashed,thin] (0,2) -- (3,2);

\node[red,above] at (1.5,1) {\footnotesize $\times$};
\node[red,below] at (1.5,3) {\footnotesize $\times$};

\end{scope}
\end{tikzpicture}~.
\eea
This formula will be checked in concrete examples in Section \ref{sec:examples}.

Let us close by noting that above we have focused on crosscap states in the untwisted Hilbert space $\cH_1$. 
In general, there also exist crosscap states $|C_a\rangle_{x, \nu}$ in the $x$-twisted Hilbert space $\cH_x$, 
labelled by $\nu = 1, \dots, N_{xa}^a$. The inner products of these crosscap states may be computed via the same 
techniques as above, though the resulting formula is somewhat more involved, and we do not include it here.

\subsection{Action of Verlinde lines on crosscap states}

Next we compute the action of Verlinde lines on the crosscap states. This is done via the following set of partial fusions and F-moves, 
\bea
\begin{tikzpicture}[scale=0.5,baseline=0]
\begin{scope}[ thick, every node/.style={sloped,allow upside down}]
   
      \draw[red] (0,0.53) to[out=80, in=90, distance=0.7 in] (2.5,0);
     \draw[red, -<-=0.08] (2.5,0) to[out=-90, in=-80, distance=0.7 in] (0,-0.53);
     
     \draw  (0,0) circle (3ex);
   \draw[thick] (-0.4,-0.4) -- (0.4,0.4);
   \draw[thick] (-0.4,0.4) -- (0.4,-0.4);
  
   \draw[blue,->-=0]  (0,0) circle (15ex);

   \node[blue, right] at (3,0) {$a$};
     \node[red, left] at (2.5,0) {$b$};
   
\end{scope}
\end{tikzpicture}
&=&
\sum_{x }\sum_{\mu=1}^{N_{ab}^x} \sqrt{d_x\over d_a d_b} \,\,\, \begin{tikzpicture}[scale=0.5,baseline=0]
\begin{scope}[ thick, every node/.style={sloped,allow upside down}]
   \draw  (0,0) circle (3ex);
   \draw[thick] (-0.4,-0.4) -- (0.4,0.4);
   \draw[thick] (-0.4,0.4) -- (0.4,-0.4);
  
       \draw[red] (0,0.53) to[out=80, in=150, distance=0.3 in] node {\opmidarrow}  (1.5,1.5);
       \draw[red] (0,-0.53) to[out=-80, in=-150, distance=0.3 in] node {\midarrow}  (1.5,-1.5);
        
          \draw[blue,-<-=0.05] (-2.8,0) to[out=90, in=110, distance=1 in] (1.5,1.5);
          \draw[blue] (-2.8,0) to[out=-90, in=-110, distance=1 in] (1.5,-1.5);
          
          \draw[] (1.5,1.5) to[out=-10, in=10, distance=0.5 in] node {\opmidarrow}  (1.5,-1.5);
        \node[blue, left] at (-3,0) {$a$};
          \node[red, below] at (0,-0.7) {$b$};
           \node[right] at (2.5,0) {$x$};
           \node[right] at (1.2,1.8) {\scriptsize$\bar\mu$};
           \node[right] at (1.2,-1.8) {\scriptsize$ \mu$};
           
           \node[right] at (1.4,-1.25) {\scriptsize$+$};
               \node[right] at (1.4,1.25) {\scriptsize$+$};
\end{scope}
\end{tikzpicture}
=\sum_{x }\sum_{\mu=1}^{N_{ab}^x} \sqrt{d_x\over d_a d_b}  \begin{tikzpicture}[scale=0.5,baseline=0]
\begin{scope}[ thick, every node/.style={sloped,allow upside down}]
   \draw  (0,0) circle (3ex);
   \draw[thick] (-0.4,-0.4) -- (0.4,0.4);
   \draw[thick] (-0.4,0.4) -- (0.4,-0.4);
  
       \draw[red] (0,0.53) to[out=80, in=150, distance=0.3 in] node {\opmidarrow}  (1.5,1.5);
       \draw[red] (0,-0.53) to[out=-80, in=-150, distance=0.3 in] node {\midarrow}  (1.5,-1.5);
        
          \draw[blue] (-0.4,0.4) to[out=135, in=110, distance=1.2 in] node {\opmidarrow}  (1.5,1.5);
          \draw[blue] (-0.4,-0.4) to[out=225, in=-110, distance=1.2 in] node {\midarrow}  (1.5,-1.5);
          
          \draw[] (1.5,1.5) to[out=-10, in=10, distance=0.5 in] node {\opmidarrow}  (1.5,-1.5);
          
          \draw[blue] (0.4,0.4) to[out=45, in=90, distance=0.4in] (2,0);
           \draw[blue, -<-=0.05] (2,0) to[out=-90, in=-45, distance=0.4in] (0.4,-0.4);
        \node[blue, left] at (2,0) {$a$};
          \node[red, below] at (0,-0.7) {$b$};
           \node[right] at (2.5,0) {$x$};
            \node[right] at (1.2,1.8) {\scriptsize$\bar\mu$};
           \node[right] at (1.2,-1.8) {\scriptsize$ \mu$};
           
            \node[right] at (1.4,-1.25) {\scriptsize$+$};
               \node[right] at (1.4,1.25) {\scriptsize$+$};
  \end{scope}
\end{tikzpicture}
\no\\
&\vphantom{.}&\hspace{-1.4 in} 
           =\sum_{x,y} \sum_{\mu=1}^{N_{ab}^x} \sum_{\nu=1}^{N_{ax}^y} \sqrt{d_y\over d_a^2 d_b} \begin{tikzpicture}[scale=0.6,baseline=0]
\begin{scope}[ thick, every node/.style={sloped,allow upside down}]
   \draw  (0,0) circle (3ex);
   \draw[thick] (-0.4,-0.4) -- (0.4,0.4);
   \draw[thick] (-0.4,0.4) -- (0.4,-0.4);
  
       \draw[red] (0,0.53) to[out=80, in=150, distance=0.3 in] node {\opmidarrow} (1.5,1.5);
       \draw[red] (0,-0.53) to[out=-80, in=-150, distance=0.3 in] node {\midarrow} (1.5,-1.5);
        
          \draw[blue] (-0.4,0.4) to[out=135, in=110, distance=1.2 in] node {\opmidarrow} (1.5,1.5);
          \draw[blue] (-0.4,-0.4) to[out=225, in=-110, distance=1.2 in] node {\midarrow} (1.5,-1.5);
          
          \draw[] (1.5,1.5) to[out=-10, in=100, distance=0.1 in] node {\opmidarrow} (2.3,0.7);
          \draw[] (1.5,-1.5) to[out=-10, in=-100, distance=0.1 in] node {\midarrow}(2.3,-0.7);
          
          \draw[blue] (0.4,0.4) to[out=45, in=180, distance=0.2in] node {\opmidarrow} (2.3,0.7);
          \draw[blue] (0.4,-0.4) to[out=-45, in=-180, distance=0.2in] node {\midarrow}(2.3,-0.7);
          
          \draw[violet] (2.3,0.7) to[out=-60, in=60, distance=0.2in]node {\opmidarrow}  (2.3,-0.7);
          
        \node[blue, left] at (1.8,0.3) {$a$};
          \node[red, below] at (0,-0.7) {$b$};
           \node[violet,right] at (2.5,0) {$y$};
            \node[right] at (2,1.3) {$x$};
            
             \node[right] at (1.2,1.8) {\scriptsize$\bar \mu$};
           \node[right] at (1.2,-1.8) {\scriptsize$ \mu$};
           
               \node[right] at (2.3,0.7) {\scriptsize$\bar\nu$};
           \node[right] at (2.3,-0.7) {\scriptsize$ \nu$};
           
            \node[right] at (1.3,-1.4) {\scriptsize$+$};
               \node[right] at (1.3,1.4) {\scriptsize$+$};
               
                \node[right,violet] at  (1.95,0.5){\scriptsize$\times$};
               \node[right,violet] at (1.95,-0.5) {\scriptsize$\times$};
  \end{scope}
\end{tikzpicture}
\,\,\,=\,\,\,\sum_{x,y}\sum_{\mu,\rho=1}^{N_{ab}^x} \sum_{\nu,\sigma=1}^{N_{ax}^y} \sqrt{d_y\over d_a^2 d_b} (P_{ab}^x)_{\mu \rho} (P_{xa}^y)_{\nu \sigma} \,\,\, 
\begin{tikzpicture}[scale=0.6,baseline=0]
\begin{scope}[ thick, every node/.style={sloped,allow upside down}]
   \draw  (0,0) circle (3ex);
   \draw[thick] (-0.4,-0.4) -- (0.4,0.4);
   \draw[thick] (-0.4,0.4) -- (0.4,-0.4);
  
       \draw[violet] (0,0.53) to[out=80, in=90, distance=0.5 in] node {\opmidarrow} (2.5,1);
       \draw[violet] (0,-0.53) to[out=-80, in=-90, distance=0.5 in] node {\midarrow} (2.5,-1);
        
        \draw[] (2.5,1) to[out=180, in=90] node {\opmidarrow} (1.5,0);
        \draw[blue] (1.5,0) to[out=-90, in=180] node {\opmidarrow} (2.5,-1);
         \draw[blue] (2.5,1) to[out=0, in=90] node {\opmidarrow} (3.5,0);
         \draw[] (3.5,0) to[out=-90, in=0] node {\opmidarrow} (2.5,-1);
           \draw[red] (3.5,0) to[out=180, in=0] node {\midarrow} (1.5,0);
        
                \node[left] at (1.8,0.7) {$x$};
          \node[violet, below] at (0,-0.9) {$y$};
           \node[red,below] at (2.5,0.1) {$b$};
            \node[blue,  right] at (3.2,0.7) {$a$};
            
            \node[right] at (2.4,1.2) {\scriptsize$\sigma$};
            \node[right] at (3.4,0) {\scriptsize$\bar \mu$};
            \node[left] at (1.6,0) {\scriptsize$\rho$};
            \node[right] at (2.4,-1.2) {\scriptsize$ \bar \nu$};
            
             \node[above,violet] at  (2.48,0.8){\scriptsize$\times$};
               \node[below,violet] at (2.48,-0.8) {\scriptsize$\times$};
                \node[above] at  (1.54,-0.2){\scriptsize$\times$};
               \node[below] at (3.49,0.2) {\scriptsize$\times$};
  \end{scope}
\end{tikzpicture}
\no\\
&\vphantom{.}&\hspace{-0.5 in}=\sum_{x,y}\sum_{\mu,\rho=1}^{N_{ab}^x} \sum_{\nu,\sigma=1}^{N_{ax}^y}(P_{ab}^x)_{\mu \rho} (P_{xa}^y)_{\nu \sigma}  (F^y_{aba})_{(x \rho \sigma)(x  \nu \mu)} \,\,\, \begin{tikzpicture}[scale=0.6,baseline=0]
\begin{scope}[ thick, every node/.style={sloped,allow upside down}]
   \draw  (0,0) circle (3ex);
   \draw[thick] (-0.4,-0.4) -- (0.4,0.4);
   \draw[thick] (-0.4,0.4) -- (0.4,-0.4);
  
  \draw[violet] (0,0.566) to[out = 45, in = -45,distance = 1.3in] node {\opmidarrow} (0,-0.566);

     \node[violet, above] at (2.2,-0.2) {$y$};

\end{scope}
\end{tikzpicture}
\eea
where in the final step we have used an F-move and (\ref{eq:basisconventions}). We thus conclude that
\bea
\label{eq:actiononcc}
\widehat a | C_b \rangle =  \sum_{x,y}\sum_{\mu,\rho=1}^{N_{ab}^x} \sum_{\nu,\sigma=1}^{N_{ax}^y}(P_{ab}^x)_{\mu \rho} (P_{xa}^y)_{\nu \sigma}  (F^y_{aba})_{(x \rho \sigma)(x  \nu \mu)} |C_y\rangle~.
\eea
Though it is not obvious from this presentation, the coefficients in this expansion must always be integers. 
When $b=1$, it is easy to see that this reduces to (\ref{eq:Paaxrel}). In the next section, we will check this formula in some concrete examples.

\section{Explicit Examples}
\label{sec:examples}

In this section we give a couple of concrete examples illustrating the formulas obtained above. 
In particular, we first consider the Fibonacci unitary modular tensor categories (UMTCs), which have trivial FS indicators and unit multiplicities, 
after which we move on to the example of Ising UMTCs, which have non-trivial FS indicators, but still unit multiplicity.

\subsection{Fibonacci UMTCs}

We begin by considering a diagonal RCFT with Fibonacci categorical symmetry. The Fibonacci category contains two simple objects $\{1, W\}$ with the fusion rule
\bea
W \times W = 1 + W~, 
\eea
 trivial FS indicator $\kappa_W = 1$, and non-trivial F-symbols, 
\bea
(F^W_{WWW})_{ij} = \left(\begin{matrix} \phi^{-1} & \phi^{-1/2} \\ \phi^{-1/2} & - \phi^{-1} \end{matrix} \right) ~,
\eea
where $\phi = {1\over 2} (1 + \sqrt{5})$ is the golden ratio. The Fibonacci fusion category has two distinct UMTC completions, differing in their braiding data. This is reflected in the modular data as 
\bea
\S = {1 \over \sqrt{1+\phi^2}} \left(\begin{matrix} 1 & \phi \\ \phi & -1 \end{matrix} \right)~, \hspace{0.5 in} \T = \ex{- 2\pi i {7 \eps \over 60}} \left(\begin{matrix} 1 & 0 \\ 0 & \ex{{4 \pi i \over 5} \eps}\end{matrix}  \right)  ~, 
\eea
with the two options corresponding to $\eps = \pm 1$.\footnote{These are realized by $(\mathfrak{g}_2)_1$ and $(\mathfrak{f}_4)_1$ WZW models, respectively.} 
The crosscap states may be computed using (\ref{eq:simplecurrentcrosscaps}), giving
\bea
\label{eq:Fibonaccicrosscaps}
|C_1\rangle &=& \eta_1\left[ - \left(1 + {2\over \sqrt{5}} \right)^{1/4} |1\rangle\!\rangle_C +  \left(1 - {2\over \sqrt{5}} \right)^{1/4} |W\rangle\!\rangle_C\right]~,
\no\\
|C_W\rangle &=& \eta_W\left[ \left(\half - {1\over2 \sqrt{5}} \right)^{1/4} |1\rangle\!\rangle_C +  \left(\half + {1\over 2\sqrt{5}} \right)^{1/4} |W\rangle\!\rangle_C\right]~.
\eea
The overall factors of $\eta_{1,W}$ are constrained by (\ref{eq:etaconstraint}) to satisfy 
\bea
\label{eq:fibonaccieta}
\eta_W = - \eta_1 P_{WW}^W~. 
\eea
The parity eigenvalues $P_{WW}^1$ and $P_{WW}^W$ are in turn constrained by (\ref{eq:PFidentity}),
but the only non-trivial constraint that this produces is $P_{WW}^1 = (P_{WW}^W)^2$. Recalling that $P_{WW}^1 = \kappa_W = 1$,
we conclude that $P_{WW}^W = \pm 1$ is unfixed by this constraint.

Using (\ref{eq:Fibonaccicrosscaps}) and the S-matrix given above, we find
\bea
\langle C_1 | \ex{-{\pi i \over 2 \tau } (L_0 + \overline{L}_0 - {\mathsf{c} \over 12})} | C_1\rangle &=& \chi_1(2 \tau) + \chi_W(2 \tau)~,
\no\\
\langle C_W | \ex{-{\pi i \over 2 \tau } (L_0 + \overline{L}_0 - {\mathsf{c} \over 12})} | C_1\rangle &=&- \eta_W \eta_1 \chi_W(2 \tau)~,
\no\\
\langle C_1 | \ex{-{\pi i \over 2 \tau } (L_0 + \overline{L}_0 - {\mathsf{c} \over 12})} | C_W\rangle &=&- \eta_W \eta_1 \chi_W(2 \tau)~,
\no\\
\langle C_W | \ex{-{\pi i \over 2 \tau } (L_0 + \overline{L}_0 - {\mathsf{c} \over 12})} | C_W\rangle &=& \chi_1(2 \tau)~,
\eea
while on the other hand
\bea
\mathrm{Tr}_{\cH_1} \cP \ex{2 \pi i \tau(L_0 + \overline{L}_0 - {\mathsf{c}\over 12})} &=& \chi_1(2 \tau) + \chi_W(2 \tau)~,
\no\\
\mathrm{Tr}_{\cH_W} \cP\, \widehat W \ex{2 \pi i \tau(L_0 + \overline{L}_0 - {\mathsf{c}\over 12})} \big|_{1} &=& P_{WW}^W \chi_W(2\tau)~,
\no\\
\mathrm{Tr}_{\cH_W} \cP \ex{2 \pi i \tau(L_0 + \overline{L}_0 - {\mathsf{c}\over 12})} \big|_{W} &=& P_{WW}^W \chi_W(2\tau)~,
\no\\
d_W^{-1}\sum_{c=1,W} \sqrt{d_c}\, P_W^{cW}\,\mathrm{Tr}_{\cH_c} \cP\, \widehat W\ex{2 \pi i \tau(L_0 + \overline{L}_0 - {\mathsf{c}\over 12})}\big|_{W} &=& \chi_1(2 \tau)~,
\eea
where in the second and fourth lines we have used the following lasso actions, 
\bea
\label{eq:Fiblassos}
\begin{tikzpicture}[scale=0.6,baseline=15]
\begin{scope}[very thick, every node/.style={sloped,allow upside down}]
%\draw[->-=0.25, ->- = 0.55, ->- = 0.9](0,0) -- (0,3);
\draw[->- = 0.7](0,0) -- (0,1);
\draw[dashed] (0,1) -- (0,2);
\draw[->- = 0.8](0,2) -- (0,3);
\draw(0,1) to[out=0,in=0,distance=0.6in] node{\opmidarrow}  (0,-1);
\draw  (0,-1) to[out=180,in=180,distance=0.8in] node{\opmidarrow}   (0,2);
\node[above] at (0,3) {$W$};
\node[right] at (1.2,0) {$W$};
\node[left] at (0.1,0.5) {$W$};
\filldraw (0,0) circle (0.3ex);
\end{scope}
\end{tikzpicture}
\,\,\,=\,\,\,
\begin{tikzpicture}[scale=0.6,baseline=15]
\begin{scope}[very thick, every node/.style={sloped,allow upside down}]
%\draw[->-=0.25, ->- = 0.55, ->- = 0.9](0,0) -- (0,3);
\draw[->- = 0.7](0,0) -- (0,3);
\filldraw (0,0) circle (0.3ex);
\node[above] at  (0,3) {$W$};
\end{scope}
\end{tikzpicture}~,
\hspace{0.5 in}
\begin{tikzpicture}[scale=0.6,baseline=15]
\begin{scope}[very thick, every node/.style={sloped,allow upside down}]
%\draw[->-=0.25, ->- = 0.55, ->- = 0.9](0,0) -- (0,3);
\draw[->- = 0.7](0,0) -- (0,1);
\draw[->-=0.7] (0,1) -- (0,2);
\draw[->- = 0.8](0,2) -- (0,3);
\draw(0,1) to[out=0,in=0,distance=0.6in] node{\opmidarrow}  (0,-1);
\draw  (0,-1) to[out=180,in=180,distance=0.8in] node{\opmidarrow}   (0,2);
\node[above] at (0,3) {$W$};
\node[right] at (1.2,0) {$W$};
\node[right] at (0,1.5) {$W$};
\node[left] at (0.1,0.5) {$W$};
\filldraw (0,0) circle (0.3ex);
\end{scope}
\end{tikzpicture}
\,\,\,=\,\,\, \phi^{-3/2}\,\,\,
\begin{tikzpicture}[scale=0.6,baseline=15]
\begin{scope}[very thick, every node/.style={sloped,allow upside down}]
%\draw[->-=0.25, ->- = 0.55, ->- = 0.9](0,0) -- (0,3);
\draw[->- = 0.7](0,0) -- (0,3);
\filldraw (0,0) circle (0.3ex);
\node[above] at  (0,3) {$W$};
\end{scope}
\end{tikzpicture}~,
\eea
which follow straightforwardly from the F-symbols. Imposing (\ref{eq:fibonaccieta}), the two computations match perfectly. 

We may also compute the action of $W$ on these crosscap states using (\ref{eq:actiononcc}), giving 
\bea
\widehat W|C_1 \rangle = |C_1 \rangle + P_{WW}^W |C_W \rangle~, \hspace{0.5in} \widehat W|C_W\rangle = P_{WW}^W |C_1 \rangle~.
\eea
This matches the results obtained from explicitly acting on the expressions in terms of the crosscap Ishibashi states. 
We see that for $P_{WW}^W = +1$ the fusion rules are realized linearly on the crosscap states, whereas for $P_{WW}^W = -1$ they are not.

\subsection{Ising UMTCs}

We next consider the Ising$^{(\nu)}$ UMTC, labelled by odd $\nu=1,3,\dots, 15$. This has the fusion rules 
\bea
\eps \times \eps =1 ~, \hspace{0.3 in} \eps \times \sigma = \sigma~, \hspace{0.3 in} \sigma \times \sigma = 1 + \eps~,
\eea
non-trivial F-symbols 
\bea
F_{\eps \sigma \eps}^\sigma = F_{\sigma \eps \sigma}^{\eps} = -1~, \hspace{0.3 in} F^\sigma_{\sigma\sigma\sigma} = {\kappa_\sigma \over \sqrt{2}} \left(\begin{matrix} 1& 1\\ 1& -1\end{matrix} \right) ~,
\eea
and FS indicators 
\bea
\kappa_1 = \kappa_\eps = 1~, \hspace{0.3 in} \kappa_\sigma = (-1)^{\nu^2 - 1 \over 8}~. 
\eea
The modular data (in the basis $\{1, \sigma, \eps\}$) is given by 
\bea
\S ={1 \over 2} \left(\begin{matrix}  1 & \sqrt{2} & 1 \\ \sqrt{2} & 0 & - \sqrt{2} \\ 1 & - \sqrt{2} & 1\end{matrix}  \right) ~, \hspace{0.5 in} \T = \ex{-2 \pi i {\mathsf{c}\over 24}}\, \mathrm{diag}(1, \ex{{\pi i \over 8}\nu},-1)~,
\eea 
where $\cc = {\nu \over 2}$ mod 8. The case of $\nu=1$ is realized by the Ising CFT, while $\nu=3$ is realized by $\mathfrak{su}(2)_2$. 

The crosscap states may be computed using (\ref{eq:simplecurrentcrosscaps}), giving
\bea
|C_1 \rangle &=& \eta_1 \left[ \sqrt{2}\, \mathrm{cos} {\pi \nu \over 8} |1\rangle\!\rangle_C + \sqrt{2}\, \mathrm{sin} {\pi \nu \over 8} |\eps\rangle\!\rangle_C\right] ~,
\no\\
|C_\eps \rangle &=& \eta_\eps \left[ \sqrt{2}\, \mathrm{sin} {\pi \nu \over 8} |1\rangle\!\rangle_C - \sqrt{2}\, \mathrm{cos} {\pi \nu \over 8} |\eps\rangle\!\rangle_C\right] ~,
\no\\
|C_\sigma\rangle &=& 2^{1/4} \eta_\sigma |\sigma \rangle\!\rangle_C~. 
\eea
Since $\eps \prec \sigma^2$, the sign $\eta_{\eps}$ is constrained by (\ref{eq:etaconstraint}) to satisfy 
\bea
\eta_\eps =  \sqrt{2}\, \eta_1 \mathrm{sin}{\pi \nu \over 4} P_{\sigma\sigma}^\eps~.
\eea
On the other hand, $P_{\sigma \sigma}^\eps$ is constrained by (\ref{eq:PFidentity}), which fixes $P_{\sigma\sigma}^\eps = P_{\sigma \sigma}^1 = \kappa_\sigma$. Unlike for Fibonacci, for the Ising UMTCs, all $P_{aa}^c$ can be fixed uniquely by the constraint (\ref{eq:PFidentity}).

We may now compute various inner products, such as 
\bea
\langle C_1 | \ex{{\pi i \over 2 \tau } (L_0 + \overline{L}_0 - {\mathsf{c} \over 12})} | C_1\rangle &=& \chi_1(2 \tau) + \chi_\eps(2 \tau) + \sqrt{2}\, \mathrm{cos}{\pi \nu \over 4} \chi_\sigma(2 \tau)~,
\no\\
\langle C_\eps | \ex{{\pi i \over 2 \tau } (L_0 + \overline{L}_0 - {\mathsf{c} \over 12})} | C_1\rangle &=& (-1)^{\nu^2 - 1 \over 8} \chi_\sigma(2 \tau)~,
\no\\
\langle C_\sigma | \ex{{\pi i \over 2 \tau } (L_0 + \overline{L}_0 - {\mathsf{c} \over 12})} | C_{1,\eps}\rangle &=& 0~,
\no\\
\langle C_{\sigma} | \ex{{\pi i \over 2 \tau } (L_0 + \overline{L}_0 - {\mathsf{c} \over 12})} | C_\sigma\rangle &=& \chi_1(2 \tau) - \chi_\eps(2 \tau)~.
\eea
These may be compared with the corresponding results on the twisted Klein bottles, and match in all cases. For example, the final inner product matches with 
\bea
\sum_{c = 1, \eps} d_\sigma^{-1} P^{c\sigma}_\sigma \mathrm{Tr}_{\cH_c} \cP\, \widehat \sigma \ex{2 \pi i \tau(L_0 + \overline{L}_0 - {\mathsf{c}\over 12})} = {2^{- \half}} \mathrm{Tr}_{\cH} \cP\, \widehat \sigma \ex{2 \pi i \tau(L_0 + \overline{L}_0 - {\mathsf{c}\over 12})} = \chi_1(2 \tau) - \chi_\eps(2 \tau)~,\hspace{0.2 in}
\eea
where we have no contributions from the trace over $\cH_\eps$ since $\cP$ restricts this trace to spin-0 primaries, namely $\mu_{{1\over 16},{1\over 16}}$, but this operator is annihilated by $\widehat \sigma$ (when treated as a map from $\cH_\eps$ to $\cH_\eps$).

We may also consider the action of Verlinde lines on these states. In particular, using (\ref{eq:actiononcc}) one finds 
\bea
\widehat \eps |C_1 \rangle &=& |C_1\rangle~, \hspace{0.3 in}\widehat \eps |C_\eps \rangle \,\,\,=\,\,\, |C_\eps\rangle~, \hspace{0.3in} \widehat \eps |C_\sigma \rangle \,\,\,=\,\,\, - |C_\sigma\rangle~, 
\\
\widehat \sigma |C_1 \rangle &=& \kappa_\sigma \left( |C_1 \rangle + | C_\eps \rangle \right)~, \hspace{0.2 in} \widehat \sigma |C_\eps \rangle \,\,\,=\,\,\, \kappa_\sigma \left( |C_1 \rangle - | C_\eps \rangle \right)~, \hspace{0.2 in} \widehat \sigma |C_\sigma\rangle = 0~,\no
\eea
which again matches the results obtained by explicitly acting on the expressions in terms of the crosscap Ishibashi states.
Note that when $\kappa_\sigma=-1$, namely for $\nu = 3,5,11,13$, the fusion rules of $\sigma$ are realized projectively on $|C_1\rangle$.
This includes the case of $\mathfrak{su}(2)_2$. 

Let us mention here that in recent work \cite{Zhang2024}, it was suggested that in the Ising model $\widehat \sigma |C_1\rangle$ provides a new crosscap state. We now see that this is not quite true---$\widehat \sigma |C_1\rangle$ is simply a sum of two previously known crosscaps $|C_1\rangle$ and $|C_\eps\rangle$. On the other hand, $|C_\sigma\rangle$ \textit{does} provide a new crosscap state for the Ising model.

\subsection*{Acknowledgements}

The authors thank Yichul Choi, Kentaro Hori, Kansei Inamura, Yu Nakayama, Zhengdi Sun, Xi Yin, Yifan Wang, and Yunqin Zheng for useful discussions, as well as Yichul Choi, Yifan Wang, and Yunqin Zheng for comments on the manuscript. JK and YK are supported by the INAMORI Frontier Program at Kyushu University. YK is also supported by JSPS KAKENHI Grant
Number 23K20046. 
\appendix 

\section{Modular Properties of Characters}
\label{app:modular}

 In this appendix we provide a brief review of basic properties of characters and the modular matrices acting on them. 
We first define the character for a representation $\mfH_a$ of the chiral algebra $\ca{A}$ by
\begin{equation}\label{eq:chi}
\chi_a(\tau) := \tr_{\mfH_a} \ex{2\pi i \tau\pa{L_0- \fr{\cc}{24}}}~. 
\end{equation}
The characters satisfy
\begin{equation}
\overline{\chi_a(\tau) } = \chi_{a^+}(-\bar{\tau})~,
\end{equation}
where $a^+$ is the charge conjugate of $a$.

The characters form a representation of $\mathrm{SL}(2,\ZZ)$ acting in the following way. 
First we introduce complex coordinates $z = x+i y$ on the torus. 
An arbitrary element $\gamma \in \mathrm{SL}(2, \ZZ)$ acts on these coordinates in the vector representation, namely 
\bea
\gamma: \hspace{0.5 in} (x,y)^T \mapsto (x',y')^T = \gamma (x,y)^T ~. 
\eea
In this representation, the generators of $\mathrm{SL}(2,\ZZ)$ may be written as\footnote{Note that $C$ acts trivially on the modular parameter $\tau$, so that the action on $\tau$ is given by the quotient $\mathrm{PSL}(2,\ZZ)$. On the coordinates themselves though, the full $\mathrm{SL}(2,\ZZ)$ acts faithfully.}
\bea
S = \left( \begin{matrix} 0 & -1 \\ 1 & 0 \end{matrix} \right) ~, \hspace{0.5 in}T  = \left( \begin{matrix} 1 & 1 \\ 0 & 1 \end{matrix} \right) ~, \hspace{0.5 in}C  = \left( \begin{matrix} -1 & 0 \\ 0 & -1\end{matrix} \right) ~,
\eea
satisfying the usual algebra 
\bea
S^2 = (ST)^3 = C~, \hspace{0.5 in} C^2 = 1~. 
\eea
The characters transform in a representation $\rho : \mathrm{SL}(2, \ZZ) \rightarrow \mathrm{GL}(d; \CC)$, with $d$ being the number of characters,
\bea
\chi_i(\tau') = \sum_{j=1}^d \rho(\gamma)_{ij}\, \chi_j(\tau)~, \hspace{0.5 in} \tau' = \gamma \tau~,
\eea
and with the action of $\gamma$ on $\tau$ given by linear fractional transformations. For convenience, we will use the following shorthand notation,
\bea
\S := \rho(S)~, \hspace{0.5 in} \T := \rho(T)~, \hspace{0.5 in} \C := \rho(C)~. 
\eea
The representation is required to satisfy a number of properties---for example, although the matrix $S$ is not symmetric, the representation $\rho$ must be chosen such that $\S$ is symmetric  ($\S_{ab} = \S_{ba}$). The representation is also constrained by unitarity ($\S^*_{ba} = \S^{-1}_{ab}$), the constraint $\S_{ab} = \S_{ab^+}^*$, and should also satisfy $\S_{1a}>0$ in a unitary theory. From (\ref{eq:chi}), one can show that $\T_{ab} = \delta_{a,b}\ex{2\pi i \pa{h_a-\fr{\cc}{24}}}$.

Let us now consider the torus partition function with the insertion of a Verlinde line $a$, 
\bea
\mathrm{Tr}_{\cH_1} \widehat a\, q^{L_0 - {\mathsf{c} \over 24 }}{\overline q}^{\overline L_0 - {\mathsf{c} \over 24 }} \hspace{0.2 in}= \hspace{0.2 in}
\begin{tikzpicture}[baseline=20,scale=0.6]
\begin{scope}[very thick, every node/.style={sloped,allow upside down}]
\shade[top color=blue, bottom color=white,opacity = 0.1] (0,0)--(3,0)--(3,3)--(0,3)--(0,0);

\draw[red] (0,1.5) -- node {\opmidarrow} (3,1.5);
\node[below, red] at (1.5,1.5) {$a$};

\draw(0,0) -- node {\midarrow} (3,0);
\draw (3,0)-- node {\dmidarrow}(3,3);
\draw (3,3)-- node {\opmidarrow}(0,3);
\draw  (0,0)-- node {\dmidarrow}(0,3);

\end{scope}
\end{tikzpicture}~,
\eea
where we have taken the convention that $\widehat a$ corresponds to a line directed towards the left. We would now like to relate this configuration to the one defining the $a$-twisted partition function, namely 
\bea
\mathrm{Tr}_{\cH_a}  q^{L_0 - {\mathsf{c} \over 24 }}{\overline q}^{\overline L_0 - {\mathsf{c} \over 24 }} \hspace{0.2 in}= \hspace{0.2 in}
\begin{tikzpicture}[baseline=20,scale=0.6]
\begin{scope}[very thick, every node/.style={sloped,allow upside down}]
\shade[top color=blue, bottom color=white,opacity = 0.1] (0,0)--(3,0)--(3,3)--(0,3)--(0,0);

\draw[red] (1.5,0) -- node {\midarrow} (1.5,3);
\node[right, red] at (1.5,1.5) {$a$};

\draw(0,0) -- node {\midarrow} (3,0);
\draw (3,0)-- node {\dmidarrow}(3,3);
\draw (3,3)-- node {\opmidarrow}(0,3);
\draw  (0,0)-- node {\dmidarrow}(0,3);

\end{scope}
\end{tikzpicture}~. 
\eea
To do so, it is clear that we must ``rotate'' the picture clockwise. More precisely, starting from the coordinate system $(x',y')$, we change to a coordinate system $(x,y)$ via $(x,y)^T = S^{-1} (x',y')^T$. Since $\rho$ is a representation, we of course have that $\rho(S^{-1}) = \S^{-1}$, and hence 
\bea
\mathrm{Tr}_{\cH_a}  q^{L_0 - {\mathsf{c} \over 24 }}{\overline q}^{\overline L_0 - {\mathsf{c} \over 24 }} &=& \mathrm{Tr}_{\cH_1} \widehat a\, {q'}^{L_0 - {\mathsf{c} \over 24 }}{\overline q'}^{\overline L_0 - {\mathsf{c} \over 24 }}  = \sum_b {\S_{ab} \over \S_{1b}} \chi_b(\tau') \chi_{b^+}(-\overline \tau')
\no\\
&=& \sum_{b,c,d} {\S_{ab} \over \S_{1b}} \S^{-1}_{bc} (\S^{-1}_{b^+ d})^* \chi_c(\tau) \chi_{d}(-\overline \tau)
\no\\
&=& \sum_{c,d} N_{a c^+}^d \chi_c(\tau) \chi_{d}(-\overline \tau)~,
\eea
where $q'= \ex{2 \pi i \tau'}$, and we have made use of the Verlinde formula  \cite{Verlinde1988}
\begin{equation}
N_{ab}^c= \sum_i \fr{\S_{a i} \S_{b i} {\S_{ci}}^*   }{\S_{1i}} ~
\end{equation}
to rewrite the expression in terms of the fusion coefficients $N_{ab}^c$. As a check of this result, note that when we consider the untwisted Hilbert space $\cH_1$, the coefficients are simply $N_{1 c^+}^d = \delta_{c^+, d}$, and we reobtain the charge-conjugation modular invariant. 
Let us also note in passing that the fusion coefficients are positive non-negative integers satisfying the following properties,
\bea
N_{ab}^c = N_{ba}^c = N_{a c^+}^{b^+}  =N_{a^+ b^+}^{c^+}~. 
\eea

Note that the convention chosen above with $\widehat b$ corresponding to a line directed towards the left is related to our convention for the lasso action in (\ref{eq:lassodiagram}) with $b$ encircling the operator counterclockwise. Indeed, by the state operator map, we have 
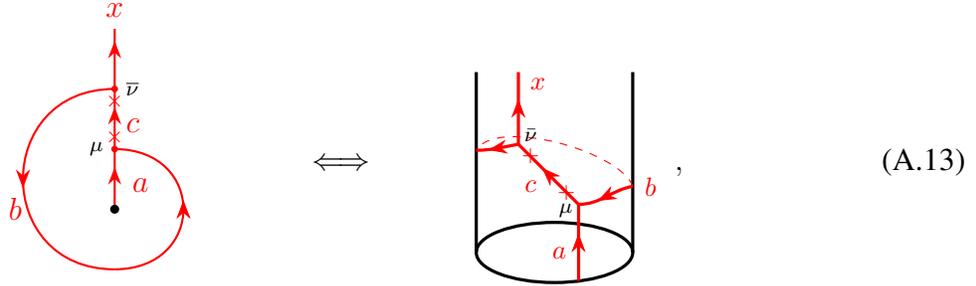
\bea
\begin{tikzpicture}[scale=0.8,baseline=15]
\begin{scope}[ thick, every node/.style={sloped,allow upside down}]
%\draw[->-=0.25, ->- = 0.55, ->- = 0.9](0,0) -- (0,3);
\draw[red, ->- = 0.7](0,0) -- (0,1);
\draw[red, ->-=0.7] (0,1) -- (0,2);
\draw[red, ->- = 0.8](0,2) -- (0,3);
\draw[red](0,1) to[out=0,in=0,distance=0.6in] node{\opmidarrow}  (0,-1);
\draw[red]  (0,-1) to[out=180,in=180,distance=0.8in] node{\opmidarrow}   (0,2);
\node[above,red] at (0,3) {$x$};
\node[left,red] at (-1.35,0) {$b$};
\node[right,red] at (0,1.4) {$c$};
\node[right,red] at (0.1,0.4) {$a$};
\node[red] at (0,1.2) {\scriptsize$\times$};
\node[red] at (0,1.8) {\scriptsize$\times$};

\filldraw[] (0,0) circle (0.3ex);
\filldraw[red] (0,1) circle (0.2ex);
\filldraw[red] (0,2) circle (0.2ex);

\node[left] at (0,1) {\scriptsize $\mu$};
\node[right] at (0,2) {\scriptsize $\overline \nu$};
\end{scope}
\end{tikzpicture} 
\hspace{0.5 in} \Longleftrightarrow \hspace{0.5 in}
\begin{tikzpicture}[scale=0.8,baseline=20]
\begin{scope}[very thick, every node/.style={sloped,allow upside down}]
    %bottom crosscap
  \draw (0,-0.5) ellipse [x radius=1.3cm, y radius=0.5cm];
  
    % center dash
 % \draw[thin, dashed] (-0.5,2.5) -- (-0.5,-1);
  %left right black line
  \draw (-1.3,2.5) -- (-1.3,-0.5);
  \draw ( 1.3,-0.5) -- ( 1.3,2.5);

      \draw[red,very thick] (0.4,-0.97) to node{\midarrow} (0.4,0.3);
      \draw[red,very thick] (0.4,0.3) to node{\midarrow} (-0.6,1.3);
      \draw[red,very thick] (-0.6,1.3) to node{\midarrow} (-0.6,2.5);
        
         \draw[red,very thick] (1.3,0.6) to[out = 190, in = 0] node{\midarrow} (0.4,0.3);
          \draw[red,very thick] (-1.3,1.2) to[out = 0, in = 190] node{\opmidarrow} (-0.6,1.3) ;
          
           \draw[red,thin,dashed] (1.3,0.6) to[out = 110, in = 60,distance=0.25in]  (-1.3,1.2) ;
        
        \node[red,left] at (0.4,-0.5) {\footnotesize$a$};
            \node[red,right] at (-0.6,2.3) {\footnotesize$x$};
             \node[red,left] at (-0.1,0.6) {\footnotesize$c$};
                \node[red,right] at ( 1.3,0.6) {\footnotesize$b$};
        
        \node[left] at (0.5,0.2) {\scriptsize$\mu$};
        \node[right] at (-0.7,1.45) {\scriptsize$\bar \nu$};
        
        \node[red] at (0.2,0.5) {\scriptsize$+$};
         \node[red] at (-0.4,1.1) {\scriptsize$+$};
         
  \end{scope}
\end{tikzpicture}~,
\eea
where we see that $b$ wraps the cylinder to the left. It is important to keep this convention in mind when we write down the generalized Cardy condition for crosscap states, as inappropriately using $\S$ instead of $\S^{-1}$ can lead to various inconsistencies.

Finally, let us close by recalling the definition of the twisted characters,
\begin{equation}
\widehat{\chi}_a(\tau) := \ex{-\pi i \pa{h_a-\fr{\cc}{24}}} \chi_a \pa{\tau+\fr{1}{2}}~.
\end{equation}
Modular transformations of these characters are described by the P-matrix, as is described further in the main text. Like the modular S-matrix, the modular P-matrix satisfies unitarity ($\P^*_{ba} = \P^{-1}_{ab}$) and symmetricity ($\P_{ab} = \P_{ba}$), as well as $\P^2 =\C$.

Note that the first row of $\P$ is necessarily real. This can be shown by using $\P^2 =\C$ and unitarity, from which it follows that $\P^* = \C \P$. 
Writing this in components and using $\C_{1b} = \delta_{1b}$, we then obtain 
\begin{equation}
\P^{*}_{1a}=(\C\,\P)_{1a}=\sum_{b}\C_{1b}\P_{ba}=\P_{1a}~,
\end{equation}
as claimed.

\section{Derivation of (\ref{eq:rearrangement})}
\label{app:computation}

In this appendix we give a brief derivation of the second equality of (\ref{eq:rearrangement}). 
We begin by nucleating a bubble of $h^{-1}$ in the $h$ line, producing a factor of $\kappa_h$, 
and then moving one of the vertices of the bubble through the crosscap, producing a second factor of $\kappa_h$, 
\bea
\begin{tikzpicture}[scale=0.7,baseline=20]
\shade[top color=blue, bottom color=white,opacity = 0.1] (0,-0.5)--(6,-0.5)--(6,2.5)--(0,2.5)--(0,-0.5);
\begin{scope}[very thick, every node/.style={sloped,allow upside down}]
%red arrow left
\draw[red] (1.5,-0.5) -- node {\opmidarrow}(1.5,1.0);
\draw[red] (1.5,1.0) -- node {\opmidarrow}(1.5,2.5);
% red arrow right
\draw[red] (4.5,2.5) -- node {\opmidarrow}(4.5,1.0);
\draw[red] (4.5,1.0) -- node {\opmidarrow}(4.5,-0.5);
% red arrow horizon
\draw[red] (1.5,1.0) -- node {\opmidarrow}(4.5,1.0);
%bottom part
\draw  (3,-0.5) --node {\midarrow}(6,-0.5);
\draw  (0,-0.5)--node {\midarrow}(3,-0.5);
%vertiiale line
\draw (6,-0.5)--node {\midarrow}(6,2.5);
\draw (0,-0.5)--node {\midarrow}(0,2.5);
%upper part
\draw(0,2.5) --node {\dmidarrow} (3,2.5);
\draw(3,2.5) --node {\dmidarrow} (6,2.5);
%dash line
\draw[thin, dashed] (3,2.5)--(3,-0.5);
% g part on the left
\node[red,left] at (1.5,1.7) {$h$};
\node[red,left] at (1.5,0.2) {$g$};
% g part on the right
\node[red,right] at (4.5,1.7) {$h$};
\node[red,right] at (4.5,0.2) {$g$};
% g on the center
\node[red,above] at (3.0,1.0) {$g h^{-1}$};

\draw[blue] (0,2.5)--(3,2.5);
\draw[blue] (0,-0.5)--(3,-0.5);

\node[red,below] at (1.5,1.1) {\footnotesize $\times$};
\node[red,below] at (4.5,1.1) {\footnotesize $\times$};
\end{scope}
\end{tikzpicture}
%MID%%%%%%%%%%%%%%%%%%%%%%%%%%%%%%%%%%%%%%%%%%%%%%%%%%%%%%%%%%%%%%%%%%
\hspace{0.1 in}=\hspace{0.05 in}
%RHS%%%%%%%%%%%%%%%%%%%%%%%%%%%%%%%%%%%%%%%%%%%%%%%%%%%%%%%%%%%%%%%%%%
\begin{tikzpicture}[scale=0.7,baseline=20]
\shade[top color=blue, bottom color=white,opacity = 0.1] (0,-0.5)--(6,-0.5)--(6,2.5)--(0,2.5)--(0,-0.5);
\begin{scope}[very thick, every node/.style={sloped,allow upside down}]
%red arrow left
\draw[red] (1.5,-0.5) -- node {\opmidarrow}(1.5,1.0);
\draw[red] (1.5,1.0) -- node {\opmidarrow}(1.5,1.75);
\draw[red] (1.5,1.75) -- node {\midarrow}(1.5,2.5);
% red arrow right
\draw[red] (4.5,2.5) -- node {\midarrow}(4.5,1.75);
\draw[red] (4.5,1.75) -- node {\opmidarrow}(4.5,1.0);
\draw[red] (4.5,1.0) -- node {\opmidarrow}(4.5,-0.5);
% red arrow horizon
\draw[red] (1.5,1.0) -- node {\opmidarrow}(4.5,1.0);
%bottom part
\draw  (3,-0.5) --node {\midarrow}(6,-0.5);
\draw  (0,-0.5)--node {\midarrow}(3,-0.5);
%vertiiale line
\draw (6,-0.5)--node {\midarrow}(6,2.5);
\draw (0,-0.5)--node {\midarrow}(0,2.5);
%upper part
\draw(0,2.5) --node {\dmidarrow} (3,2.5);
\draw(3,2.5) --node {\dmidarrow} (6,2.5);
%dash line
\draw[thin, dashed] (3,2.5)--(3,-0.5);
% g part on the left
\node[red,left] at (1.5,2.1) {$h^{-1}$};
\node[red,left] at (1.5,1.4) {$h$};
\node[red,left] at (1.5,0.2) {$g$};
% g part on the right
\node[red,right] at (4.5,2.1) {$h^{-1}$};
\node[red,right] at (4.5,1.4) {$h$};
\node[red,right] at (4.5,0.2) {$g$};
% g on the center
\node[red,above] at (3.0,1.0) {$g h^{-1}$};

\draw[blue] (0,2.5)--(3,2.5);
\draw[blue] (0,-0.5)--(3,-0.5);

\filldraw[red] (1.5, 1.75) circle (0.2 ex);
\filldraw[red] (4.5, 1.75) circle (0.2 ex);

\node[red,below] at (1.5,1.1) {\footnotesize $\times$};
\node[red,below] at (4.5,1.1) {\footnotesize $\times$};
\end{scope}
\end{tikzpicture}
~. 
\eea
We now perform an F-move on the right-hand side, with the internal line being the identity line running between the right $h$-$h^{-1}$ vertex and the central $gh^{-1}$ line,
and use the second equation of (\ref{eq:basisconventions}) to collapse a bubble to obtain,
\bea
\begin{tikzpicture}[scale=0.7,baseline=20]
\shade[top color=blue, bottom color=white,opacity = 0.1] (0,-0.5)--(6,-0.5)--(6,2.5)--(0,2.5)--(0,-0.5);
\begin{scope}[very thick, every node/.style={sloped,allow upside down}]
%red arrow left
\draw[red] (1.5,-0.5) -- node {\opmidarrow}(1.5,1.0);
\draw[red] (1.5,1.0) -- node {\opmidarrow}(1.5,1.75);
\draw[red] (1.5,1.75) -- node {\midarrow}(1.5,2.5);
% red arrow right
\draw[red] (4.5,2.5) -- node {\midarrow}(4.5,1.75);
\draw[red] (4.5,1.75) -- node {\opmidarrow}(4.5,1.0);
\draw[red] (4.5,1.0) -- node {\opmidarrow}(4.5,-0.5);
% red arrow horizon
\draw[red] (1.5,1.0) -- node {\opmidarrow}(4.5,1.0);
%bottom part
\draw  (3,-0.5) --node {\midarrow}(6,-0.5);
\draw  (0,-0.5)--node {\midarrow}(3,-0.5);
%vertiiale line
\draw (6,-0.5)--node {\midarrow}(6,2.5);
\draw (0,-0.5)--node {\midarrow}(0,2.5);
%upper part
\draw(0,2.5) --node {\dmidarrow} (3,2.5);
\draw(3,2.5) --node {\dmidarrow} (6,2.5);
%dash line
\draw[thin, dashed] (3,2.5)--(3,-0.5);
% g part on the left
\node[red,left] at (1.5,2.1) {$h^{-1}$};
\node[red,left] at (1.5,1.4) {$h$};
\node[red,left] at (1.5,0.2) {$g$};
% g part on the right
\node[red,right] at (4.5,2.1) {$h^{-1}$};
\node[red,right] at (4.5,1.4) {$h$};
\node[red,right] at (4.5,0.2) {$g$};
% g on the center
\node[red,above] at (3.0,1.0) {$g h^{-1}$};

\draw[blue] (0,2.5)--(3,2.5);
\draw[blue] (0,-0.5)--(3,-0.5);

\filldraw[red] (1.5, 1.75) circle (0.2 ex);
\filldraw[red] (4.5, 1.75) circle (0.2 ex);

\node[red,below] at (1.5,1.1) {\footnotesize $\times$};
\node[red,below] at (4.5,1.1) {\footnotesize $\times$};
\end{scope}
\end{tikzpicture}%MID%%%%%%%%%%%%%%%%%%%%%%%%%%%%%%%%%%%%%%%%%%%%%%%%%%%%%%%%%%%%%%%%%%
\hspace{0.1 in}=\,\,\, (F_{gh^{-1}, h, h^{-1}}^{gh^{-1}})^{-1}\hspace{0.05 in}
%RHS%%%%%%%%%%%%%%%%%%%%%%%%%%%%%%%%%%%%%%%%%%%%%%%%%%%%%%%%%%%%%%%%%%
\begin{tikzpicture}[scale=0.7,baseline=20]
\shade[top color=blue, bottom color=white,opacity = 0.1] (0,-0.5)--(6,-0.5)--(6,2.5)--(0,2.5)--(0,-0.5);
\begin{scope}[very thick, every node/.style={sloped,allow upside down}]
%red arrow left
\draw[red] (1.5,-0.5) -- node {\opmidarrow}(1.5,1.0);
\draw[red] (1.5,1.0) -- node {\opmidarrow}(1.5,1.75);
\draw[red] (1.5,1.75) -- node {\midarrow}(1.5,2.5);
% red arrow right
\draw[red] (4.5,2.5) -- node {\midarrow}(4.5,1.0);
\draw[red] (4.5,1.0) -- node {\opmidarrow}(4.5,-0.5);
% red arrow horizon
\draw[red] (1.5,1.0) -- node {\opmidarrow}(4.5,1.0);
%bottom part
\draw  (3,-0.5) --node {\midarrow}(6,-0.5);
\draw  (0,-0.5)--node {\midarrow}(3,-0.5);
%vertiiale line
\draw (6,-0.5)--node {\midarrow}(6,2.5);
\draw (0,-0.5)--node {\midarrow}(0,2.5);
%upper part
\draw(0,2.5) --node {\dmidarrow} (3,2.5);
\draw(3,2.5) --node {\dmidarrow} (6,2.5);
%dash line
\draw[thin, dashed] (3,2.5)--(3,-0.5);
% g part on the left
\node[red,left] at (1.5,2.1) {$h^{-1}$};
\node[red,left] at (1.5,1.4) {$h$};
\node[red,left] at (1.5,0.2) {$g$};
% g part on the right
\node[red,right] at (4.5,1.8) {$h^{-1}$};
\node[red,right] at (4.5,0.2) {$g$};
% g on the center
\node[red,above] at (3.0,1.0) {$g h^{-1}$};

\draw[blue] (0,2.5)--(3,2.5);
\draw[blue] (0,-0.5)--(3,-0.5);

\filldraw[red] (1.5, 1.75) circle (0.2 ex);
\filldraw[red] (4.5, 1.75) circle (0.2 ex);

\node[red,below] at (1.5,1.1) {\footnotesize $\times$};
\node[red,left] at (4.5,1) {\footnotesize $\times$};
\end{scope}
\end{tikzpicture}~. 
\eea
Another F-move with the internal line being $g h^{-1}$ then gives 
\bea
 (F_{gh^{-1}, h, h^{-1}}^{gh^{-1}})^{-1}
 \begin{tikzpicture}[scale=0.7,baseline=20]
\shade[top color=blue, bottom color=white,opacity = 0.1] (0,-0.5)--(6,-0.5)--(6,2.5)--(0,2.5)--(0,-0.5);
\begin{scope}[very thick, every node/.style={sloped,allow upside down}]
%red arrow left
\draw[red] (1.5,-0.5) -- node {\opmidarrow}(1.5,1.0);
\draw[red] (1.5,1.0) -- node {\opmidarrow}(1.5,1.75);
\draw[red] (1.5,1.75) -- node {\midarrow}(1.5,2.5);
% red arrow right
\draw[red] (4.5,2.5) -- node {\midarrow}(4.5,1.0);
\draw[red] (4.5,1.0) -- node {\opmidarrow}(4.5,-0.5);
% red arrow horizon
\draw[red] (1.5,1.0) -- node {\opmidarrow}(4.5,1.0);
%bottom part
\draw  (3,-0.5) --node {\midarrow}(6,-0.5);
\draw  (0,-0.5)--node {\midarrow}(3,-0.5);
%vertiiale line
\draw (6,-0.5)--node {\midarrow}(6,2.5);
\draw (0,-0.5)--node {\midarrow}(0,2.5);
%upper part
\draw(0,2.5) --node {\dmidarrow} (3,2.5);
\draw(3,2.5) --node {\dmidarrow} (6,2.5);
%dash line
\draw[thin, dashed] (3,2.5)--(3,-0.5);
% g part on the left
\node[red,left] at (1.5,2.1) {$h^{-1}$};
\node[red,left] at (1.5,1.4) {$h$};
\node[red,left] at (1.5,0.2) {$g$};
% g part on the right
\node[red,right] at (4.5,1.8) {$h^{-1}$};
\node[red,right] at (4.5,0.2) {$g$};
% g on the center
\node[red,above] at (3.0,1.0) {$g h^{-1}$};

\draw[blue] (0,2.5)--(3,2.5);
\draw[blue] (0,-0.5)--(3,-0.5);

\filldraw[red] (1.5, 1.75) circle (0.2 ex);
\filldraw[red] (4.5, 1.75) circle (0.2 ex);

\node[red,below] at (1.5,1.1) {\footnotesize $\times$};
\node[red,left] at (4.5,1) {\footnotesize $\times$};
\end{scope}
\end{tikzpicture}
%MID%%%%%%%%%%%%%%%%%%%%%%%%%%%%%%%%%%%%%%%%%%%%%%%%%%%%%%%%%%%%%%%%%%
\hspace{0.1 in}&=&\,\,\, (F_{gh^{-1}, h, h^{-1}}^{gh^{-1}})^{-1}F_{g,h^{-1},h}^g \hspace{0.05 in}
%RHS%%%%%%%%%%%%%%%%%%%%%%%%%%%%%%%%%%%%%%%%%%%%%%%%%%%%%%%%%%%%%%%%%%
\begin{tikzpicture}[scale=0.7,baseline=20]
\begin{scope}[very thick, every node/.style={sloped,allow upside down}]
\shade[top color=blue, bottom color=white,opacity = 0.1] (0,-0.5)--(0,2.5)--(6,2.5)--(6,-0.5)--(0,-0.5);
%red curved
\draw[red] (4.5,2.5) to[out=-90,in=-90,distance=0.6 in ]node {\opmidarrow}(1.5,2.5); 
\draw[red] (4.5,2.5) to[out=-90,in=45 ]node {\midarrow} (4.15, 1.75); 
\draw[red] (1.5,2.5) to[out=-90,in=135 ]node {\opmidarrow} (1.85, 1.75) ; 

\draw[red] (1.5,-0.5) to[out=90,in=90,distance=0.6 in ] node {\opmidarrow}(4.5,-0.5); 

%bottom part
\draw  (3,-0.5) --node {\midarrow}(6,-0.5);
\draw  (0,-0.5)--node {\midarrow}(3,-0.5);
%vertiiale line
\draw (6,-0.5)--node {\midarrow}(6,2.5);
\draw (0,-0.5)--node {\midarrow}(0,2.5);
%upper part
\draw(0,2.5) --node {\dmidarrow} (3,2.5);
\draw(3,2.5) --node {\dmidarrow} (6,2.5);
%dash line
\draw[thin, dashed] (3,2.5)--(3,-0.5);
% g part on the left
\node[red,left] at (1.6,2.1) {$h^{-1}$};
\node[red,left] at (1.7,0) {$g$};
% g part on the right
\node[red,right] at (4.4,2.1) {$h^{-1}$};
\node[red,right] at (4.4,0) {$g$};

\node[red,right] at (2.6,1.8) {$h$};
% g on the center
%\node[red,above] at (3.0,1.0) {$g_1 g_2^{-1}$};

\draw[blue] (0,2.5)--(3,2.5);
\draw[blue] (0,-0.5)--(3,-0.5);

\filldraw[red] (4.15, 1.75) circle (0.3 ex);
\filldraw[red] (1.85, 1.75) circle (0.3 ex);

\end{scope}
\end{tikzpicture}
\no\\
%MID%%%%%%%%%%%%%%%%%%%%%%%%%%%%%%%%%%%%%%%%%%%%%%%%%%%%%%%%%%%%%%%%%%
\hspace{0.1 in}&\vphantom{.}&\hspace{-1.5 in}=\,\,\, (F_{gh^{-1}, h, h^{-1}}^{gh^{-1}})^{-1}F_{g,h^{-1},h}^g \hspace{0.05 in}
%RHS%%%%%%%%%%%%%%%%%%%%%%%%%%%%%%%%%%%%%%%%%%%%%%%%%%%%%%%%%%%%%%%%%%
\begin{tikzpicture}[scale=0.7,baseline=20]
\begin{scope}[very thick, every node/.style={sloped,allow upside down}]
\shade[top color=blue, bottom color=white,opacity = 0.1] (0,-0.5)--(0,2.5)--(6,2.5)--(6,-0.5)--(0,-0.5);
%red curved
\draw[red] (4.5,2.5) to[out=-90,in=-90,distance=0.6 in ]node {\opmidarrow}(1.5,2.5); 
\draw[red] (1.5,-0.5) to[out=90,in=90,distance=0.6 in ] node {\opmidarrow}(4.5,-0.5); 
%bottom part
\draw  (3,-0.5) --node {\midarrow}(6,-0.5);
\draw  (0,-0.5)--node {\midarrow}(3,-0.5);
%vertiiale line
\draw (6,-0.5)--node {\midarrow}(6,2.5);
\draw (0,-0.5)--node {\midarrow}(0,2.5);
%upper part
\draw(0,2.5) --node {\dmidarrow} (3,2.5);
\draw(3,2.5) --node {\dmidarrow} (6,2.5);
%dash line
\draw[thin, dashed] (3,2.5)--(3,-0.5);
% g part on the left
\node[red,left] at (1.6,1.7) {$h$};
\node[red,left] at (1.7,0) {$g$};
% g part on the right
\node[red,right] at (4.4,1.7) {$h$};
\node[red,right] at (4.4,0) {$g$};

% g on the center
%\node[red,above] at (3.0,1.0) {$g_1 g_2^{-1}$};

\draw[blue] (0,2.5)--(3,2.5);
\draw[blue] (0,-0.5)--(3,-0.5);

\end{scope}
\end{tikzpicture}~.
\eea
Finally, from the following pentagon identity, 
\bea
\begin{tikzpicture}[baseline=0,scale = 0.5, baseline=-10,rotate=180]
\draw [very thick] (0,0) to (0,-2);
\draw [very thick] (0,0) to (2.5,2.5);
\draw [very thick] (0,0) to (-0.84,0.84);
\draw [very thick,dashed] (-0.833333,0.833333) to (-1.66667,1.66667);
\draw [very thick] (-1.6,1.6) to (-2.5,2.5);

\draw [very thick] (-0.75,0.75) to (1.75-0.75,1.75+0.75);
\draw [very thick] (-1.5,1.5) to (1-1.5,1+1.5);
\node[above] at (0,-2) {$gh^{-1}$};
\node[below] at (-2.5,2.5) {$h^{-1}$};
\node[below] at (1.75-0.75,1.75+0.65) {$h^{-1}$};
\node[below] at (1-1.5,1+1.5) {$h$};
\node[below] at (2.5,2.5) {$g$};
\node[right] at (-0.1,0) {$h^{-1}$};
\node[right] at (-0.77,0.76) {$1$};

\begin{scope}[xshift=5in, yshift = 2in]
\draw [very thick] (0,0) to (0,-2);
\draw [very thick] (0,0) to (2.5,2.5);
\draw [very thick] (-1.25,1.25) to (-2.5,2.5);
\draw [very thick,dashed] (0,0) to (-1.25,1.25);
\draw [very thick] (1.5,1.5) to (-1+1.5,1+1.5);
\draw [very thick] (-1.5,1.5) to (1-1.5,1+1.5);
\node[above] at (0,-2) {$gh^{-1}$};
\node[below] at (-2.5,2.5) {$h^{-1}$};
\node[below] at (1.75-0.75,1.75+0.65) {$h^{-1}$};
\node[below] at (1-1.5,1+1.5) {$h$};
\node[below] at (2.5,2.5) {$g$};
\node[right] at (-0.5,0.3) {$1$};
\node[left] at (0.5,0.3) {$g h^{-1}$};
\end{scope}

\begin{scope}[xshift=10in, yshift = 0in]
\draw [very thick] (0,0) to (0,-2);
\draw [very thick] (0,0) to (2.5,2.5);
\draw [very thick] (0,0) to (-2.5,2.5);
\draw [very thick] (0.75,0.75) to (-1.75+0.75,1.75+0.75);
\draw [very thick] (1.5,1.5) to (-1+1.5,1+1.5);
\node[above] at (0,-2) {$g h^{-1}$};
\node[below] at (-2.5,2.5) {$h^{-1}$};
\node[below] at (1.75-0.75,1.75+0.65) {$h^{-1}$};
\node[below] at (1-1.5,1+1.5) {$h$};
\node[below] at (2.5,2.5) {$g$};
\node[left] at (0.2,0) {$g $};
\node[left] at (0.85,0.78) {$g h^{-1}$};
\end{scope}

\begin{scope}[xshift=7in, yshift = -1in]
\draw [very thick] (0,0) to (0,-2);
\draw [very thick] (0,0) to (2.5,2.5);
\draw [very thick] (0,0) to (-2.5,2.5);
\draw [very thick,dashed] (0.75,0.75) to (-0.875+0.75,0.875+0.75);
\draw [very thick] (-0.875+0.75,0.875+0.75) to (-1.75+0.75,1.75+0.75);
\draw [very thick] (-0.125,1.625) to (0.875-0.125,0.875+1.625);
\node[above] at (0,-2) {$g h^{-1}$};
\node[below] at (-2.5,2.5) {$h^{-1}$};
\node[below] at (1.75-0.75,1.75+0.65) {$h^{-1}$};
\node[below] at (1-1.5,1+1.5) {$h$};
\node[below] at (2.5,2.5) {$g$};
\node[left] at (0.2,0) {$g$};
\node[right] at (0.5,0.9) {$1$};
\end{scope}

\begin{scope}[xshift=3in, yshift = -1in]
\draw [very thick] (0,0) to (0,-2);
\draw [very thick] (0,0) to (2.5,2.5);
\draw [very thick] (0,0) to (-2.5,2.5);
\draw [very thick,dashed] (-0.75,0.75) to (0.875-0.75,0.875+0.75);
\draw [very thick] (0.875-0.75,0.875+0.75) to (1.75-0.75,1.75+0.75);
\draw [very thick] (0.125,1.625) to (-0.875+0.125,0.875+1.625);
\node[above] at (0,-2) {$g h^{-1}$};
\node[below] at (-2.5,2.5) {$h^{-1}$};
\node[below] at (1.75-0.75,1.75+0.65) {$h^{-1}$};
\node[below] at (1-1.5,1+1.5) {$h$};
\node[below] at (2.5,2.5) {$g$};
\node[right] at (-0.2,0) {$h^{-1}$};
\node[left] at (-0.6,0.9) {$1$};
\end{scope}

\draw[thick,stealth-] (2,-2) -- (5,-3);
\draw[thick,stealth-] (11,-3) -- (14.5,-3);
\draw[thick,stealth-] (20,-3) -- (23,-2);
\draw[thick,stealth-] (2,4) -- (8.5,6);
\draw[thick,stealth-] (16,6) -- (6.5+16,4);

\node[right] at (4.5, -3.5) {$\kappa_h$};
\node[above] at (12.75, -3) {$1$};
\node[right] at (23.5, -3.5) {$F_{g h^{-1} h}^g$};
\node[above] at (4.5, 7) {$1$};
\node[above] at (20, 7) {$F_{gh^{-1}, h, h^{-1}}^{gh^{-1}}$};

\end{tikzpicture}
\eea
we find that 
\bea
 (F_{gh^{-1}, h, h^{-1}}^{gh^{-1}})^{-1}F_{g,h^{-1},h}^g = \kappa_h~, 
\eea
which gives the right-hand side of (\ref{eq:rearrangement}).

Note that this result can be thought of as a special case of the following generalization of the completeness relation in (\ref{eq:basisconventions}), 
\bea
\label{eq:generalizedcompl}
\begin{tikzpicture}[baseline={([yshift=-1ex]current bounding box.center)},vertex/.style={anchor=base,
    circle,fill=black!25,minimum size=18pt,inner sep=2pt},scale=0.5]
   \draw[->-=0.5,thick] (0,0) -- (0,4);
   \draw[-<-=0.5,thick] (2,0) -- (2,4);
     \node[below] at (0,-0.2) {$a$};
     \node[below] at (2,0) {$b$};
    \end{tikzpicture} 
\hspace{0.2 in }= \hspace{0.2 in }\kappa_b \sum_{c}  \sum_{\nu =1}^{N^a_{bc}}  \sqrt{d_c \over d_a d_b}
\begin{tikzpicture}[baseline={([yshift=-1ex]current bounding box.center)},vertex/.style={anchor=base,
    circle,fill=black!25,minimum size=18pt,inner sep=2pt},scale=0.6]
   \draw[->-=0.7,thick] (0,0) -- (1,1);
   \draw[-<-=0.7,thick] (2,0) -- (1,1);
    \draw[->-=0.6,thick] (1,1) -- (1,3);
     \draw[->-=0.7,thick] (1,3) -- (0,4);
       \draw[-<-=0.7,thick] (1,3) -- (2,4);
     \node[below] at (0,-0.2) {$a$};
     \node[below] at (2,0) {$b$};
      \node[above] at (0,4) {$a$};
     \node[above] at (2,4) {$b$};
      \node[right] at (1,2) {$c$};
      \node[above] at (0.75,0.4) {\footnotesize$+$};
      \node[below] at (0.75,3.6) {\footnotesize$+$};
      
      \node[left] at (1,1.1) {\scriptsize$\overline\nu$};
       \node[left] at (1,2.8) {\scriptsize$\nu$};
    \end{tikzpicture}~,
\eea
with $\kappa_b$ the FS indicator of $b$.
This can be derived as follows. We begin by using the definition of the $\kappa_b$ to re-express the 
fusion trivalent junctions in (\ref{eq:Homspacebasis}) as 
\begin{equation}
     \begin{tikzpicture}[baseline={([yshift=-1ex]current bounding box.center)},vertex/.style={anchor=base,
    circle,fill=black!25,minimum size=18pt,inner sep=2pt},scale=0.4]
   \draw[->-=0.5,thick] (-2,-2) -- (0,0);
   \draw[->-=0.5,thick] (2,-2) -- (0,0);
   \draw[->-=0.7,thick] (0,0)--(0,2) ;
     \node[below] at (-2,-2.2) {$a$};
     \node[below] at (2,-2) {$b$};
     \node[above] at (0,2) {$c$};
     \node[above] at (0,-0.2) {\footnotesize$\times$};
     \node[right] at (0,0) {\scriptsize$\mu$};
    \end{tikzpicture}
    = \kappa_b
   \begin{tikzpicture}[baseline={([yshift=-1ex]current bounding box.center)},vertex/.style={anchor=base,
    circle,fill=black!25,minimum size=18pt,inner sep=2pt},scale=0.4]
   \draw[->-=0.5,thick] (-2,-2) -- (0,0);
   \draw[->-=0.2,-<-=0.5,->-=0.85,thick] (5,-2) -- (0,0);
   \draw[->-=0.7,thick] (0,0)--(0,2) ;
     \node[below] at (-2,-2.2) {$a$};
     \node[below] at (1.5,1) {$b$};
     \node[below] at (2.5,-1.2) {$b^+$};
     \node[below] at (4.5,-2) {$b$};
     \node[above] at (0,2) {$c$};
     \node[above] at (0,-0.2) {\footnotesize$\times$};
     \node[right] at (0,0.3 ) {\scriptsize$\mu$};
     \filldraw[black] (2,-0.8) circle (1mm); 
     \filldraw[black] (3.5,-1.4) circle (1mm); 
     \draw[dashed] (2,-0.8) to [out = -90, in = -90,distance = 0.8 cm]   (-0.8,-0.8);

    \end{tikzpicture}
    \label{eq:Bsymbollike}~.
\end{equation}
Then performing a G-move with internal line being the identity line shown above, and shrinking the resulting bubble using (\ref{eq:basisconventions}), we obtain
\bea
   \begin{tikzpicture}[baseline={([yshift=-1ex]current bounding box.center)},vertex/.style={anchor=base,
    circle,fill=black!25,minimum size=18pt,inner sep=2pt},scale=0.4]
   \draw[->-=0.5,thick] (-2,-2) -- (0,0);
   \draw[->-=0.5,thick] (2,-2) -- (0,0);
   \draw[->-=0.7,thick] (0,0)--(0,2) ;
     \node[below] at (-2,-2.2) {$a$};
     \node[below] at (2,-2) {$b$};
     \node[above] at (0,2) {$c$};
     \node[above] at (0,-0.2) {\footnotesize$\times$};
     \node[right] at (0,0) {\scriptsize$\mu$};
    \end{tikzpicture} =\:\:\kappa_b \sqrt{\frac{d_a d_b}{d_c}} \sum_{\nu=1}^{N_{b^{+}c}^a}\,\, (F_{abb^{+}}^a)_{(c {\mu\nu})1}
     \begin{tikzpicture}[baseline={([yshift=-1ex]current bounding box.center)},vertex/.style={anchor=base,
    circle,fill=black!25,minimum size=18pt,inner sep=2pt},scale=0.4]
   \draw[->-=0.5,thick] (-2,-2) -- (0,0);
   \draw[-<-=0.8,->-=0.3,thick] (2,-2) -- (0,0);
   \draw[->-=0.7,thick] (0,0)--(0,2) ;
     \node[below] at (-2,-2.2) {$a$};
     \node[below] at (2,-2) {$b$};
     \node[below] at (0.5,-0.75) {$b^{+}$};
     \node[above] at (0,2) {$c$};
     \node[above] at (-0.5,-1.05) {\footnotesize$+$};
     \filldraw[black] (1.1,-1.1) circle (1mm);
     \node[left] at (0,0) {\scriptsize$\bar\nu$};
    \end{tikzpicture}~;
        \label{eq:definitionofBsymbol1}
    \eea
     likewise for the splitting junctions,
    \bea
    \begin{tikzpicture}[baseline={([yshift=-1ex]current bounding box.center)},vertex/.style={anchor=base,
    circle,fill=black!25,minimum size=18pt,inner sep=2pt},scale=0.35]
   \draw[->-=0.6,thick] (0,-2) -- (0,0);
   \draw[->-=0.6,thick]  (0,0) -- (2,2);
   \draw[->-=0.6,thick] (0,0)--(-2,2) ;
     \node[above] at (2,2) {$b$};
     \node[above] at (-2,2) {$a$};
     \node[below] at (0,-2) {$c$}; 
       \node[right] at (0,-0.2) {\scriptsize$\overline \mu$};
      \node[] at (0,-0.4) {\footnotesize$\times$};
    \end{tikzpicture} = \kappa_b \sqrt{\frac{d_a d_b}{d_c}}\sum_{\nu=1}^{N_{b^{+}c}^a}\,\, (F^a_{abb^+})^{-1}_{1(c{\mu\nu})}
       \begin{tikzpicture}[baseline={([yshift=-1ex]current bounding box.center)},vertex/.style={anchor=base,
    circle,fill=black!25,minimum size=18pt,inner sep=2pt},scale=0.35]
   \draw[->-=0.6,thick] (0,-2) -- (0,0);
   \draw[-<-=0.4,->-=0.95,thick]  (0,0) -- (2,2);
   \draw[->-=0.6,thick] (0,0)--(-2,2) ;
     \node[above] at (0.5,0.5) {$b^{+}$};
     \node[above] at (2,2) {$b$};
     \node[above] at (-2,2) {$a$};
     \node[below] at (0,-2) {$c$};
     \filldraw[black] (1.1,1.1) circle (1mm);
       \node[below] at (-0.4,0.2) {\scriptsize$ \nu$};
      \node[] at (-0.4,0.4) {\footnotesize$+$};
    \end{tikzpicture} ~.
    \label{eq:definitionofBsymbol2}
\eea
Using this and the usual completeness relation, we may then show (\ref{eq:generalizedcompl}) as follows, 
\bea
\begin{tikzpicture}[baseline={([yshift=-1ex]current bounding box.center)},vertex/.style={anchor=base,
    circle,fill=black!25,minimum size=18pt,inner sep=2pt},scale=0.5]
   \draw[->-=0.5,thick] (0,0) -- (0,4);
   \draw[-<-=0.5,thick] (2,0) -- (2,4);
     \node[below] at (0,-0.2) {$a$};
     \node[below] at (2,0) {$b$};
    \end{tikzpicture} 
&=& \:\:\kappa_b\:\:
    \begin{tikzpicture}[baseline={([yshift=-1ex]current bounding box.center)},vertex/.style={anchor=base,
    circle,fill=black!25,minimum size=18pt,inner sep=2pt},scale=0.4]
   \draw[->-=0.5,thick] (0,0) -- (0,6);
   \draw[-<-=0.2,->-=0.55,-<-=0.9,thick] (2,0) -- (2,6);
     \node[left] at (0,-0.2) {$a$};
     \node[right] at (2,0) {$b$};
     \node[right] at (2,3) {$b^+$};
     \node[right] at (2,5.5) {$b$};
    %  \node[left] at (0,5.5) {$a$};
    %  \node[left] at (0,3) {$a^{\:\:\:}$};
  % \filldraw[black] (0,2) circle (1mm);   
  % \filldraw[black] (0,4) circle (1mm); 
  \filldraw[black] (2,2) circle (1mm);   
  \filldraw[black] (2,4) circle (1mm); 
  % \draw[thin, dashed] (0,2) -- (2,2);
  % \draw[thin, dashed] (0,4) -- (2,4);
    \end{tikzpicture} 
    =\kappa_b \sum_{c} \sum_{\mu=1}^{N^c_{ab^{+}}} \sqrt{d_c \over d_a d_b} \,\,\,
    \begin{tikzpicture}[baseline={([yshift=-1ex]current bounding box.center)},vertex/.style={anchor=base,
    circle,fill=black!25,minimum size=18pt,inner sep=2pt},scale=0.5]
   \draw[->-=0.7,thick] (0,0) -- (1,1);
   \draw[-<-=0.3,->-=0.8,thick] (3,-1) -- (1,1);
    \draw[->-=0.6,thick] (1,1) -- (1,3);
     \draw[->-=0.7,thick] (1,3) -- (0,4);
       \draw[->-=0.3,-<-=0.8,thick] (1,3) -- (3,5);
     \node[below] at (0,-0.2) {$a$};
     \node[below] at (1.5,0.3) {$b^+$};
     \node[below] at (2.5,-0.5) {$b$};
      \node[above] at (0,4) {$a$};
     \node[above] at (1.5,3.7) {$b^+$};
     \node[above] at (2.5,4.5) {$b$};
      \node[right] at (1,2) {$c$};
      \node[above] at (1,0.8) {\footnotesize$\times$};
      \node[below] at (1,3.2) {\footnotesize$\times$};
      \filldraw[black] (2,4) circle (1mm); 
      \filldraw[black] (2,0) circle (1mm); 
      \node[left] at (1,1.1) {\scriptsize$\mu$};
       \node[left] at (1,2.8) {\scriptsize$\overline\mu$};
    \end{tikzpicture}
    \no\\
  &=& \kappa_b \sum_{c} \sum_{\mu=1}^{N^c_{ab^{+}}} \sum_{\nu, \rho=1}^{N^a_{bc}} \sqrt{d_c \over d_a d_b} {\frac{d_a d_b}{d_c}}  (F_{ab^+b}^a)_{(c {\mu\nu})1} (F_{ab^+b}^a)^{-1}_{1(c {\mu\rho})} \,\,\,
    \begin{tikzpicture}[baseline={([yshift=-1ex]current bounding box.center)},vertex/.style={anchor=base,
    circle,fill=black!25,minimum size=18pt,inner sep=2pt},scale=0.5]
   \draw[->-=0.7,thick] (0,0) -- (1,1);
   \draw[-<-=0.2,->-=0.55,-<-=0.9,thick] (4,-2) -- (1,1);
    \draw[->-=0.6,thick] (1,1) -- (1,3);
     \draw[->-=0.7,thick] (1,3) -- (0,4);
       \draw[-<-=0.2,->-=0.55,-<-=0.9,thick] (1,3) -- (4,6);
     \node[below] at (0,-0.2) {$a$};
     \node[below] at (1.5,0.3) {$b$};
     \node[below] at (2.5,-0.6) {$b^{+}$};
     \node[above] at (3.8,-1.2) {$b$};
      \node[above] at (0,4) {$a$};
     \node[above] at (1.5,3.7) {$b$};
     \node[above] at (2.5,4.5) {$b^{+}$};
     \node[below] at (3.8,5.5) {$b$};
      \node[right] at (1,2) {$c$};
      \node[above] at (0.8,0.3) {\footnotesize$+$};
      \node[below] at (0.8,3.7) {\footnotesize$+$};
      \filldraw[black] (2,4) circle (1mm); 
      \filldraw[black] (3,5) circle (1mm); 
      \filldraw[black] (2,0) circle (1mm); 
      \filldraw[black] (3,-1) circle (1mm); 
      \node[left] at (1,1.1) {\scriptsize$\overline\nu$};
       \node[left] at (1,2.8) {\scriptsize$\rho$};
    \end{tikzpicture}\\
&=&\kappa_b \sum_{c}  \sum_{\nu =1}^{N^a_{bc}}  \sqrt{d_c \over d_a d_b}
\begin{tikzpicture}[baseline={([yshift=-1ex]current bounding box.center)},vertex/.style={anchor=base,
    circle,fill=black!25,minimum size=18pt,inner sep=2pt},scale=0.6]
   \draw[->-=0.7,thick] (0,0) -- (1,1);
   \draw[-<-=0.7,thick] (2,0) -- (1,1);
    \draw[->-=0.6,thick] (1,1) -- (1,3);
     \draw[->-=0.7,thick] (1,3) -- (0,4);
       \draw[-<-=0.7,thick] (1,3) -- (2,4);
     \node[below] at (0,-0.2) {$a$};
     \node[below] at (2,0) {$b$};
      \node[above] at (0,4) {$a$};
     \node[above] at (2,4) {$b$};
      \node[right] at (1,2) {$c$};
      \node[above] at (0.85,0.4) {\footnotesize$+$};
      \node[below] at (0.85,3.6) {\footnotesize$+$};
      
      \node[left] at (1,1.1) {\scriptsize$\overline\nu$};
       \node[left] at (1,2.8) {\scriptsize$\nu$};
    \end{tikzpicture}
\eea
where in the final step we have used (\ref{eq:FFsumidentity}) together with the definition of the FS indicator.

\bibliographystyle{ytamsalpha} 
\def\arxivfont{\rm}
%\baselineskip=.95\baselineskip
\bibliography{ref.bib}

\end{document}